\documentclass[12pt]{article}%
\usepackage[nosort]{cite}
\usepackage{graphicx}
\usepackage{multicol}
\usepackage{amsfonts}
\usepackage{amssymb}
\usepackage{amsmath}
\usepackage{heck}
\usepackage{afterpage}
\usepackage{setspace}
\usepackage{verbatim}
\usepackage{color}
\usepackage{longtable}
\usepackage{float}
\usepackage{subcaption}
\usepackage{epsfig}
\usepackage{enumerate}
\usepackage{epstopdf}
\usepackage[enableskew, vcentermath]{youngtab}
\usepackage{adjustbox}
\usepackage{multirow}
\usepackage{tikz}
\usepackage[margin=1in]{geometry}
\usepackage{titletoc}
\usepackage{hyperref}
\usepackage{mathdots}
\usepackage{changepage}
\usepackage{xfrac}
\usepackage{color, colortbl}
\usepackage[utf8]{inputenc}
\usepackage{numprint}
\npdecimalsign{.}
\nprounddigits{3}
\usepackage[mathscr]{euscript}
\usepackage{yfonts}
\providecommand{\U}[1]{\protect\rule{.1in}{.1in}}
\pdfoutput=1
\newsavebox{\mysavebox}

\hypersetup{colorlinks,citecolor=black,filecolor=black,linkcolor=black,urlcolor=black}
\usetikzlibrary{decorations.markings}

\numberwithin{equation}{section}

\usetikzlibrary{chains}
\allowdisplaybreaks
\tikzset{node distance=2em, ch/.style={circle,draw,on chain,inner sep=2pt},chj/.style={ch,join},every path/.style={shorten >=4pt,shorten <=4pt},line width=1pt,baseline=-1ex}

\newcommand{\ba}{\begin{eqnarray}}
\newcommand{\ea}{\end{eqnarray}}

\newcommand{\cN}{\mathcal{N}}
\newcommand{\cO}{\mathcal{O}}

\DeclareMathOperator{\SU}{\mathit{SU}}
\DeclareMathOperator{\SO}{\mathit{SO}}

\newcommand{\be}{\begin{equation}}
\newcommand{\ee}{\end{equation}}
\tikzstyle{startstop} = [rectangle, rounded corners, minimum width=3cm, minimum height=1cm,text centered, draw=black, fill=blue!10]
\tikzstyle{startstop} = [rectangle, rounded corners, minimum width=3cm, minimum height=1cm,text centered, draw=black, fill=blue!10]
\tikzstyle{io} = [trapezium, trapezium left angle=70, trapezium right angle=110, minimum width=3cm, minimum height=1cm, text centered, draw=black, fill=blue!30]
\tikzstyle{process} = [rectangle, minimum width=3cm, minimum height=1cm, text centered, draw=black, fill=orange!30]
\tikzstyle{decision} = [diamond, minimum width=3cm, minimum height=1cm, text centered, draw=black, fill=green!30]
\tikzstyle{arrow} = [thick,->,>=stealth]
\tikzset{->-/.style={decoration={
  markings,
  mark=at position #1 with {\arrow[scale=2.4]{>}}},postaction={decorate}}}
\makeatletter \@addtoreset{equation}{section} \makeatother

\definecolor{LightCyan}{rgb}{0.88,1,1}

\begin{document}

\date{August 2018}

\title{Nilpotent Networks and 4D RG Flows}

\institution{PENN}{\centerline{${}^{1}$Department of Physics and Astronomy, University of Pennsylvania, Philadelphia, PA 19104, USA}}

\institution{UNC}{\centerline{${}^{2}$Department of Physics, University of North Carolina, Chapel Hill, NC 27599, USA}}

\authors{Fabio Apruzzi\worksat{\PENN , \UNC}\footnote{e-mail: {\tt fabio.apruzzi@gmail.com}},
Falk Hassler\worksat{\PENN,\UNC}\footnote{e-mail: {\tt fhassler@unc.edu}},\\[4mm]
Jonathan J. Heckman\worksat{\PENN}\footnote{e-mail: {\tt jheckman@sas.upenn.edu}},
and Thomas B. Rochais\worksat{\PENN}\footnote{e-mail: {\tt thb@sas.upenn.edu}}}

\abstract{Starting from a general $\mathcal{N} = 2$ SCFT, we study the network of
$\mathcal{N} = 1$ SCFTs obtained from relevant deformations by nilpotent mass parameters.
We also study the case of flipper field deformations where the mass parameters are promoted to a
chiral superfield, with nilpotent vev. Nilpotent elements of semi-simple algebras
admit a partial ordering connected by a corresponding directed graph. We
find strong evidence that the resulting fixed points are connected by a similar network of
4D RG flows. To illustrate these general concepts, we also present a full list of nilpotent deformations
in the case of explicit $\mathcal{N} = 2$ SCFTs, including the case of a single D3-brane probing a
$D$- or $E$-type F-theory 7-brane, and 6D $(G,G)$ conformal matter compactified on a $T^2$, as
described by a single M5-brane probing a $D$- or $E$-type singularity. We also
observe a number of numerical coincidences of independent interest, including a collection of
theories with rational values for their conformal anomalies, as well as a surprisingly nearly constant
value for the ratio $a_{\mathrm{IR}} / c_{\mathrm{IR}}$ for the entire network of flows associated with a given UV $\mathcal{N} = 2$ SCFT.
The $\texttt{arXiv}$ submission also includes the full dataset of theories which can be accessed with
a companion \texttt{Mathematica} script.}

\maketitle

\setcounter{tocdepth}{2}

\tableofcontents


\newpage

\section{Introduction \label{sec:INTRO}}

Conformal field theories (CFTs)\ play a central role in physics. Deformations
which drive one fixed point to another also provide important insights into
more general quantum field theories.

Even so, it is often difficult to establish the existence of fixed points, let alone
determine deformations to new ones. Common techniques include combinations of methods
from supersymmetry, string compactification, holography, and / or the conformal bootstrap.

Part of the issue with understanding relevant perturbations of CFTs is that
(by definition)\ they grow deep in the infrared. From this
perspective, it is perhaps not surprising that comparatively short flows where
there is only a small drop in the number of degrees of freedom (as measured by
various anomalies)\ are often easier to study.

One way to understand long flows is to break them up into a sequence of nearby
short flows. This strategy has recently been used to make surprisingly sharp
statements in the study of 6D\ supersymmetric
RG\ flows \cite{Heckman:2015ola, Cordova:2015fha, Heckman:2015axa,
Cremonesi:2015bld, Heckman:2016ssk, Mekareeya:2016yal, Heckman:2018pqx}. In particular, the
mathematical partial ordering of nilpotent orbits in flavor symmetry algebras
automatically defines a hierarchy of 6D\ RG\ flows \cite{Heckman:2016ssk,
Mekareeya:2016yal, Heckman:2018pqx}. For a recent review of 6D superconformal field theories,
see reference \cite{Heckman:2018jxk}.

In this paper we ask whether the same mathematical
structure leads to an improved understanding of RG\ flows in lower-dimensional
systems. The specific class of theories we study are $\mathcal{N}=1$
deformations of 4D $\mathcal{N}=2$ SCFTs. For the UV\ theories under
consideration, we assume the existence of a flavor symmetry algebra 
$\mathfrak{g}_{\text{flav}}$, which a priori could be composed of 
several simple factors:%
\begin{equation}
\mathfrak{g}_{\text{flav}}=\mathfrak{g}_{\text{flav}}^{(1)}\times
...\times\mathfrak{g}_{\text{flav}}^{(n)}%
\end{equation}
for $\mathfrak{g}_{\text{flav}}^{(i)}$ a simple Lie algebra. Associated with
this flavor symmetry are a collection of mass parameters $m_{\text{adj}}$, and 
corresponding dimension two mesonic operators $\mathcal{O}_{\text{adj}}$ 
transforming in the adjoint representation\footnote{More canonically, one can
view the mass parameters as elements in the dual $\mathfrak{g}_{\text{flav}}^{\ast}$.},
which can be used to activate relevant deformations to new conformal fixed
points in the IR via superpotential deformations:%
\begin{equation}
\delta W=\text{Tr}_{\mathfrak{g}_{\text{flav}}}\left(  m_{\text{adj}}%
\cdot\mathcal{O}_{\text{adj}}\right)  . \label{plainmass}%
\end{equation}
Promoting the mass parameters to a chiral superfield $M_{\text{adj}}$
transforming in the adjoint representation of $\mathfrak{g}_{\text{flav}}$, we can
consider the related deformations associated with expanding around background
vevs for these ``flipper fields:''
\begin{equation}
\delta W=\text{Tr}_{\mathfrak{g}_{\text{flav}}}\left(  (m_{\text{adj}%
}+M_{\text{adj}})\cdot\mathcal{O}_{\text{adj}}\right)  , \label{MSdef}%
\end{equation}
where now, we interpret the mass deformation $m_{\text{adj}}=\left\langle
M_{\text{adj}}\right\rangle $ as a background vev.

The key point we shall be exploiting in this work is that given a flavor
symmetry Lie algebra $\mathfrak{g}_{\text{flav}}$, there is a partial ordering
available for nilpotent elements, as defined by the orbit of an element under
the adjoint action of the algebra. Given nilpotent elements $\mu,\nu
\in\mathfrak{g}_{\text{flav}}$, we say that $\mu\prec\nu$ when Orbit$(\mu
)\subset\overline{\text{Orbit}(\nu)}$. Since the mass parameters
$m_{\text{adj}}$ transform in the adjoint, this sets up a conjectural relation
between relevant deformations, as in lines (\ref{plainmass}) and (\ref{MSdef})
and 4D\ RG\ flows. Intuitively, as the size of the orbit increases, the number
of degrees of freedom which pick up a mass also increases, leading to a longer
flow into the infrared.

Another quite interesting feature of nilpotent mass deformations is that at
least in the case where we have a plain mass deformation as in line
(\ref{plainmass}), the Seiberg-Witten curve of the UV\ $\mathcal{N}=2$ theory
descends to an $\mathcal{N} = 1$ curve of the deformed $\mathcal{N}=1$ theory which
fixes the relative scaling dimensions of various operators \cite{Heckman:2010qv}.
The fact that it is still singular provides evidence of an $\mathcal{N} = 1$ fixed point.

One of our aims in this work will be to provide substantial evidence that this network
of nilpotent orbits defines a corresponding hierarchy of 4D\ RG\ flows. For the most part,
this involves a mild generalization of the procedure proposed in \cite{Heckman:2010fh}, studied in detail in
\cite{Heckman:2010qv} (see also \cite{Cecotti:2010bp}) and further extended in references \cite{Gadde:2013fma,
Agarwal:2013uga, Agarwal:2014rua, Agarwal:2015vla, Maruyoshi:2016tqk, Maruyoshi:2016aim, Agarwal:2016pjo},
and applied in various model building contexts in references \cite{Heckman:2011hu, Heckman:2011bb, Heckman:2012nt, Heckman:2012jm, Heckman:2015kqk, DelZotto:2016fju}.

The appearance of a nilpotent element $\mu$ implies the existence of
an $\mathfrak{su}(2)\subset\mathfrak{g}_{\text{flav}}$ subalgebra, with
generators $\mu$, $\mu^{\dag}$ and $[\mu,\mu^{\dag}]$. Labelling the
associated generator of the Cartan subalgebra for this$\mathfrak{\ su}(2)$
subalgebra as $T_{3}$, the infrared R-symmetry is given by a linear combination of the form
(see e.g. \cite{Heckman:2010qv}):%
\begin{equation}
R_{\mathrm{IR}}=R_{\mathrm{UV}}+\left(  \frac{t}{2}-\frac{1}{3}\right)  J_{\mathcal{N}=2}%
-tT_{3}+\underset{i}{\sum}t_{i}F_{i},
\end{equation}
where $R_{\mathrm{UV}}$ and $R_{\mathrm{IR}}$ respectively denote the UV\ and IR\ R-symmetry
(treated as an $\mathcal{N}=1$ theory), $J_{\mathcal{N}=2}$ is an additional
$U(1)$ symmetry which is always present in an $\mathcal{N}=2$ SCFT when
interpreted as an $\mathcal{N}=1$ theory. The last set of terms refers to the
possibility of additional $U(1)$'s, including those which emerge in the
infrared. The IR\ R-symmetry is then fixed via the procedure of a-maximization
over the parameters $t$ and $t_{i}$, as in reference
\cite{Intriligator:2003jj}.\footnote{In practice it is
often necessary to make additional assumptions about
these emergent symmetries to actually carry out concrete calculations.}

In the absence of these emergent $U(1)$'s, we find strong evidence that the
partially ordered set defined by the nilpotent elements of a Lie algebra
exactly aligns with the corresponding hierarchy of 4D\ RG\ flows. For example,
the conformal anomalies $a_{\mathrm{IR}}$ and $c_{\mathrm{IR}}$ decrease along such
trajectories, and anomalies involving flavor currents (with generators
suitably normalized) also decrease along such flows.

Far more non-trivial is that \textit{even in the presence} of emergent $U(1)$'s, there
is still such a partial ordering of 4D theories, as dictated by the nilpotent
cone of the Lie algebra. This is considerably more subtle and requires a case
by case analysis. For this reason, we focus on explicit examples.

One class of theories already studied in \cite{Heckman:2010qv} for plain mass deformations,
and with some masses promoted to chiral superfields in \cite{Maruyoshi:2016tqk, Maruyoshi:2016aim}
involves nilpotent mass deformations of the $\mathcal{N}=2$ theories defined by a D3-brane
probing an F-theory 7-brane with constant axio-dilaton. This includes the
$H_{0}$, $H_{1}$, $H_{2}$ Argyres-Douglas theories \cite{Argyres:1995jj, Argyres:1995xn},
the $E_{6}$, $E_{7}$, $E_{8}$ Minahan Nemeschansky theories \cite{Minahan:1996fg, Minahan:1996cj},
and $\mathcal{N}=2$ $SU(2)$ gauge theory with four flavors and corresponding $SO(8)$ flavor symmetry
(namely $D_{4}$) \cite{Seiberg:1994aj}. The string theory interpretation of nilpotent deformations
is also quite interesting, as they are associated with T-brane configurations
of 7-branes (see e.g. \cite{Aspinwall:1998he, Donagi:2003hh, Cecotti:2010bp, Anderson:2013rka, Collinucci:2014taa, Collinucci:2014qfa, Bena:2016oqr, Marchesano:2016cqg, Anderson:2017rpr, Bena:2017jhm, Marchesano:2017kke, Cvetic:2018xaq}),
namely they leave intact the Weierstrass model of the associated F-theory geometry,
but nevertheless deform the physical theory.

Here, we systematically study all possible nilpotent deformations for the $D$-
and $E$-series theories, systematically sweeping out the corresponding network
of 4D\ RG\ flows (we do not consider the $H$-series in any detail since they have only a few
nilpotent deformations). An interesting feature of these examples is that only the
Coulomb branch operator sometimes appears to drop below the unitarity bound,
and even this happens only for the largest nilpotent orbits. In such cases, we
see no evidence that the fixed point does not exist (since the underlying
geometry is still singular), and instead find it most plausible that the Coulomb
branch operator decouples as a free field, with a corresponding emergent $U(1)$ acting on only
this operator, as per the procedure advocated in \cite{Kutasov:2003iy, Intriligator:2003mi}.

We also study nilpotent mass deformations of
4D\ $\mathcal{N}=2$ conformal matter, namely the compactification of
6D\ conformal matter \cite{DelZotto:2014hpa, Heckman:2014qba} on a $T^{2}$. Here, we
consider the case where there is a
$G_{L}\times G_{R}$ flavor symmetry with $G_{L}=G_{R} = G$ given by $SO(8)$, $E_{6}$,
$E_{7}$, or $E_{8}$. The 4D\ anomaly polynomials for
these theories were computed in \cite{Ohmori:2015pua, Ohmori:2015pia}.
The Seiberg-Witten and Gaiotto curves for these models are known, both via mirror symmetry
\cite{DelZotto:2015rca}, and via its relation to compactifications of class $\mathcal{S}$
theories \cite{Ohmori:2015pua, Ohmori:2015pia}.

Nilpotent mass deformations of 4D conformal matter involve specifying a pair
of nilpotent elements, one for each flavor symmetry factor. In this case, the
string theory interpretation involves a pair of 7-branes intersecting along
the common $T^{2}$. Such nilpotent deformations involve activating background
values for gauge fields of the corresponding 7-branes.

This already leads to many new $\mathcal{N}=1$ fixed
points and the partial ordering for the product Lie algebra predicts a
corresponding hierarchy of 4D\ fixed points. We present strong evidence that
this is the case, again sweeping over all pairs of nilpotent orbits, and for
each one computing the corresponding values of various IR\ anomalies, checking
there is a corresponding decrease along a given trajectory in the nilpotent cone.

One issue which shows up in these cases is that in sufficiently long flows,
mesonic operators often decouple. This in turn signals that such operators
cannot be used to trigger further flows.
A priori, this could mean that the network of connections in the
nilpotent cone may have links which do not produce 4D RG flows.
Even though we have not found a single example where this actually occurs, we leave
a systematic analysis of this possibility for future work.

\begin{figure}[t!]
\begin{center}
\includegraphics[trim={0cm 0cm 0cm 0cm},clip,scale=0.5]{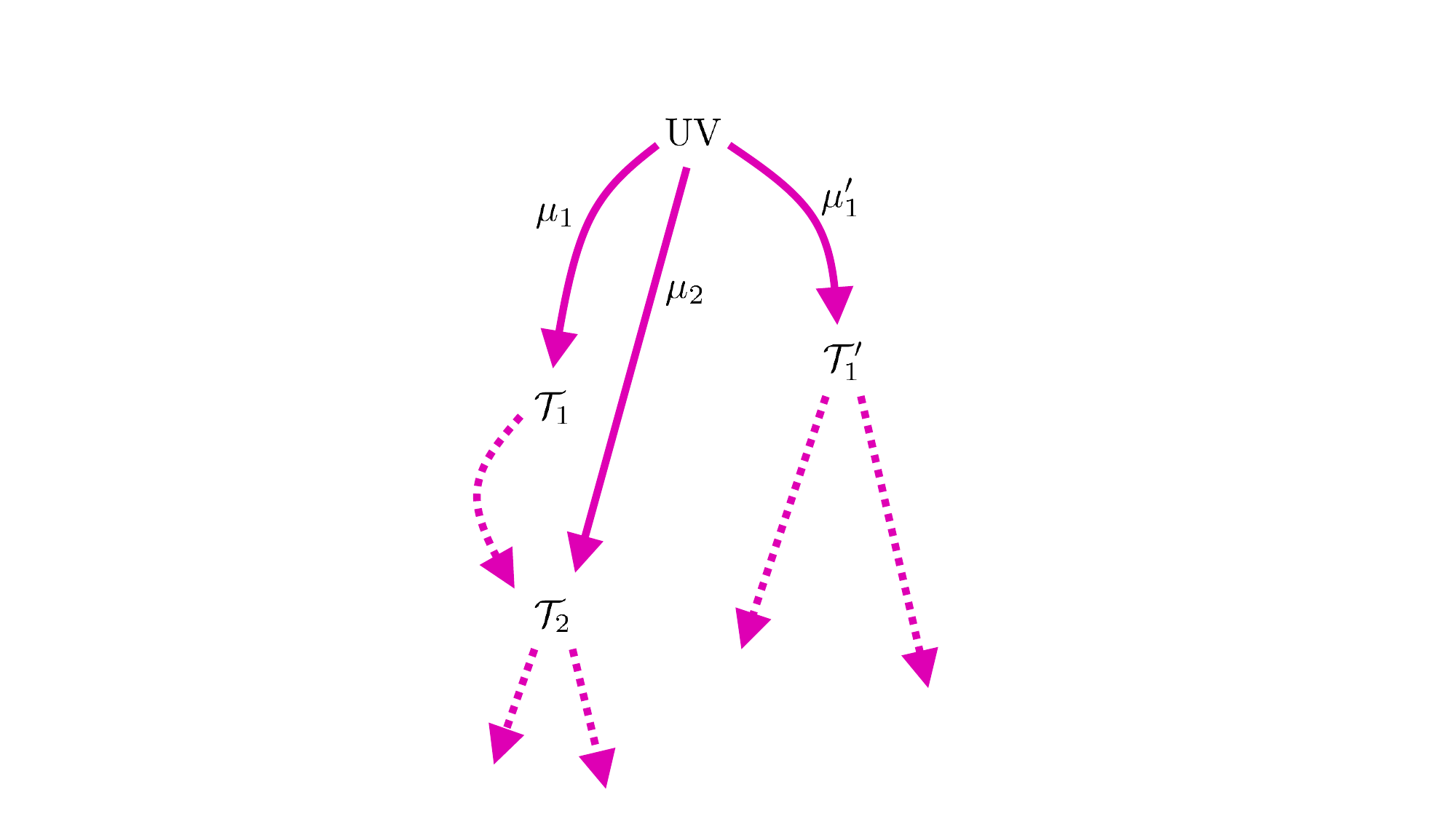}
\end{center}
\caption{Depiction of the network of 4D RG flows generated by elements
of the nilpotent cone. Starting from a UV $\mathcal{N} = 2$ fixed point,
each nilpotent orbit in the flavor symmetry algebra
determines a candidate $\mathcal{N} = 1$ fixed point. Additionally, the network of
connections between nilpotent orbits also motivates the existence of additional
flows between these $\mathcal{N} = 1$ fixed points.}
\label{fig:flowsketch}
\end{figure}

With this set of theories in hand, additional numerical studies are amenable
to treatment, though the list of theories is so large that we have chosen to
collect the full dataset in an accompanying \texttt{Mathematica} package
available for download with our \texttt{arXiv} submission. For example, by
sweeping over all theories, we find several examples of theories where the
conformal anomalies $a_{\mathrm{IR}}$ and $c_{\mathrm{IR}}$ are rational numbers. In some cases
such as reference \cite{Maruyoshi:2016tqk, Maruyoshi:2016aim},
this was interpreted as evidence for an emergent $\mathcal{N}=2$
supersymmetry in the infrared, and we find another example of this type for a
deformation of the $E_{7}$ Minahan-Nemeschansky theory. It is not
clear to us whether there is $\mathcal{N}=2$ enhancement in all cases,
but certainly the list of such rational theories we find suggests additional
structure is present. Another numerical curiosity we observe is that for a given
choice of UV $\mathcal{N} = 2$ SCFT, the value
of the ratio:%
\begin{equation}
\frac{a_{\mathrm{IR}}}{c_{\mathrm{IR}}}\simeq \text{constant} \pm O(1\% - 5\%)
\end{equation}
is nearly constant over all nilpotent deformations, in line with the
observation made in reference \cite{Maruyoshi:2018nod} for a different set of theories.

The rest of this paper is organized as follows. First, in section \ref{sec:GENERAL}
we analyze for a general $\mathcal{N}=2$ theory with flavor
symmetries, the structure of the $\mathcal{N}=1$ theories obtained via both
plain mass deformations and their extension to flipper field deformations. In particular,
we analyze the network of 4D RG flows predicted by the nilpotent cone.
Section \ref{sec:INHERIT} discusses the structure of IR fixed points assuming no
operators decouple, and section \ref{sec:EMERGE} discusses the structure of theories
in the presence of emergent IR symmetries. In section \ref{sec:D3} we discuss nilpotent
deformations of D3-brane probes of $D$- and $E$-type 7-branes and in section \ref{sec:CM} we
discuss nilpotent deformations of 4D $\mathcal{N} = 2$ conformal matter. We
conclude in section \ref{sec:CONC}. Some additional review material, as well as technical details and instructions on
how to use the companion \texttt{Mathematica} files are presented in the Appendices.

\section{Nilpotent Deformations: Generalities \label{sec:GENERAL}}

In this section we discuss some general features of nilpotent mass
deformations of $\mathcal{N}=2$ SCFTs. Throughout, we assume the existence of
a continuous flavor symmetry algebra which may consist of several simple
factors:%
\begin{equation}
\mathfrak{g}_{\mathrm{UV}} \equiv \mathfrak{g}_{\text{flav}}=\mathfrak{g}_{\text{flav}}^{(1)}\times
...\times\mathfrak{g}_{\text{flav}}^{(n)}. \label{flavaflav}%
\end{equation}
We assume either that there are no abelian factors in the UV, or more
generally, that the only non-vanishing anomalies involving flavor symmetry
currents involve precisely two insertions of the same kind (which is automatic
in the traceless non-abelian case). Note that we can then also allow the
appearance of abelian symmetry factors, provided they satisfy this condition.

We assume adjoint valued mass parameters $m_{\text{adj}}$, 
and corresponding dimension two mesonic operators $\mathcal{O}_{\text{adj}}$ 
which serve as coordinates on the Higgs branch of moduli space. 
Note that there could be non-trivial chiral ring relations for these operators,
as can often happen when there is more than one simple Lie algebra factor for $\mathfrak{g}_{\mathrm{UV}}$. Since we
will couch our analysis in terms of basic properties of symmetry breaking patterns,
our analysis will not depend on such detailed knowledge of the UV theory.

It will prove useful to view our $\mathcal{N} = 2$ SCFT as an $\mathcal{N} = 1$ SCFT 
with additional symmetries. Along these lines, we recall that the $\mathcal{N} = 2$ SCFT 
has an $SU(2) \times U(1)$ R-symmetry. Labelling the generator
of the Cartan subalgebra for the $SU(2)$ factor by $I_{3}$ with eigenvalues
$\pm1/2$ in the fundamental representation, and $R_{\mathcal{N} = 2}$ for the $U(1)$
factor normalized so that the complex scalar of a free $\mathcal{N}=2$ vector
multiplet has charge $+2$, the $\mathcal{N}=1$ R-symmetry is given by the
linear combination (see e.g \cite{Tachikawa:2009tt, Heckman:2010qv}):%
\begin{equation}
R_{\mathrm{UV}}=\frac{1}{3}R_{\mathcal{N}=2}+\frac{4}{3}I_{3}.
\end{equation}
There is another linear combination which we can form which is a global
symmetry of the UV\ theory. We label this as:%
\begin{equation}
J_{\mathcal{N}=2}=R_{\mathcal{N}=2}-2I_{3}.
\end{equation}
See table \ref{tab:N2} for the charge assignments of Coulomb branch operators and 
mesonic operators which serve as coordinates on the Higgs branch.
\begin{table}[t]
  \centering
  \begin{tabular}{ |c||c|c|c|c| }
 \hline
 & $\cO_{\text{adj}}$ & $Z_i$  \\  \hline 
 $R_{\mathrm{UV}}$ & $4/3$ & $2/3$ $\Delta_{\mathrm{UV}}(Z_i)$\\  \hline
 $J_{\cN=2}$ & $-2$ & $2\Delta_{\mathrm{UV}}(Z_i)$  \\  \hline
 $R_{\cN=2}$ & $0$ & $2 \Delta_{\mathrm{UV}}(Z_i)$\\  \hline
 $I_{3}$ & $1$ & $0$  \\  \hline
  \end{tabular}
  \caption{Charge assignments for the mesons $\cO_{\text{adj}}$ and Coulomb branch parameters $Z_i$ in the UV theory.}
  \label{tab:N2}
\end{table}

The Higgs branch is parameterized by dimension two operators transforming in
the adjoint representation of $\mathfrak{g}_{\text{flav}} \equiv \mathfrak{g}_{\mathrm{UV}}$,
which we denote by $\mathcal{O}_{\text{adj}}$. The mass parameters $m_{\text{adj}}$ which pair
with these operators transform in the adjoint representation of $\mathfrak{g}%
_{\text{flav}}$. 

We consider both the case of a plain mass deformation:%
\begin{equation}
\delta W_{\text{plain}}=\text{Tr}_{\mathfrak{g}_{\text{flav}}}\left(  m_{\text{adj}}%
\cdot\mathcal{O}_{\text{adj}}\right)  , \label{plainagain}%
\end{equation}
as well as the flipper field deformations
associated with promoting the mass parameters to a dynamical chiral superfield in the adjoint
of the flavor symmetry which mixes with the original interacting theory:%
\begin{equation}
\delta W_{\text{flip}}=\text{Tr}_{\mathfrak{g}_{\text{flav}}}\left(  (m_{\text{adj}%
}+M_{\text{adj}})\cdot\mathcal{O}_{\text{adj}}\right)  .
\end{equation}
We shall often first deal with the case of plain mass deformations, since
flipper field deformations are a mild extension of this case (though the
resulting IR\ physics can be quite different, see e.g. \cite{Gadde:2013fma, 
Agarwal:2015vla, Maruyoshi:2016tqk, Maruyoshi:2016aim}).
An important feature of our analysis is that the general structure of symmetries and anomalies enables us
to give a uniform analysis of RG\ flows for many such relevant deformations.

Though it may be difficult to explicitly construct, we know that the IR\ physics
on the Coulomb branch is controlled by a Seiberg-Witten curve
\cite{Seiberg:1994rs, Seiberg:1994aj}, and
mass deformations enter as flavor symmetry neutral combinations constructed
from the holomorphic Casimir invariants of $\mathfrak{g}_{\text{flav}}$. In
the special case of an $\mathcal{N}=2$ SCFT, all mass deformations have been
switched off and this curve will exhibit singularities, as required to have
massless degrees of freedom at the origin of the Coulomb branch.

We will in particular be interested in nilpotent deformations. For the
classical algebras, these can always be presented in terms of an explicit
nilpotent matrix, which upon conjugation by a complexified symmetry generator
can always be taken to be proportional to a matrix in Jordan normal form. For example,
in $\mathfrak{su}(4)$ we have:%
\begin{equation}
\left[
\begin{array}
[c]{cccc}%
0 & m_{1\overline{2}} & 0 & 0\\
0 & 0 & m_{2\overline{3}} & 0\\
0 & 0 & 0 & m_{3\overline{4}}\\
0 & 0 & 0 & 0
\end{array}
\right]  \sim m \times \left[
\begin{array}
[c]{cccc}%
0 & 1 & 0 & 0\\
0 & 0 & 1 & 0\\
0 & 0 & 0 & 1\\
0 & 0 & 0 & 0
\end{array}
\right]  .
\end{equation}
The labelling scheme for the classical $\mathfrak{su}$, $\mathfrak{sp}$ and
$\mathfrak{so}$ algebras are dictated by its presentation as a direct sum of
nilpotent Jordan blocks. These blocks in turn define a partition of an integer
which we write as $[\mu_{1}^{a_{1}},...,\mu_{k}^{a_{k}}]$ with $\mu
_{1}>...>\mu_{k}>0$ and $a_{i}$ the multiplicity. In the case of
$\mathfrak{su(}N)$, each partition of the integer $N$ defines a nilpotent
orbit. In the case of $\mathfrak{so}(2N)$, there are some additional
restrictions on partitions of $2N$, namely we require every even number in a partition to appear an
even number of times. Similar considerations hold for $\mathfrak{sp}(N)$ and $\mathfrak{so}(2N + 1)$.
In the case of the exceptional algebras, we instead
label the nilpotent orbit by its embedding in some subalgebra of the
larger parent algebra, which is known as the Bala-Carter label.

Now, one of the very interesting features of nilpotent mass deformations is
that all holomorphic Casimir invariants (by definition)\ must vanish, and so
the presentation of the singular geometry is exactly the same as the
$\mathcal{N}=2$ theory. In contrast to the $\mathcal{N}=2$ case, however, this
does not mean it is possible to read absolute scaling dimensions of operators
from the curve (see reference \cite{Argyres:1995xn} for the analysis of $\mathcal{N}=2$
theories), but instead only the relative scaling dimensions of operators \cite{Heckman:2010qv}.
Nevertheless, the appearance of a singular curve provides one indication
that we are still dealing with a conformal field theory, albeit one with
reduced supersymmetry.

Assuming the existence of such a fixed point, there is a partial ordering of
nilpotent orbits which suggests a physical ordering of theories. Given a pair
of nilpotent elements $\mu$ and $\nu$, we say that $\mu\prec\nu$ when
Orbit$(\mu)\subset\overline{\text{Orbit}(\nu)}$, where the overline denotes
the Zariski closure of the orbit in $\mathfrak{g}_{\text{flav}}$.

Physically, the bigger the orbit, the more degrees of freedom have picked up a
mass. So, it is natural to expect bigger orbits to be deeper in
the infrared. Moreover, for each of the simple Lie algebras, there is a
classification of all possible nilpotent orbits, and the associated
containment relations for these choices. This partially ordered set and its
interconnections defines a directed graph, namely the Hasse diagram of the
nilpotent cone. Returning to our example of explicit nilpotent matrices in $\mathfrak{su}(4)$,
for example, we can see a clear hierarchy:%
\begin{equation}
\left[
\begin{array}
[c]{cccc}%
0 & 0 & 0 & 0\\
0 & 0 & 0 & 0\\
0 & 0 & 0 & 0\\
0 & 0 & 0 & 0
\end{array}
\right]  \prec\left[
\begin{array}
[c]{cccc}%
0 & m_{1\overline{2}} & 0 & 0\\
0 & 0 & 0 & 0\\
0 & 0 & 0 & 0\\
0 & 0 & 0 & 0
\end{array}
\right]  \prec\left[
\begin{array}
[c]{cccc}%
0 & m_{1\overline{2}} & 0 & 0\\
0 & 0 & m_{2\overline{3}} & 0\\
0 & 0 & 0 & m_{3\overline{4}}\\
0 & 0 & 0 & 0
\end{array}
\right]  .
\end{equation}

\begin{figure}[t!]
\begin{center}
\includegraphics[trim={0cm 5cm 0cm 5cm},clip,scale=0.5]{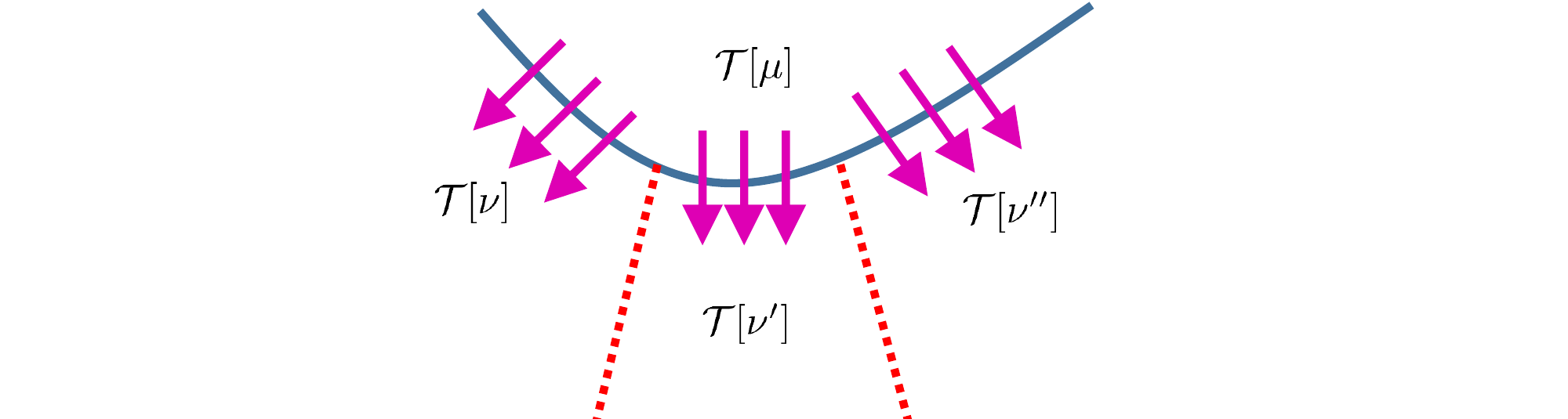}
\end{center}
\caption{Depiction of the deformations from one nilpotent orbit to another. Here, we label a theory
by a choice of nilpotent orbit $\mathcal{T}[\mu]$, and subsequent deformations
deeper down in the nilpotent cone to theories $\mathcal{T}[\nu]$, $\mathcal{T}[\nu^{\prime}]$
and $\mathcal{T}[\nu^{\prime \prime}]$. These physical paths to new orbits are parameterized
by the remnants of the original mesonic operators. An important subtlety with this picture is that
as we proceed from the UV to the IR, various mesonic operators may decouple, severing some of the
candidate links between theories. In explicit examples, however, we have not observed this
pathological behavior.}
\label{fig:slicesketch}
\end{figure}

It is tempting to also interpret this diagram as a
collection of candidate RG\ flows between $\mathcal{N}=1$ fixed points.
Given a sequence of theories $\mathcal{T}_{\mathrm{UV}}\rightarrow
...\rightarrow\mathcal{T}_{i}\rightarrow\mathcal{T}_{i+1}\rightarrow...$, and
associated nilpotent orbits $\varnothing\prec ... \prec \mu_{i}\prec\mu_{i+1} \prec ...$, we can ask
whether there is a flow directly from the intermediate $\mathcal{N}=1$ fixed
point $\mathcal{T}_{i}$ to $\mathcal{T}_{i+1}$. Indeed, we can
subtract the two deformations of the original parent theory:%
\begin{equation}
\delta W_{i \rightarrow i+1}=\text{Tr}_{\mathfrak{g}_{\text{flav}}}\left(  \left(  \mu
_{i+1}-\mu_{i}\right)  \cdot\mathcal{O}_{\text{adj}}\right)  , \label{defdef}%
\end{equation}
which is itself a relevant deformation of the UV\ fixed point theory. Assuming
that the operators necessary to perform such a deformation do not decouple in
theory $\mathcal{T}_{i}$, this strongly indicates that each link in the
directed graph defined by the Hasse diagram also defines a flow between
$\mathcal{N}=1$ fixed points. Carrying out a systematic analysis of this is
somewhat subtle, especially when operators start to decouple in long flows,
but this at least shows that the structure of the nilpotent cone leads to a
rich network of 4D\ RG\ flows.
See figure \ref{fig:slicesketch} for a depiction of the flows generated by these
mesonic operators.

Let us now make more precise the sense in which operator deformations such as
those of line (\ref{defdef}) lead to perturbations of one fixed point to
another. Along these lines, we start in some theory $\mathcal{T}[\mu]$, as
characterized by Orbit$(\mu)$. Given a nilpotent element, the Jacobson-Morozov
theorem guarantees the existence of a homomorphism $\mathfrak{su}%
(2)\rightarrow\mathfrak{g}_{\mathrm{UV}}$, and we label the generators of this algebra
by $T_{3}$, $T_{+}$ and $T_{-}$ in the obvious notation. Decomposing the
adjoint representation into irreducible representations of this $\mathfrak{su}%
(2)$ subalgebra, we get:%
\begin{equation}
V_{\text{adj}}=\underset{j}{%
{\displaystyle\bigoplus}
}(j),
\end{equation}
where we allow each spin $j$ to come with some multiplicity. The highest spin
states of each representation specify the deformations of the nilpotent orbit.
Indeed, a convenient way to compute the dimension of the orbit is via the
formula:%
\begin{equation}
\dim\text{Orbit}(\mu)=\dim V_{\text{adj}}-\dim V_{0}-\dim V_{1/2},
\end{equation}
where here, we have decomposed the states of the adjoint representation under
the $T_{3}$ grading:%
\begin{equation}
V_{\text{adj}}=\underset{s}{%
{\displaystyle\bigoplus}
}V_{s}.
\end{equation}

In the physical theory, these top spin states are distinguished by their role
in the breaking pattern of the flavor symmetry. More formally, we begin with
the $\mathcal{N}=1$ current supermultiplet for the flavor symmetry of the
original theory $\mathcal{J}_{A}$, with $A$ an index in the adjoint
representation. In the unbroken phase, we have the conservation rule:%
\begin{equation}
\overline{D}^{2}\mathcal{J}_{A}=0\text{.}%
\end{equation}
We can also track what becomes of this relation in the broken phase (after the
mass deformation has been switched on). Since $\mathcal{J}_{A}$
transforms in the adjoint representation of $\mathfrak{g}_{\mathrm{UV}}$, we can decompose it into
representations of this $\mathfrak{su}(2)$ subalgebra, so we
label it by a choice of spin $j$, and $T_{3}$ charge $s$, namely
$\mathcal{J}_{j,s}$. In the broken phase, the current is not conserved, since
it is explicitly broken by our mass deformation. We can follow the standard
Noether procedure to see the source of the current non-conservation.
Introducing a \textquotedblleft pion\textquotedblright\ chiral superfield
$\Lambda$ which parameterizes the flavor symmetry generators, we can send:%
\begin{equation}
\mathcal{O}_{\text{adj}}\rightarrow e^{i\Lambda}\mathcal{O}_{\text{adj}%
}e^{-i\Lambda}.
\end{equation}
Then, the superpotential deformation transforms as:%
\begin{equation}
\delta W \rightarrow\text{Tr}_{\mathfrak{g}_{\mathrm{UV}}}(m_{\text{adj}}\cdot
e^{i\Lambda}\mathcal{O}_{\text{adj}}e^{-i\Lambda}),
\end{equation}
so since $m_{\text{adj}}$ can, without loss of generality, be taken to be the
raising operator of the $\mathfrak{su}(2)_{D}$ subalgebra, we learn
that we instead have (see e.g. \cite{Xie:2016hny, Maruyoshi:2016tqk}):%
\begin{equation}
-\frac{1}{4}\overline{D}^{2}\mathcal{J}_{j,s}=\mathcal{O}_{j,s-1}\text{.}
\label{D2J}%
\end{equation}
Note in particular the relative shift in the $T_{3}$ charge $s$.

As explained in \cite{Xie:2016hny, Maruyoshi:2016tqk},
this relation tells us that in the perturbed chiral ring
relations, operators which are not the highest spin states can pair with
components of the current multiplet, forming a long multiplet. Said
differently, in the chiral ring, the operators appearing on the righthand side
of equation (\ref{D2J}) are automatically set to zero (since they appear as $\overline{D}^2$ of something else),
and do not parameterize vacua of the deformed theory. This leaves us with just the highest spin
states, namely $\mathcal{O}_{j,j}$ for the various spin $j$ representations.
Indeed, all other mesons with $\mathcal{O}_{j,s}$ for $s<j$ can be expressed
in terms of the $\mathcal{O}_{j,j}$ using the field equations \cite{Xie:2016hny,
Maruyoshi:2016tqk, Maruyoshi:2016aim, Benvenuti:2017lle}.

In particular, we see that any further deformations of the nilpotent orbit,
namely a candidate flow from theory $\mathcal{T}_{i}$ to a theory
$\mathcal{T}_{i+1}$, will involve precisely these directions.
Provided no such operators decouple as we flow from the
UV\ to the IR, this shows that the directed graph defined by
the Hasse diagram is also a network of RG\ flows. The caveat to this statement
is that it could indeed happen that some operators decouple as we flow from
the UV\ to the IR. Indeed, as we will shortly explain, for a given
$\mathfrak{su}(2)$ representation, the highest spin states
have lowest scaling dimension.

To study this and related issues in more detail, it is of course helpful to
have an explicit example where the underlying
theory is described by a Lagrangian. In subsequent sections we will present a
more general analysis which does not rely on the existence of a Lagrangian.

\subsection{Illustrative Lagrangian Example}

We now illustrate some of the above considerations for a UV $\mathcal{N} = 2$
SCFT which has a Lagrangian description. Most of the other examples
we consider do not admit a convenient presentation of this sort, and so we will instead need to
rely on more general abstract considerations.

The example we consider is $\mathcal{N}=2$ $SU(2)$ gauge theory
with four flavors in the fundamental representation. Some nilpotent mass deformations for this theory
were considered previously in \cite{Heckman:2010qv}, so we refer the interested reader there for
additional background. Our main interest here will be to characterize every
possible nilpotent orbit of the parent $\mathfrak{so}(8)$ flavor symmetry
algebra, and to discuss the explicit structure of the broken symmetry generators.

From the definition of the theory, there is a manifest $\mathfrak{su}(4)$
flavor symmetry which rotates the fields. In $\mathcal{N}=1$ language, we
specify four chiral superfields $q$ in the
$(\mathbf{2},\mathbf{4)}$ of $\mathfrak{su}(2)_{\text{gauge}}\times\mathfrak{su}(4)_{\text{flav}}$, and
four chiral superfields $\widetilde{q}$ in the $(\mathbf{2},\overline
{\mathbf{4}})$ of $\mathfrak{su}(2)_{\text{gauge}}\times\mathfrak{su}(4)_{\text{flav}}$.
There is also a coupling to the adjoint valued chiral superfield
associated with the $\mathfrak{su}(2)_{\text{gauge}}$
$\mathcal{N}=2$ vector multiplet:%
\begin{equation}
W_{\mathcal{N}=2}=\sqrt{2}\widetilde{q}_{\overline{f}}\varphi q^{f},
\end{equation}
where the sum on $f=1,...,4$ runs over the flavors of the model, and
we suppress $\mathfrak{su}(2)_{\text{gauge}}$ indices. This
presentation allows us to explicitly track nilpotent mass deformations
associated with the $\mathfrak{su}(4)$ symmetry algebra, as in reference \cite{Heckman:2010qv}.

Though convenient, this presentation obscures the fact that there is actually
an $\mathfrak{so}(8)$ flavor symmetry. We can assemble the $q$ and
$\widetilde{q}$ into an eight-dimensional representation of $SO(8)$, and
instead treat our field content as a half hypermultiplet transforming in the
$(\mathbf{2},\mathbf{8}_{s})$ of $\mathfrak{su}(2)_{\text{gauge}}\times\mathfrak{so}(8)_{\text{flav}}$.
Labelling the associated holomorphic chiral superfield by $Q^{i}$ with
$i=1,...,8$, we introduce a conjugate spinor of $SO(8)$ $Q_{i}^{c}$ which
canonically pairs with this field so that the superpotential can then be written
as:%
\begin{equation}
W_{\mathcal{N}=2}=\sqrt{2}Q_{i}^{c}\varphi Q^{i},
\end{equation}
where again, we suppress the $\mathfrak{su}(2)_{\text{gauge}}$ indices.
The associated mesons can be written as:%
\begin{equation}
\mathcal{O}^{A}=\left(  \rho^{A}\right)  _{j}^{i}Q_{i}^{c}Q^{j},
\end{equation}
with $\rho^{A}$ the explicit matrix representatives acting on the
$\mathbf{8}_{s}$, and $A$ an adjoint index of $SO(8)$. In this language,
nilpotent mass deformations can be viewed as specific choices for the
$\rho^{A}$ (upon complexification of the flavor symmetry algebra).
\begin{figure}[ptb]
\centering
\begin{tikzpicture}[]
\node[circle,draw=black, fill=white, inner sep=0.25cm,minimum size=5pt] (UV) {UV};
\node[below=of UV] (p1)  {$[1^8]$};
\node[below=of p1] (p2)  {$[2^2,1^4]$};
\node[below=of p2] (p3)  {$[3,1^5]$};
\node[left=of p3]  (p3l) {$[2^4]^I$};
\node[right=of p3] (p3r) {$[2^4]^{II}$};
\node[below=of p3] (p4)  {$[3,2^2,1]$};
\node[below=of p4] (p5)  {$[3^2,1^2]$};
\node[below=of p5] (p6)  {$[5,1^3]$};
\node[left=of p6]  (p6l) {$[4^2]^I$};
\node[right=of p6] (p6r) {$[4^2]^{II}$};
\node[below=of p6] (p7)  {$[5,3]$};
\node[below=of p7] (p8)  {$[7,1]$};
\node[circle,draw=black, fill=white, inner sep=0.25cm,minimum size=5pt] (IR) [below=of p8] {IR};
\node[at={(-6,0|-p1)},anchor=west] (r1) {$r=0$};
\node[at={(-6,0|-p2)},anchor=west] (r2) {$r=1$};
\node[at={(-6,0|-p3)},anchor=west] (r3) {$r=2$};
\node[at={(-6,0|-p4)},anchor=west] (r4) {$r=3$};
\node[at={(-6,0|-p5)},anchor=west] (r5) {$r=4$};
\node[at={(-6,0|-p6)},anchor=west] (r6) {$r=10$};
\node[at={(-6,0|-p7)},anchor=west] (r7) {$r=12$};
\node[at={(-6,0|-p8)},anchor=west] (r8) {$r=28$};
\node[at={(+4,0|-p1)},anchor=west] (a1) {$a_{\mathrm{IR}}=\frac{23}{24}$};
\node[at={(+4,0|-p2)},anchor=west] (a2) {$a_{\mathrm{IR}}=0.797$};
\node[at={(+4,0|-p3)},anchor=west] (a3) {$a_{\mathrm{IR}}=0.710$};
\node[at={(+4,0|-p4)},anchor=west] (a4) {$a_{\mathrm{IR}}=0.652$};
\node[at={(+4,0|-p5)},anchor=west] (a5) {$a_{\mathrm{IR}}=0.608$};
\node[at={(+4,0|-p6)},anchor=west] (a6) {$a_{\mathrm{IR}}=0.474$};
\node[at={(+4,0|-p7)},anchor=west] (a7) {$a_{\mathrm{IR}}=0.451$};
\node[at={(+4,0|-p8)},anchor=west] (a8) {$a_{\mathrm{IR}}=0.366$};
\draw[->, line width=1pt] (p1.south) -- (p2.north);
\draw[->, line width=1pt] (p2.south) -- (p3.north);
\draw[->, line width=1pt] (p2.west) -- (p3l.north);
\draw[->, line width=1pt] (p2.east) -- (p3r.north);
\draw[->, line width=1pt] (p3.south) -- (p4.north);
\draw[->, line width=1pt] (p3l.south) -- (p4.west);
\draw[->, line width=1pt] (p3r.south) -- (p4.east);
\draw[->, line width=1pt] (p4.south) -- (p5.north);
\draw[->, line width=1pt] (p5.south) -- (p6.north);
\draw[->, line width=1pt] (p5.west) -- (p6l.north);
\draw[->, line width=1pt] (p5.east) -- (p6r.north);
\draw[->, line width=1pt] (p6.south) -- (p7.north);
\draw[->, line width=1pt] (p6l.south) -- (p7.west);
\draw[->, line width=1pt] (p6r.south) -- (p7.east);
\draw[->, line width=1pt] (p7.south) -- (p8.north);
\end{tikzpicture}
\caption{The network of RG flows induced by nilpotent plain mass
deformations for $\mathcal{N} = 2$ Super Yang-Mills with $SU(2)$ gauge group and four flavors.
This theory has an $SO(8)$ flavor symmetry in the UV. This network is identical to the Hasse diagram
of the Lie algebra $\mathfrak{so}(8)$. The parameter $r = 2 \mathrm{Tr}_{\mathfrak{so}(8)} (T_3 T_3)$
is the embedding index for the homomorphism $\mathfrak{su}(2) \rightarrow \mathfrak{so}(8)$ defined
by a nilpotent orbit. The value of the
conformal anomaly $a_{\mathrm{IR}}$ decreases, as expected. These flows are determined using the method
described in sections \ref{sec:INHERIT} and \ref{sec:EMERGE}.}
\label{RGflow}%
\end{figure}
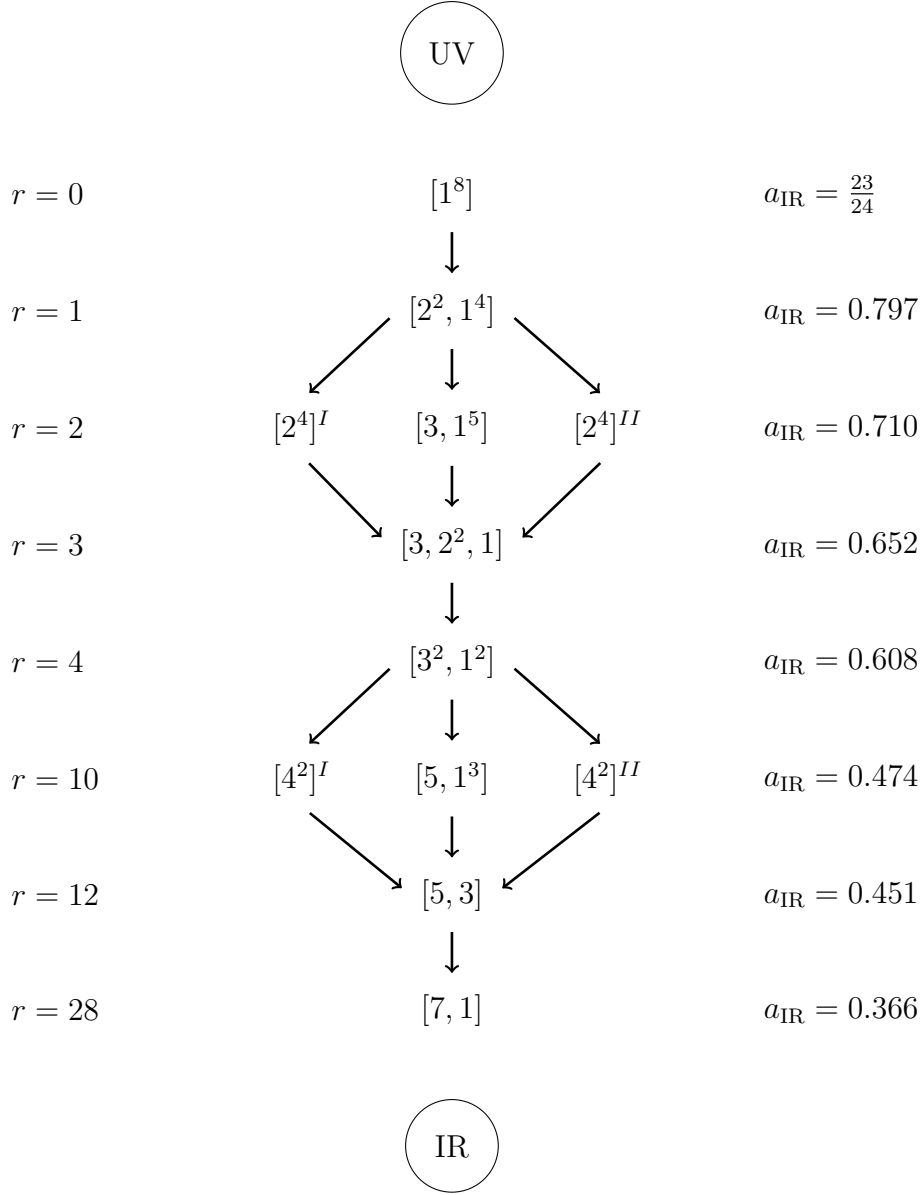

Figure \ref{RGflow} illustrates the resulting network of nilpotent orbits and
RG\ flows in this specific case. We also display the value of $a_{\mathrm{IR}}$ as we
pass from the UV\ to the IR. The specific method used to calculate the
IR\ R-charges is essentially the same as in reference \cite{Heckman:2010qv}, and we will
discuss it in greater detail in sections \ref{sec:INHERIT} and
\ref{sec:EMERGE}.

Another important aspect of this example is that we can also explicitly track
the structure of the broken symmetry currents. To do so, we observe
that the Lagrangian density for the $SO(8)$ theory is, in $\mathcal{N}=1$
language, given by:%
\begin{equation}
\mathcal{L}_{\mathcal{N}=2}=\mathcal{L}_{\text{gauge}}+\int d^{2}\theta
\,d^{2}\bar{\theta}\ Q_{i}^{\dagger}e^{V}Q^{i}+\int d^{2}\theta
\ W_{\mathcal{N}=2}+h.c.\,,
\end{equation}
with $V$ the $SU(2)$ $\mathcal{N} = 1$ vector multiplet. Here, $\mathcal{L}_{\text{gauge}}$
includes the remaining contributions to the $\mathcal{N}=2$ vector multiplet, namely
the kinetic terms for the vector multiplet and adjoint valued chiral superfield.

By varying the action with respect to $Q^{i}$, we obtain the following
equation of motion:
\begin{equation}
- \frac{1}{4} \overline{D}^{2} Q_{i}^{\dagger}e^{V} + 2 \sqrt{2}(Q^{c})_{i}\varphi\, = 0.
\end{equation}
For the theory with no mass deformations, we have the on-shell F-term constraint:
\begin{equation}
(Q^{c})_{i}\varphi=0.
\end{equation}
Using the on shell equations of motion, we observe that the flavor current in the UV:
\begin{equation}
\mathcal{J}_{A}=(\rho_{A})^{j}{}_{i}(Q^{c})_{j}^{\dagger}%
e^{V}Q^{i}\,,
\end{equation}
is actually conserved, namely $\overline{D}^{2}\mathcal{J}_{A}=0$.

Next, we add the superpotential deformation:
\begin{equation}
W_{D}=m^{j}{}_{i}(Q^{c})_{j}Q^{i}.
\end{equation}
The current $\mathcal{J}_{A}$ is no longer conserved, because of this explicit breaking term.
To see what happens, consider following the Noether procedure with flavor transformation:
\begin{equation}
\delta_{\text{flav}}Q^{i}=\epsilon_{A}(\rho^{A})^{i}{}_{j}Q^{j}.
\end{equation}
This yields:
\begin{equation}
-\frac{1}{4}\overline{D}^{2}\mathcal{J}_{A}= (Q^{c})_{i}m^{i}{}_{j}(\rho_{A})^{j}{}_{l}%
Q^{l}\,.
\end{equation}
$m^{i}_{j}$ is the raising operator of the $\mathfrak{su}(2)_{D}$ subalgebra 
and expressing the adjoint index $A$ in terms of spin $j$ and $T_3$ eigenvalue 
results exactly in equation (\ref{D2J}). As already mentioned, an analogous 
procedure also works for non-Lagrangian theories (see
e.g. \cite{Xie:2016hny, Maruyoshi:2016tqk, Maruyoshi:2016aim}).

\section{Inherited Infrared Symmetries\label{sec:INHERIT}}

In this section we turn to an analysis of the 4D $\mathcal{N}=1$ fixed points
generated by nilpotent mass deformations, focussing on the structure of the
symmetries inherited from the original UV $\mathcal{N}=2$ SCFT. Our aim will
be to understand both the structure of the infrared R-symmetry, as well as
global symmetries preserved by a nilpotent mass deformation. Additionally, we
compute the anomalies associated with these symmetries.

One technical assumption we make in this section is that there are no emergent
abelian symmetries. When emergent symmetries are present, as necessarily
occurs when some operators decouple, it is necessary to track which operators
have dimension coming close to the unitarity bound. This requires a more case
by case treatment of the nilpotent deformation in question, and is best
handled by way of explicit cases.

We begin by treating the case of plain mass deformations and then turn to the
case of flipper field deformations. After this, we show that under mild assumptions on the
values of $a_{\mathrm{UV}}$ and $c_{\mathrm{UV}}$ that various numerical quantities are
strictly monotonic along directed paths through the Hasse diagram of nilpotent orbits.

\subsection{Plain Mass Deformations}

Suppose, then, that we introduce a nilpotent mass deformation of a 4D
$\mathcal{N}=2$ SCFT. This initiates an explicit breaking
pattern of the $SU(2) \times U(1)$ R-symmetry of the UV theory, 
as well as well as the flavor symmetries
$\mathfrak{g}_{\mathrm{UV}}$. By definition, there is
a generator $T_{3}$ in the Cartan subalgebra such that the operator
Tr$_{\mathfrak{g}_{\mathrm{UV}}}\left(  \mu\cdot\mathcal{O}_{\text{adj}%
}\right)$ has $T_{3}$ charge $-1$. What this means is that a linear
combination of $T_{3}$ and $J_{\mathcal{N}=2}$ will remain unbroken along the
entire flow to the infrared.

In addition to these symmetries, there are of course all the generators of
$\mathfrak{g}_{\mathrm{UV}}$ which commute with our nilpotent orbit. This defines
another flavor symmetry algebra $\mathfrak{g}_{\mathrm{IR}}$ which may also include
various abelian symmetry factors.

Assuming that we indeed flow to a new fixed point in the infrared with
$\mathcal{N}=1$ supersymmetry, the infrared R-symmetry will be a linear
combination of all available abelian symmetries:%
\begin{equation}
R_{\mathrm{IR}}=R_{\mathrm{UV}}+t_{J}J_{\mathcal{N}=2}-tT_{3}+t_{\text{other}}T_{\text{other}},
\end{equation}
where $T_{\text{other}}$ is shorthand for all other abelian symmetries
inherited from the UV.

Now, for our plain mass deformation to be a relevant perturbation, it follows
that the IR R-charge of this operator deformation is fixed to be $+2$. Since
Tr$_{\mathfrak{g}_{\text{flav}}}\left(  \mu\cdot\mathcal{O}_{\text{adj}%
}\right)  $ has charges $R_{\mathrm{UV}}=+4/3$, $J_{\mathcal{N} = 2}=-2$, $T_{3}=-1$ and is neutral
under $T_{\text{other}}$, we learn that the IR\ R-symmetry is actually
constrained to be:%
\begin{equation}
R_{\mathrm{IR}}=R_{\mathrm{UV}}+\left(  \frac{t}{2}-\frac{1}{3}\right)  J_{\mathcal{N}=2}%
-tT_{3}+t_{\text{other}}T_{\text{other}},
\end{equation}
where to fix the remaining parameters $t$ and $t_{\text{other}}$, we must
resort to a-maximization \cite{Intriligator:2003jj}, namely we calculate the trial value of the
conformal anomaly $a_{\text{trial}}(t,t_{\text{other}})$ as a function of $t$
and $t_{\text{other}}$:%
\begin{equation}
a_{\text{trial}}(t,t_{\text{other}})=\frac{3}{32}\left(  3\text{Tr}R_{\mathrm{IR}}%
^{3}(t,t_{\text{other}})-\text{Tr}R_{\mathrm{IR}}(t,t_{\text{other}})\right)  ,
\end{equation}
and find the local maximum with respect to these parameters.

Since we are assuming the absence of emergent symmetries in the infrared, we
can use anomaly matching to express various IR\ quantities in terms of
UV\ data. In particular, we shall have need to reference the anomalies:%
\begin{align}
a_{\mathrm{UV}}  &  =\frac{3}{32}\left(  3\text{Tr}R_{\mathrm{UV}}^{3}-\text{Tr}R_{\mathrm{UV}}\right) \\
c_{\mathrm{UV}}  &  =\frac{1}{32}\left(  9\text{Tr}R_{\mathrm{UV}}^{3}-5\text{Tr}R_{\mathrm{UV}}\right)
\\
k_{\mathrm{UV}}\times\delta^{AB}  &  =-6\text{Tr}\left(  R_{\mathrm{UV}}J_{\text{flav}}%
^{A}J_{\text{flav}}^{B}\right)  ,
\end{align}
in the obvious notation.

Let us first establish that $t_{\text{other}}$ actually vanishes. To this end,
we note that since we have assumed below line (\ref{flavaflav}) that the
anomalies involving the UV flavor symmetries always involve precisely two
insertions of the \textit{same} flavor symmetry,\footnote{Indeed, recall that
the \textquotedblleft other\textquotedblright\ in $t_{\text{other}}$ is
shorthand for labelling possibly multiple abelian symmetry factors. This means
there could be mixed terms between these factors. If all these abelian factors
descend from a non-abelian symmetry, such mixed anomalies automatically
vanish, but it could a priori still be present for abelian symmetries
inherited from the UV\ theory. This is the main reason the assumption below
line (\ref{flavaflav}) is required.} the only way for $t_{\text{other}}$ to
make an appearance in $a_{\text{trial}}$ is through a mixed anomaly with a
symmetry generator of the $SU(2)\times U(1)$ R-symmetry of the $\mathcal{N}=2$
SCFT. Since the dependence on $t_{\text{other}}$ has only quadratic
dependence, the local maximum necessarily has $t_{\text{other}}=0$. Hence, the
infrared R-symmetry is actually given by the linear combination:%
\begin{equation}
R_{\mathrm{IR}}=R_{\mathrm{UV}}+\left(  \frac{t}{2}-\frac{1}{3}\right)  J_{\mathcal{N}=2}%
-tT_{3},
\end{equation}
with $t$ to be fixed by a-maximization.

This analysis was already carried out in reference \cite{Heckman:2010qv} for a specific class of
deformations, but the generalization to our case follows formally the same
steps. The only change is that now, we need to pay attention to the appearance
of possibly multiple UV symmetry factors\ in:%
\begin{equation}
\mathfrak{g}_{\mathrm{UV}}=\mathfrak{g}_{\mathrm{UV}}^{(1)}\times...\times\mathfrak{g}%
_{\mathrm{UV}}^{(n)},
\end{equation}
so we need to label the $RFF$ anomaly for each such factor:%
\begin{equation}
\text{Tr}\left(  R_{\mathrm{UV}}J_{A_i}^{(i)}J_{B_i}^{(i)}\right)  =-\frac{k_{\mathrm{UV}}^{(i)}}%
{6}\delta_{A_i B_i}.
\end{equation}
Since we can decompose our $T_{3}$ generator as a direct sum for each simple
factor:%
\begin{equation}
T_{3}=T_{3}^{(1)}\oplus...\oplus T_{3}^{(n)}.
\end{equation}
The value of $a_{\text{trial}}(t)$ is given by:%
\begin{equation}
a_{\text{trial}}(t)=\frac{3}{32}\left[  \left(  36a_{\mathrm{UV}}-27c_{\mathrm{UV}}-\frac{9}%
{4}\underset{i=1}{\overset{n}{\sum}}k_{\mathrm{UV}}^{(i)}r^{(i)}\right)  t^{3}%
+(-72a_{\mathrm{UV}}+36c_{\mathrm{UV}})t^{2}+(48a_{\mathrm{UV}}-12c_{\mathrm{UV}})t\right]  , \label{atrialplain}%
\end{equation}
where in obtaining this formula we have used the structure of anomalies as
dictated by the UV$\ \mathcal{N}=2$ theory.
Here, $r^{(i)}$ refers to the embedding index for the generator
$T_{3}^{(i)}$ in $\mathfrak{g}_{\mathrm{UV}}^{(i)}$:%
\begin{equation}
r^{(i)}\equiv 2 \text{Tr}_{\mathfrak{g}_{\mathrm{UV}}^{(i)}}\left(  T_{3}^{(i)}%
T_{3}^{(i)}\right)  ,
\end{equation}
see Appendix \ref{app:EMBED} for details.

The local maximum of $a_{\text{trial}}(t)$ is then given by the critical
point:%
\begin{equation}
t_{\ast}=\frac{4}{3}\times\frac{8a_{\mathrm{UV}}-4c_{\mathrm{UV}}-\sqrt{4c_{\mathrm{UV}}^{2}%
+(4a_{\mathrm{UV}}-c_{\mathrm{UV}})\underset{i=1}{\overset{n}{\sum}}k_{\mathrm{UV}}^{(i)}r^{(i)}}%
}{16a_{\mathrm{UV}}-12c_{\mathrm{UV}}-\underset{i=1}{\overset{n}{\sum}}k_{\mathrm{UV}}^{(i)}r^{(i)}}.
\label{tstarplain}%
\end{equation}
With this in hand, we can evaluate the anomalies of our candidate infrared
fixed point. In the case of the flavor symmetry anomalies, the structure
depends on the remaining flavor symmetry generators associated with each
semi-simple factor, and we denote these unbroken symmetry currents by
$J_{A_{i}}^{(i)}$. In terms of the parameter $t_{\ast}$, the IR\ values of
these anomalies are:%
\begin{align}\label{IRanomalies}
a_{\mathrm{IR}}  &  =\frac{3}{32}\left[  \left(  36a_{\mathrm{UV}}-27c_{\mathrm{UV}}-\frac{9}%
{4}\underset{i=1}{\overset{n}{\sum}}k_{\mathrm{UV}}^{(i)}r^{(i)}\right)  t_{\ast}
^{3}+(-72a_{\mathrm{UV}}+36c_{\mathrm{UV}})t_{\ast}^{2}+(48a_{\mathrm{UV}}-12c_{\mathrm{UV}})t_{\ast}\right] \\
c_{\mathrm{IR}}  &  =\frac{1}{32}\left[  \left(  108a_{\mathrm{UV}}-81c_{\mathrm{UV}}-\frac{27}%
{4}\underset{i=1}{\overset{n}{\sum}}k_{\mathrm{UV}}^{(i)}r^{(i)}\right)  t_{\ast}%
^{3}+(-216a_{\mathrm{UV}}+108c_{\mathrm{UV}})t_{\ast}^{2}+(96a_{\mathrm{UV}}+12c_{\mathrm{UV}})t_{\ast}\right]
\end{align}
and:
\begin{equation}
K_{\mathrm{IR}}^{(i)} = \frac{3}{2}k_{\mathrm{UV}}^{(i)}\times t_{\ast},
\end{equation}
In the above, we have introduced the anomaly coefficient $K_{\mathrm{IR}}^{(i)}$:%
\begin{equation}
\text{Tr}\left(  R_{\mathrm{IR}}J_{A_i}^{(i)}J_{B_i}^{(i)}\right)  =-\frac{K_{\mathrm{IR}}^{(i)}}%
{6}\delta_{A_i B_i},
\end{equation}
where we take the same normalization of all Lie algebra generators as
inherited from the parent UV\ symmetry. In a given simple factor in the IR,
there could be several subalgebras:%
\begin{equation}
\mathfrak{h}_{1}^{(i)}\times...\times\mathfrak{h}_{m_{i}}^{(i)}\subset
\mathfrak{g}_{\mathrm{IR}}^{(i)}\subset\mathfrak{g}_{\mathrm{UV}}^{(i)}\text{,}%
\end{equation}
each with a different embedding index. We can of course take generators
normalized with respect to these unbroken flavor symmetries to define the more
standard quantity via the embedding index:%
\begin{equation}
k_{l_{i} , IR}^{(i)}=\text{Ind}(\mathfrak{h}_{l_{i}}^{(i)}\rightarrow\mathfrak{g}%
_{\mathrm{UV}}^{(i)})\times K_{\mathrm{IR}}^{(i)}\text{.} \label{kvsK}%
\end{equation}
The physically more meaningful quantity is $k_{\mathrm{IR}}^{(i)}$, though it is often
more straightforward to evaluate $K_{\mathrm{IR}}^{(i)}$.

\subsubsection{Operator Scaling Dimensions}

Having determined the infrared R-symmetry, we can now extract the scaling
dimensions for a number of operators. It is helpful to organize this analysis
according to the representation content of the subalgebra $\mathfrak{g}%
_{\mathrm{IR}}\times\mathfrak{su}(2)_{D}$, where $\mathfrak{su}(2)_{D}$ is the 
subalgebra implicitly defined by a choice of nilpotent
orbit. For example, since the mesons transform in the adjoint representation
of $\mathfrak{g}_{\mathrm{UV}}$, there is a corresponding decomposition into
representations:%
\begin{align}
\mathfrak{g}_{\mathrm{UV}}  &  \supset\mathfrak{g}_{\mathrm{IR}}\times\mathfrak{su}%
(2)_{D}\\
\text{adj}(\mathfrak{g}_{\mathrm{UV}})  &  \rightarrow\underset{a}{%
{\displaystyle\bigoplus}
}\left(  R_{(a)},j_{(a)}\right)  ,
\end{align}
where on the righthand side we implicitly sum over irreducible representations
of $\mathfrak{g}_{\mathrm{IR}}\times\mathfrak{su}(2)_{D}$ which appear in the
decomposition of the adjoint. More generally, given operators in some
representation of $\mathfrak{g}_{\mathrm{UV}}$, we can always decompose into
irreducible representations of $\mathfrak{g}_{\mathrm{IR}}\times\mathfrak{su}%
(2)_{D}$.

Supposing then that we have a UV operator transforming in a spin $j$
representation of $\mathfrak{su}(2)_{D}$, we get operators of $T_{3}$ charge
$j,j-1,...,-j$, and we can calculate their scaling dimension in the IR\ theory using
our infrared R-symmetry:%
\begin{equation}
\Delta_{\mathrm{IR}}=\frac{3}{2}\left(  R_{\mathrm{UV}}+\left(  \frac{t_{\ast}}{2}-\frac{1}%
{3}\right)  J_{\mathcal{N}=2}-t_{\ast}T_{3}\right)  .
\end{equation}
In the specific case of a Coulomb branch scalar $Z$, we know that since it has
vanishing $I_{3}$ charge, we have $3R_{\mathrm{UV}}(Z)=J_{\mathcal{N}=2}\left(
Z\right)  $, and $T_{3}(Z)=0$ (as it is neutral under all of $\mathfrak{g}%
_{\mathrm{UV}}$), so we immediately obtain:%
\begin{equation}
\Delta_{\mathrm{IR}}\mathcal{(}Z)=\frac{3}{2}t_{\ast}\times\Delta_{\mathrm{UV}}(Z).
\label{opcoul}%
\end{equation}
In the case of a mesonic operator $\mathcal{O}_{j,s}$ transforming in a spin
$j$ representation of $\mathfrak{su}\left(  2\right)  _{D}$, with
$T_{3}$ charge $s$, the scaling dimension in the IR\ is:%
\begin{equation}
\Delta_{\mathrm{IR}}\left(  \mathcal{O}_{j,s}\right)  =3-\frac{3}{2}t_{\ast}(1+s).
\label{opmes}%
\end{equation}

\subsubsection{Monotonicity}

With these results in place, we now show that various numerical quantities are
indeed monotonic as we proceed to larger orbits in the nilpotent cone. We will
also establish this numerically by \textquotedblleft brute
force\textquotedblright\ when we turn to an analysis of explicit
$\mathcal{N}=2$ theories.

To begin, we recall from reference \cite{Hofman:2008ar, Hofman:2016awc} that there is the
Hofman-Maldacena bound on the ratio $a_{\mathrm{UV}}/c_{\mathrm{UV}}$ for any $\mathcal{N} = 2$ SCFT:
\begin{equation}
\frac{1}{2}\leq\frac{a_{\mathrm{UV}}}{c_{\mathrm{UV}}}\leq\frac{5}{4}. \label{HofMalda}%
\end{equation}
We now use this general bound to establish some monotonicity results for
nilpotent mass deformations.

Now, as we proceed to larger orbits, the size of the corresponding embedding
indices necessarily increases. Introducing the parameter:%
\begin{equation}
\mathcal{K}\equiv\underset{i=1}{\overset{n}{\sum}}k_{\mathrm{UV}}^{(i)}r^{(i)},
\end{equation}
we observe that this quantity always increases as we proceed down a directed
path in the Hasse diagram. To establish various monotonicity results, it thus
suffices to evaluate their response as we vary $\mathcal{K}$.

First of all, we can consider the parameter $t_{\ast}$ given by equation (\ref{tstarplain}),
treated as a function of $\mathcal{K}$. If we introduce the Hofman-Maldacena
bounds, as well as the constraints from unitarity 
$a_{\mathrm{UV}},c_{\mathrm{UV}},k_{\mathrm{UV}}^{(i)}>0$, we immediately find (as can be checked
explicitly using \texttt{Mathematica}) that the derivative:%
\begin{equation}
\frac{\partial t_{\ast}}{\partial\mathcal{K}}<0,
\end{equation}
so in particular, $t_{\ast}$ always decreases along a flow. Moreover, since
the Coulomb branch operators are all proportional to $t_{\ast}$, we also learn
that these dimensions are also always strictly decreasing.

One can also perform a similar analysis for the parameter $a_{\mathrm{IR}}$ as a
function of $\mathcal{K}$. In addition to the numerical bounds already introduced,
we also require $t_{\ast}>0$, which in turn requires $16a_{\mathrm{UV}}-12c_{\mathrm{UV}}%
-\mathcal{K}>0$. Curiously enough, we find that in order for this quantity to
decrease monotonically, we need to impose a slightly stronger condition than
that of line (\ref{HofMalda}) for the lower bound:%
\begin{equation}
\frac{3}{4}\leq\frac{a_{\mathrm{UV}}}{c_{\mathrm{UV}}}\leq\frac{5}{4}. \label{sharpee}%
\end{equation}

The most conservative interpretation of this sharper requirement is that as we
pass to larger orbits, we should expect some operators to decouple, in
which case the expressions used for $t_{\ast}$ and $a_{\mathrm{IR}}$ would need to be
modified anyway. We will indeed see examples of this type, though we hasten to
add that in the explicit models we consider, the sharper condition of line
(\ref{sharpee}) is actually satisfied.

\subsection{Flipper Field Deformations}

Having dealt with the case of plain mass deformations, we now turn to
flipper field deformations of an $\mathcal{N}=2$ SCFT. Recall that this
involves promoting the mass parameters of the $\mathcal{N}=2$ theory to an
adjoint valued chiral superfield, and switching on a background vev:%
\begin{equation}
\delta W=\text{Tr}_{\mathfrak{g}_{\text{flav}}}\left(  (m_{\text{adj}%
}+M_{\text{adj}})\cdot\mathcal{O}_{\text{adj}}\right)  .
\end{equation}
Again, we confine our analysis to the case where this vev is a nilpotent mass deformation.

Since we are activating a breaking pattern which is identical to the case of
the plain mass deformation, much of the analysis of the previous section will
carry over unchanged. The primary issue is that now, we need to track the
additional modifications to the infrared R-symmetry which come from having
these additional fields transforming in the adjoint representation.

From the perspective of the UV theory, we have two decoupled SCFTs, namely the
original $\mathcal{N}=2$ fixed point, and a decoupled free chiral multiplet.
Consequently, there is a $U(1)$ flavor symmetry with generator $T_{\text{flip}}$ which
acts on each flipper field, so that it has charge $+1$.
The trial infrared R-symmetry is then a general linear combination of the
form:
\begin{equation}
R_{\mathrm{IR}}^{\text{flip}}(t) = R_{\mathrm{IR}}^{\text{plain}}(t)+t_{\text{flip}}T_{\text{flip}}
\end{equation}
where we have also left implicit the sum over all flippers.
Here, the trial infrared R-symmetry in the case of a plain mass deformation
is:
\begin{equation}
R_{\mathrm{IR}}^{\text{plain}}(t)=R_{\mathrm{UV}}+\left(  \frac{t}{2}-\frac{1}{3}\right)
J_{\mathcal{N}=2}-tT_{3}.
\end{equation}

Now, upon decomposing into representations of $\mathfrak{su}(2)_{D}$, we see that
all flipper fields will deform the theory via operators such as $M_{j,-s}%
\mathcal{O}_{j,s}$. If we first activate the plain mass deformation, and then
couple to the flipper fields, we see that since the operators $\mathcal{O}_{j,j}$
with the highest spin have the lowest scaling dimension, then these are the
operators which actually drive a new flow \cite{Maruyoshi:2016tqk, Maruyoshi:2016aim}. 
For this to be so, we require a constraint on the infrared R-charge assignments 
(see e.g. \cite{Gadde:2013fma, Benvenuti:2017lle}):%
\begin{equation}
R_{\mathrm{IR}}(M_{j,-j})+R_{\mathrm{IR}}(\mathcal{O}_{j,j})=2, \label{flippar}%
\end{equation}
so the new trial IR\ R-symmetry is:%
\begin{equation}
R_{\mathrm{IR}}^{\text{flip}}(t)=R_{\mathrm{IR}}^{\text{plain}}(t)+\left(  t-\frac{2}{3}\right)
T_{\text{flip}}.
\end{equation}

We can also calculate the new trial $a_{\text{trial}}^{\text{flip}}(t)$ by
breaking up the trace over states into those coming from the original
$\mathcal{N}=2$ theory, and those coming from the flipper fields which actually
participate in the flow. Doing so, we get:%
\begin{equation}
a_{\mathrm{trial}}^{\mathrm{flip}}\left(  t\right)  =a_{\mathrm{trial}%
}^{\text{plain}}(t)+\sum_{j_{(a)}}{\left[  \frac{3}{32}\left(  3\left(
R_{\mathrm{IR}}^{\text{flip}}(M_{j_{(a)},-j_{(a)}})-1\right)  ^{3}-\left(
R_{\mathrm{IR}}^{\text{flip}}(M_{j_{(a)},-j_{(a)}})-1)\right)  \right)
\right]  ,}%
\end{equation}
where in the first term, $a_{\text{\textrm{trial}}}^{\text{plain}}(t)$ is
the same quantity as in line (\ref{atrialplain}), and in the second set of
terms, we sum over all highest spin states which appear in the branching rules
for the $\mathfrak{su}(2)_{D}$ subalgebra. The R-charge for each
such flipper field is evaluated with respect to the original R-symmetry of the
plain mass deformation case, namely:%
\begin{equation}
R_{\mathrm{IR}}^{\text{plain}}(M_{j_{(a)},-j_{(a)}})=\frac{2}{3}%
+j_{(a)}\times t.
\end{equation}

Maximizing over the parameter $t$ appearing in $a_{\mathrm{trial}%
}^{\mathrm{flip}}\left(  t\right)  $, we again obtain the infrared R-symmetry,
and can read off the scaling dimensions of operators, much as before. By a
similar token, we can also read off the new value of the conformal anomaly
$c_{\mathrm{IR}}^{\mathrm{flip}}$. Collecting these expressions here, we have \cite{Giacomelli:2017ckh}:%
\begin{align}
a_{\mathrm{IR}}^{\mathrm{flip}}  &  =a_{\mathrm{IR}}^{\text{plain}}(t_{\ast
})+\sum_{j_{(a)}}{\left[  \frac{3}{32}\left(  3\left(  R_{\mathrm{IR}%
}^{\text{flip}}(M_{j_{(a)},-j_{(a)}})-1\right)  ^{3}-\left(  R_{\mathrm{IR}%
}^{\text{flip}}(M_{j_{(a)},-j_{(a)}})-1)\right)  \right)  \right]  }\\
c_{\mathrm{IR}}^{\mathrm{flip}}  &  =a_{\mathrm{IR}}^{\text{plain}}(t_{\ast
})+\sum_{j_{(a)}}{\left[  \frac{1}{32}\left(  9\left(  R_{\mathrm{IR}%
}^{\text{flip}}(M_{j_{(a)},-j_{(a)}})-1\right)  ^{3}-5\left(
R_{\mathrm{IR}}^{\text{flip}}(M_{j_{(a)},-j_{(a)}})-1)\right)  \right)
\right]  ,}%
\end{align}
in the obvious notation.

With the infrared R-symmetry in hand, we can also evaluate the new anomalies
involving the flavor symmetry. Since the flipper fields also transform in
irreducible representations of $\mathfrak{g}_{\mathrm{IR}}$, the IR\ flavor symmetry, we need to
take into account the specific branching rules associated with the
decomposition of the adjoint representation. With notation as in line
(\ref{kvsK}), we have:
\begin{equation}
k_{\mathrm{IR}}(\mathfrak{h}_{l_i}^{(i)})=\text{Ind}(\mathfrak{h}_{l_i}%
^{(i)}\rightarrow\mathfrak{g}_{\mathrm{UV}}^{(i)})\times K_{\mathrm{IR}}(\mathfrak{g}%
_{\mathrm{UV}})+6\sum_{j_{(a)}}(1-(1 - T_{3}(M_{j_{(a)},-j_{(a)}}))t_{\ast
})\mathrm{Ind}(\rho_{a}(\mathfrak{h}_{l_i}^{(i)})).
\end{equation}
Here, $\mathrm{Ind}(\rho_{a}(\mathfrak{h}_{l}^{(i)}))$ indicates the
index of the representation associated with a given flipper field for the
flavor symmetry algebra $\mathfrak{h}_{l_i}^{(i)}$.

Much as in the case of the plain mass deformations, we can read off the
scaling dimensions of our operators. The operator scaling dimensions for the
Coulomb branch scalars and mesonic operators are basically the same as in
lines (\ref{opcoul}) and (\ref{opmes}) except that now we use a modified value
for $t_{\ast}$ due to the coupling to flipper fields. In the case of the
flipper fields, we can read off the scaling dimensions of those that actually
participate in a flow via equation (\ref{flippar}). For those flipper fields which
do not actually participate in a flow, we instead have a collection of
decoupled free fields. In what follows, we shall ignore these contributions,
focussing exclusively on the interacting fixed point.

\section{Emergent Symmetries and Operator Decoupling\label{sec:EMERGE}}

In our analysis so far, we have assumed that there are no emergent symmetries
in the infrared. Our aim in this section will be to discuss some general
features of when to expect emergent symmetries in the case of nilpotent mass
deformations. We turn to specific UV\ theories in the following sections.
Turning the discussion around, the mathematical ordering of nilpotent orbits
provides some helpful clues on the nature of these candidate fixed points.

Now, one way such emergent symmetries can show up is when various operators
start to decouple. Assuming that a fixed point is
really present, if we assume the absence of emergent symmetries and find
the pathological behavior that some operator has dimension below the
unitarity bound, then it is an indication
that this operator has actually decoupled.
The minimal procedure of reference \cite{Kutasov:2003iy}
prescribes that we introduce an additional $U(1)$ flavor symmetry which only
acts on the offending operator. From our starting point of an $\mathcal{N}=2$
theory, the main thing we will be able to check is the scaling dimension of
the Coulomb branch and mesonic operators of the UV\ parent theory.

Another related possibility is that the IR\ theory actually enhances to an
$\mathcal{N}=2$ supersymmetric theory in the infrared. This can occur, for
example, in the case of flipper field deformations \cite{Maruyoshi:2016tqk, Maruyoshi:2016aim},
and recently a set of general sufficient conditions for such behavior to occur were proposed in
\cite{Giacomelli:2018ziv}. A necessary (but insufficient) condition to have such an enhancement is that the various anomalies of the IR fixed point all become
rational numbers rather than the algebraic numbers present for a more general
nilpotent mass deformation.
There are however known counter-examples that have rational anomalies but no SUSY enhancement to $\cN=2$ \cite{Evtikhiev:2017heo}.

Our plan in this section will be to setup some general diagnostics for
symmetry enhancement in the case of nilpotent mass deformations. First, we
consider the decoupling of Coulomb branch operators, and then we turn to the
decoupling of mesonic operators. After this we discuss some special cases
associated with rational values for the anomalies.
Finally, we discuss some preliminary aspects of how the partial ordering
implied by a Hasse diagram lines up with the physical RG\ flows.

\subsection{Decoupling of Coulomb Branch Operators}

Suppose then, that we perform our initial a-maximization procedure, and,
assuming the absence of any emergent $U(1)$'s, we calculate the scaling
dimension of a Coulomb branch operator $Z$. According to our general formula
from line (\ref{opcoul}), we have:%
\begin{equation}
\Delta_{\mathrm{IR}}\mathcal{(}Z)=\frac{3}{2}t_{\ast}\times\Delta_{\mathrm{UV}}(Z).
\end{equation}
If this yields a value less than one, but we still expect the presence of an
IR\ fixed point, this is a strong indication that this operator has actually
decoupled (and so has dimension exactly one). 
By inspection of our expression for the parameter $t_{\ast}$ we see
that this occurs whenever the embedding index becomes sufficiently large.

Assuming this is the only operator to decouple, it is also straightforward to
calculate the new infrared R-symmetry. Following Appendix B of
\cite{Intriligator:2003mi}, we have:
\begin{align}
a_{\mathrm{IR}}^{\mathrm{new}}(t)  &  =a_{\mathrm{IR}}^{\mathrm{old}}%
(t)+\frac{3}{32}\big[\left(  3\left(  R_{\mathrm{old}}(Z)+t_{Z}-1\right)
^{3}-3\left(  R_{\mathrm{old}}(Z)-1\right)  ^{3}\right) \nonumber\\
&  -\left(  \left(  R_{\mathrm{old}}(Z)+t_{Z}-1\right)  -\left(
R_{\mathrm{old}}(Z)-1\right)  \right)  \big]
\end{align}
for $a$ in the IR. Here, $t_{Z}$ denotes the charge of $Z$ under the emergent
$U(1)$ which only acts on this operator. Performing $a$-maximization with respect to
$t_{Z}$ then yields
\begin{equation}
R_{\mathrm{new}}(Z)\equiv R_{\mathrm{old}}(Z)+t_{Z}=\frac{2}{3}\,.
\end{equation}
At this point, we see that adding the emergent $U(1)$ indeed corrects the
scaling dimension of the offending operator to one, and it decouples.
Substituting in this result, along with the fact that
$R_{\mathrm{old}}(Z)=t\times\Delta_{\mathrm{UV}}(Z)$ implies
\begin{equation}
a_{\mathrm{IR}}^{\mathrm{new}}(t)=a_{\mathrm{IR}}^{\mathrm{old}}(t)-\frac
{3}{32}\left[  3\left(  \Delta_{\mathrm{UV}}(Z)t-1\right)  ^{3}-\left(
\Delta_{\mathrm{UV}}(Z)t-1\right)  \right]  +\frac{1}{48}\,.
\end{equation}
Now, we perform the second part of $a$-maximization by taking the partial
derivative of $a_{\mathrm{IR}}^{\mathrm{new}}(t)$ with respect to $t$ and
setting it equal to zero. For the new value
\begin{align}
  t^{\mathrm{new}}_* &= -\frac{4}{3 \left(48 a_{\mathrm{UV}} - 36 c_{\mathrm{UV}} - 3 k_{\mathrm{UV}} r - 4 \Delta_{\mathrm{UV}}^3\right) } \Big(
  -24 a_{\mathrm{UV}} + 12 c_{\mathrm{UV}} + 3 \Delta_{\mathrm{UV}}^2 \nonumber \\
  &+ \Big\{36 c_{\mathrm{UV}}^2 + 36 a_{\mathrm{UV}} k_{\mathrm{UV}} r - 6 k_{\mathrm{UV}} r \Delta_{\mathrm{UV}} + 48 a_{\mathrm{UV}} \left(-2 + \Delta_{\mathrm{UV}}\right) \left(-1 + \Delta_{\mathrm{UV}}\right) \Delta_{\mathrm{UV}} + \Delta_{\mathrm{UV}}^4 \nonumber \\
    & \ \ - 3 c_{\mathrm{UV}} \left(3 k_{\mathrm{UV}} r + 4 \Delta_{\mathrm{UV}} \left(6 + \left(-6 + \Delta_{\mathrm{UV}}\right) \Delta_{\mathrm{UV}}\right)\right) \Big\}^{1/2} \Big)
\end{align}
we find a maximum of $a_{\mathrm{IR}}^{\mathrm{new}}$. Note that we use the
abbreviation $\Delta_{\mathrm{UV}}$ for $\Delta_{\mathrm{UV}}(Z)$ in this
equation to increase the brevity. One can check that the second derivative of the trial
$a_{\mathrm{IR}}^{\text{new}}(t)$ is indeed negative definite at the critical point, so we do get a local
maximum.

Let us summarize the central charges after decoupling the offending operator:
\begin{align}
a_{\mathrm{IR}}^{\mathrm{new}}  &  =a_{\mathrm{IR}}^{\mathrm{old}%
}(t^{\mathrm{new}}_{\ast})-\frac{3}{32}\left[  3\left(  \Delta_{\mathrm{UV}%
}(Z)t^{\mathrm{new}}_{\ast}-1\right)  ^{3}-\left(  \Delta_{\mathrm{UV}%
}(Z)t^{\mathrm{new}}_{\ast}-1\right)  \right]  +\frac{1}{48}\\
c_{\mathrm{IR}}^{\mathrm{new}}  &  =c_{\mathrm{IR}}^{\mathrm{old}%
}(t^{\mathrm{new}}_{\ast})-\frac{1}{32}\left[  9\left(  \Delta_{\mathrm{UV}%
}(Z)t^{\mathrm{new}}_{\ast}-1\right)  ^{3}-5\left(  \Delta_{\mathrm{UV}%
}(Z)t^{\mathrm{new}}_{\ast}-1\right)  \right]  +\frac{1}{24}\\
K_{\mathrm{IR}}^{\mathrm{new}}  &  =K_{\mathrm{IR}}^{\mathrm{old}%
}(t^{\mathrm{new}}_{\ast})\,,
\end{align}
where $a_{\mathrm{IR}}^{\mathrm{old}}$, $c_{\mathrm{IR}}^{\mathrm{old}}$, and
$K_{\mathrm{IR}}^{\mathrm{old}}$ are the central charges which were computed
without the emergent $U(1)$. We emphasize that $K_{\mathrm{IR}}$ does not receive
any additional contributions besides $K_{\mathrm{IR}}^{\mathrm{old}%
}(t^{\mathrm{new}}_{\ast})$ due to the fact that $Z$ is not charged under the
flavor symmetry. Thus, removing the contribution from such operators does not
directly affect the flavor central charge, just indirectly by modifying the
value of $t_{\ast}$.

\subsection{Decoupling of Mesonic Operators}

Let us now turn to the possible decoupling of mesonic operators. When we turn
to specific examples, we find that this does not occur for the probe D3-brane theories,
but does occur for 4D conformal matter theories.

We first treat the case of plain mass deformations, and then turn to the case
of flipper field deformations. Returning to our general formula for the
operator scaling dimensions (in the absence of emergent $U(1)$'s), we see from
equation (\ref{opmes}) that the scaling dimension of an operator
$\mathcal{O}_{j,s}$ is:%
\begin{equation}
\Delta_{\mathrm{IR}}\left(  \mathcal{O}_{j,s}\right)  =3-\frac{3}{2}t_{\ast}(1+s).
\end{equation}
So, the bigger the spin of the operator under the $\mathfrak{su}%
(2)_{D}$ subalgebra, the smaller the scaling dimension.
This is counteracted to some extent by the decreasing value of $t_{\ast}%
$, though in practice, it is still true that as we descend to larger nilpotent
orbits, more mesonic operators start to decouple. For a given spin $j$
representation of $\mathfrak{su}(2)_{D}$, it is hopefully clear that
the highest spin state with $s=j$ will have lowest candidate scaling
dimension, so if this operator has scaling dimension above the unitarity
bound, the remaining operators in the same $\mathfrak{su}(2)_{D}$
multiplet will also be above the bound.

On the other hand, if the highest spin operator falls below the unitarity bound, we can again posit that
it decouples, with a single emergent $U(1)$ which acts
only on this operator. Now, in
addition to the highest spin operator $\mathcal{O}_{j,j}$, there are
often other values of $s$ in the same multiplet which might also appear to violate
the unitarity bound. Note, however, that via our previous discussion of the
broken flavor symmetry generators and the relation of equation (\ref{D2J}):%
\begin{equation}
-\frac{1}{4}\overline{D}^{2}\mathcal{J}_{j,s}= \mathcal{O}%
_{j,s-1}\text{,}%
\end{equation}
we know that components of the flavor current and the mesons pair up in
long multiplets. As a result, we again only need to apply our procedure
for the \textquotedblleft top spin\textquotedblright\ operators of a given
$\mathfrak{su}(2)_{D}$ multiplet.

Once again, reference \cite{Intriligator:2003mi} tells us that all we need to do is
remove the contribution from the offending operator $\mathcal{O}_{i}$ as follows:
\begin{align}
a_{\mathrm{IR}}^{\mathrm{new}}(t)  &  =a_{\mathrm{IR}}^{\mathrm{old}}%
(t)+\sum_{i}\frac{3}{32}\big[\left(  3\left(  R_{\mathrm{old}}(\mathcal{O}%
_{i})+t_{\mathcal{O}_{i}}-1\right)  ^{3}-3\left(  R_{\mathrm{old}}%
(\mathcal{O}_{i})-1\right)  ^{3}\right) \nonumber\\
&  -\left(  \left(  R_{\mathrm{old}}(\mathcal{O}_{i})+t_{\mathcal{O}_{i}%
}-1\right)  -\left(  R_{\mathrm{old}}(\mathcal{O}_{i})-1\right)  \right)
\big]\,.
\end{align}
Naively, one would take the index $i$ in this equation to run over all mesons which
appear to have dimension below the unitarity bound. However, our discussion of the deformed
symmetry current near line (\ref{D2J}) shows that only the highest spin component of each
$\mathfrak{su}(2)_{D}$ multiplet actually participates in the chiral ring of the IR fixed point.

The procedure of $a$-maximization with respect to
$t_{\mathcal{O}_{i}}$ then yields
\[
R_{\mathrm{new}}(\mathcal{O}_{i})\equiv R_{\mathrm{old}}(\mathcal{O}%
_{i})+t_{\mathcal{O}_{i}}=\frac{2}{3}\,.
\]
Again, we see that all bad $\mathcal{O}_{i}$ decouple. The value of $t_{\ast}$
is determined by $a$-maximization of $a_{\mathrm{IR}}^{\mathrm{new}}(t)$
and the corresponding anomalies are:
\begin{align}
a_{\mathrm{IR}}^{\mathrm{new}}  &  =a_{\mathrm{IR}}^{\mathrm{old}}(t)-\sum
_{i}\frac{3}{32}\left[  3\left(  R_{\mathrm{old}}(\mathcal{O}_{i})-1\right)
^{3}-\left(  R_{\mathrm{old}}(\mathcal{O}_{i})-1\right)  \right]  +\frac
{1}{48}\\
c_{\mathrm{IR}}^{\mathrm{new}}  &  =c_{\mathrm{IR}}^{\mathrm{old}}(t)-\sum
_{i}\frac{1}{32}\left[  9\left(  R_{\mathrm{old}}(\mathcal{O}_{i})-1\right)
^{3}-5\left(  R_{\mathrm{old}}(\mathcal{O}_{i})-1\right)  \right]  +\frac
{1}{24}\,.
\end{align}
We can also give a general formula for the
new $k_{\mathrm{IR}}(\mathfrak{h}_{l_i}^{(i)})$ after we 
decouple all the offending mesons:
\begin{equation}
k_{\mathrm{IR}}(\mathfrak{h}_{l_i}^{(i)})=\text{Ind}(\mathfrak{h}_{l_i}%
^{(i)}\rightarrow\mathfrak{g}_{\mathrm{UV}}^{(i)})\times K_{\mathrm{IR}}(\mathfrak{g}_{\mathrm{UV}}^{(i)}) - 6\sum_{a}(1-(1 + T_{3}(\mathcal{O}_{a}))t_{\ast
})\mathrm{Ind}(\rho_{a}(\mathfrak{h}_{l_i}^{(i)})),
\end{equation}
where $\mathrm{Ind}(\rho_{a}(\mathfrak{h}_{l}^{(i)}))$ is the index of the irreducible
representation under which $\mathcal{O}_{i}$ transforms,
and $t_{\ast}$ is the fixed value of the maximization parameter at the last step when there are no
unitarity bound violations anymore.

Consider next the case of mesonic operators which decouple in the
flipper field deformations. As noted in \cite{Benvenuti:2017lle}, when an operator decouples,
one can introduce an additional \textquotedblleft flipping
field\textquotedblright\ which couples to this field. Doing this is equivalent
to the standard procedure of introducing an additional $U(1)$ anyway. Let us
see how this works in detail.

With each $M$, there comes an additional $U(1)$ symmetry in the UV theory.
Coupling the mesons to the $M$'s protects them from dropping below the
unitarity bound in the IR. From another point of view, the process of removing one
of the previously offending $\mathcal{O}$'s is equivalent to adding a coupling to
$M$, as explained in \cite{Benvenuti:2017lle}. Compared to the plain mass
deformation the new UV $U(1)$ is equivalent to the emergent $U(1)$ that we
would have to introduce by hand, once a meson drops below the unitarity bound.
Hence, for all flipper field deformations we do not need to worry about any
of the mesons decoupling or how it might affect the anomalies. This is
automatically being taken care of by the $M$'s. In fact as explained in
\cite{Benvenuti:2017lle}, the mesons $\mathcal{O}$ are zero in the chiral
ring, and therefore there are no unitarity violations associated to them. In
the following, we describe this intriguing mechanism in more detail from
another point of view.

The analysis involves essentially the same equations
as already presented in section \ref{sec:INHERIT}, which we
present here for convenience of the reader.
Recall that with flipper field deformations, we have a free chiral superfield $M$ in the
adjoint of $\mathfrak{g}_{\mathrm{UV}}$ coupled to $\mathcal{O}_{\text{adj}}$ via $\delta W =
\mathrm{Tr}_{\mathfrak{g}_{\mathrm{UV}}} (M_{\text{adj}} \cdot\mathcal{O}_{\text{adj}})$,
with a background value $\langle M_{\text{adj}} \rangle = m_{\text{adj}}$ our nilpotent mass term.
There is automatically an extra $U(1)$ symmetry
for each $M_{j_{(a)},-j_{(a)}}$ in the UV. The first part of the
trial IR R-charge is fixed by the plain mass deformation term $\mathrm{Tr}%
_{\mathfrak{g}_{\mathrm{UV}}}(m_{adj} \cdot \mathcal{O}_{\text{adj}})$. In the UV the $M_{j_{(a)},-j_{(a)}}$ are free multiplets
and they are charged under an extra $U(1)$. We call the generator
corresponding the this extra $U(1)$ $T_{\text{flip}}$. The charge of the
fluctuation of $M$ is normalized to $T_{\text{flip}}(M) = 1$, and nothing else is
charged under it. Moreover we know that $T_{3}(M_{j_{(a)},-j_{(a)}})=-
T_{3}(\mathcal{O}_{(j_{(a)},j_{(a)})})=-j_{(a)}$. Now, we have to take this
additional symmetry into account while computing the trial IR R-charge
\begin{equation}
\label{RirU1}R_{\mathrm{IR}}=R_{\mathrm{UV}}+\left(  \frac{t}{2}-\frac{1}%
{3}\right)  J_{\mathcal{N}=2} - t T_{3} + t_{\text{flip}} T_{\text{flip}}\,.
\end{equation}
Applying this relation to the superpotential deformation $\delta W$, we find
\begin{equation}
R^{\mathrm{new}}_{\mathrm{IR}}(\delta W) =R^{\mathrm{old}}_{\mathrm{IR}%
}(\mathcal{O}_{j_{(a)},j_{(a)}})+ t_{\text{flip}} + \frac{2}{3} - t T_{3}(M_{(j_{(a)}%
,-j_{(a)})})
\end{equation}
So, we have:
\begin{align}
\label{RirU1s} &  R^{\mathrm{old}}_{\mathrm{IR}}(\mathcal{O})+ t_{\text{flip}} + \frac{2}{3}
- t T_{3}(M_{j_{(a)},-j_{(a)}})= 2\\
\implies &  t_{\text{flip}} =\frac{4}{3}-R^{\mathrm{old}}_{\mathrm{IR}}(\mathcal{O})+ t
T_{3}(M_{j_{(a)},-j_{(a)}})=t-\frac{2}{3}%
\end{align}

This implies an additional contribution to $a_{\mathrm{IR}}=a_{\mathrm{IR}%
}^{old}+\delta a_{\mathrm{IR}}$ as follows: {\small
\begin{align}
\label{dair}\delta a_{\mathrm{IR}}  &  = \frac{3}{32} \left[  3\left(
t_{\text{flip}} T_{\text{flip}}(M) - t_{\ast} T_{3}(M) + R_{\mathrm{UV}}(M) - 1\right)
^{3}-\left(  t_{\text{flip}} T_{\text{flip}}-t_{\ast} T_{3}(M) + R_{\mathrm{UV}(M) - 1}\right)  \right] \nonumber\\
&  =\frac{3}{32} \left[  3(t_{\text{flip}} T_{\text{flip}}(M)- t_{\ast} T_{3}(M))^{3}-3(t_{\text{flip}} T_{\text{flip}}(M) - t_{\ast} T_{3}(M))^{2}+\frac{2}{9} \right]
\nonumber\\
&  = \frac{3}{32}\left[  3\left(  \frac{4}{3}-R_{\mathrm{IR}%
}(\mathcal{O})\right)  ^{3}-3\left(  \frac{4}%
{3}-R_{\mathrm{IR}}(\mathcal{O})\right)
^{2}+\frac{2}{9} \right] \nonumber\\
&  = -\frac{3}{32}\left[  3\left(  R_{\mathrm{IR}}%
(\mathcal{O})-1\right)  ^{3}-\left(  R_{\mathrm{IR}}(\mathcal{O})-1\right)  \right]  .
\end{align}
} As a result we can see that adding an additional $U(1)$ through the above
coupling is equivalent to removing the contribution from the ``bad'' operators
directly. This is why the flipper fields automatically rescue
the mesons whenever they would naively drop below the unitarity bound had this
coupling not been there. These additional coupling terms are identical to the ones that we were forced
to add whenever one of the mesons dropped below the unitarity bound before adding flipper fields.

Another quicker approach which builds upon equation \eqref{dair} is to make
use of the fact that $R(M)+R(\mathcal{O})=2$ so that we get:
\begin{align}
\label{dairquick}\delta a_{\mathrm{IR}}  &  = -\frac{3}{32}\left[  3\left(
2-R_{\mathrm{IR}}(M)-1\right)  ^{3}-\left(  2-R_{\mathrm{IR}}(M)-1\right)
\right] \nonumber\\
&  = -\frac{3}{32}\left[  3\left(  -R_{\mathrm{IR}}(M)+1\right)  ^{3}-\left(
-R_{\mathrm{IR}}(M)+1\right)  \right] \nonumber\\
&  = \frac{3}{32}\left[  3\left(  R_{\mathrm{IR}}(M)-1\right)  ^{3}-\left(
R_{\mathrm{IR}}(M)-1\right)  \right]  .
\end{align}
Therefore, adding directly the contribution from the $M$'s is equivalent to
removing the contribution from the ``bad'' $\mathcal{O}$'s. This recovers
our expressions for $a_{\mathrm{IR}}$ and $c_{\mathrm{IR}}$ up to the presence of
free chiral multiplets that do not couple.

As a result, none of the mesons in the flipper deformed theories can drop below
the unitarity bound because they are all automatically rescued by the
$M$'s to which they couple.

\subsection{Rational Theories}

One of the interesting features of the ``brute force'' sweeps we perform in later sections
reveals that in some cases, the anomalies are all rational numbers, even though
a priori, we should only expect algebraic numbers as per the procedure of a-maximization.
We refer to such IR fixed points as rational theories.
Clearly, this suggests some additional emergent structure in the infrared, and in some
favorable circumstances, this can also be identified with the appearance of enhanced $\mathcal{N} = 2$
supersymmetry, as in the case of the Maruyoshi-Song deformations \cite{Maruyoshi:2016tqk, Maruyoshi:2016aim}.
In the specific examples we consider, we find that this can happen both with and without operators decoupling,
and both for plain mass deformations and flipper field deformations, see Appendix \ref{completeTables} for details.

There has very recently been some progress in
understanding some additional sufficient criteria for $\mathcal{N}=2$
enhancements \cite{Giacomelli:2018ziv}. The main idea in this analysis is that whenever we
encounter a flavor singlet operator of the IR\ theory, we need to be able to
interpret as a scalar operator parameterizing a direction of the Coulomb
branch. This is not the case in our rational theories, but it is also
unclear whether there is any additional supersymmetry enhancement. We leave a full treatment
of possible enhancements in these theories for future work.

\subsection{Ordering of RG Flows}

As we can see, there is no clean expression that describes
$a_{\mathrm{IR}}$ as a function of the embedding index, once we take into
account operators that decouple in the IR. One might rightfully worry that $a_{\mathrm{IR}}$
would not necessarily be a simple monotonically decreasing function of $r$
anymore. However, we observe empirically that the RG flows continue to follow
the trajectory of paths through the Hasse diagram, even
after introducing emergent $U(1)$'s and flipper field operators. This is
explicitly shown in the explicit examples we consider.

We close this section with two important remarks:

\begin{enumerate}
\item If no operator drops below the unitarity bound, the theories are
guaranteed to follow the flow pattern specified by the Hasse diagrams.

\item In all of the other cases studied in this paper, even when operators
decouple, we \textit{still} observe that the RG flows respect the partial ordering of
nilpotent orbits. So, while the RG flows could have a weaker ordering than the
mathematical ordering (if the wrong mesons hit the unitarity bound) we see
that they do not appear to violate the partial ordering of nilpotent orbits.
\end{enumerate}

\section{D3-Brane Probe Theories \label{sec:D3}}

In the previous sections we introduced a general procedure for treating
nilpotent mass deformations. In this section, we turn to a systematic analysis
of all such deformations for the $\mathcal{N}=2$ theories defined by a
D3-brane probing a 7-brane with $D_{4}$, $E_{6}$, $E_{7}$ or $E_{8}$ flavor
symmetry. In what follows we do not include the contribution
from the decoupled free hypermultiplet with scalars parameterizing motion of the D3-brane
parallel to the 7-brane.

Some examples of nilpotent mass deformations for these theories were analyzed
in \cite{Heckman:2010qv}, as well as \cite{Maruyoshi:2016tqk}.
In the F-theory interpretation where we wrap
the 7-brane on a surface $\mathcal{S}_{\text{GUT}}$, we have a partially
twisted gauge theory with a $(0,1)$-connection and an adjoint valued $(2,0)$
form $\Phi_{(2,0)}$ \cite{Beasley:2008dc} (see also \cite{Bershadsky:1997zs, Donagi:2008ca}).
In terms of the associated F-theory geometry,
deformations of $\Phi_{(2,0)}$ with non-vanishing Casimir invariant
translate to complex structure deformations of the associated elliptically
fibered Calabi-Yau fourfold. The nilpotent case is especially interesting
because it is essentially \textquotedblleft invisible\textquotedblright\ to
the complex geometry of the model. We can then view the mass parameters
$m_{\text{adj}}$ as background values for $\Phi_{(2,0)}$
\cite{Heckman:2010fh, Heckman:2010qv}, and
the particular case of a nilpotent mass deformation
defines a T-brane configuration \cite{Aspinwall:1998he, Donagi:2003hh, Cecotti:2010bp, Anderson:2013rka, Collinucci:2014taa, Collinucci:2014qfa, Bena:2016oqr, Marchesano:2016cqg, Anderson:2017rpr, Bena:2017jhm, Marchesano:2017kke, Cvetic:2018xaq}.

From this perspective, it is also natural to view the flipper field deformation
as promoting the zero mode of $\Phi_{(2,0)}$ to a dynamical field. This is actually
somewhat subtle in the context of a full F-theory compactification, because
making $\Phi_{(2,0)}$ dynamical requires us to wrap the 7-brane on a compact
K\"{a}hler surface, which also introduces dynamical gauge fields (zero modes
from the (0,1) connection can be eliminated by choosing a suitable surface and
background vector bundle). However, by introducing a sufficiently large number
of additional spectator fields which also interact with this gauge field, we
can always take a limit where this gauge theory is infrared free (in contrast
to the case typically assumed in decoupling limits from gravity).

In both the case of plain mass deformations as well as its extension to
flipper field deformations, we see that the IR\ fixed points defined by the
D3-brane provide additional insight into the structure of T-brane
configurations in F-theory.

Let us now turn to an analysis of the fixed points in these theories. Much as
in the earlier sections of this paper, it is helpful to split our analysis up
into the cases of plain mass deformations and flipper field deformations. We
also discuss in detail the special case of rational theories, which suggest
additional structure in the IR. This includes all the previous $\mathcal{N}=2$
enhancement theories found in \cite{Maruyoshi:2016tqk}, as well as another one which comes about
from deformations of the $E_{7}$ Minahan-Nemeschansky theory (see also \cite{CartaGiacomelliSavelli}).

\subsection{Summary of UV $\mathcal{N}=2$ Fixed Points}

In this section we briefly summarize some aspects of the $\mathcal{N}=2$
theories. We first list the anomalies and scaling
dimensions of the Coulomb branch operator $Z$. These
values can be found in \cite{Aharony:2007dj} and are
summarized in table \ref{UVscaling} for later convenience:\footnote{While it
is entirely possible to study nilpotent deformations of the Argyres-Douglas
theories they are too simple to be of interest. However, for convenience we do
list their UV values in table \ref{UVscaling}.}
\begin{table}[H]
  \centering
  {\renewcommand{\arraystretch}{1.2}
\begin{tabular}{ |c||c|c|c|c|c|c|c| }
 \hline
 $G$ & $H_0$ & $H_1$ & $H_2$ & $D_4$ & $E_6$  & $E_7$ & $E_8$ \\  \hline 
 $\Delta_{\mathrm{UV}}$($Z$) & $\frac{6}{5}$ &$\frac{4}{3}$ &$\frac{3}{2}$ &2 & 3 & 4 & 6 \\  \hline
 $a_{\mathrm{UV}}$ &  $\frac{43}{120}$ &$\frac{11}{24}$ &$\frac{7}{12}$ &$\frac{23}{24}$ & $\frac{41}{24}$  & $\frac{59}{24}$ & $\frac{95}{24}$ \\  \hline
 $c_{\mathrm{UV}}$ &  $\frac{11}{30}$ &$\frac{1}{2}$ &$\frac{2}{3}$ &$\frac{7}{6}$ & $\frac{13}{6}$  & $\frac{19}{6}$ & $\frac{31}{6}$ \\  \hline
 $k_{\mathrm{UV}}$ &  $\frac{12}{5}$ &$\frac{8}{3}$ &3 &4 & 6  & 8 & 12 \\  \hline
\end{tabular}}
\caption{Scaling dimensions and anomalies of rank 1 4D $\cN=2$ SCFTs.}
\label{UVscaling}
\end{table}

From there the anomalies and scaling dimensions in the IR can directly be
computed from the previously derived equations. The only necessary information
is the embedding index of the $\mathfrak{su}(2)_D$ subalgebra defined by the 
nilpotent orbit. Since we only
have one flavor symmetry factor, the Cartan matrix is uniquely
specified by the nilpotent orbit one wants to consider. Then it is only a
matter of evaluating the formulae of sections \ref{sec:INHERIT} and \ref{sec:EMERGE}.

\subsection{Plain Mass Deformations}

It is noteworthy that for all of the rank one probe D3-brane theories,
the mesons never appear to decouple. However,
$\Delta_{\mathrm{IR}}(Z)$ sometimes does decouple when
the value of $r$ becomes too large. In general the unitarity bound for the
operator $Z$ is violated whenever:
\begin{align}
r  &  \geq5 &  &  \text{for }\SO(8)\nonumber\\
r  &  \geq19 &  &  \text{for }E_{6}\nonumber\\
r  &  \geq40 &  &  \text{for }E_{7}\nonumber\\
r  &  \geq107 &  &  \text{for }E_{8}\,.
\end{align}

There are a large number of possible nilpotent deformations. Due to the size
of the resulting tables we only list our results for flavor symmetry $D_{4}$
and all rational results for the exceptional groups. Rational coefficients are
of particular interest as they suggest additional structure present in the IR.
When comparing our results with the subset of cases studied
in \cite{Heckman:2010qv} we find perfect agreement aside from the last column of table 5
which contains the correct value of $t_{\ast}$ but a minor typo
for the values of $a_{\mathrm{IR}}$ and $c_{\mathrm{IR}}$.

The complete list of all the possible deformations can be accessed via a
\texttt{Mathematica} routine summarized in Appendix~\ref{completeTables}. Due to
the very large amount of data we only list here the rational results for the
exceptional groups in Appendix \ref{completeTables}.

\begin{figure}[ptb]
\centering
\begin{subfigure}[b]{0.45\textwidth}
\includegraphics[width=\textwidth]{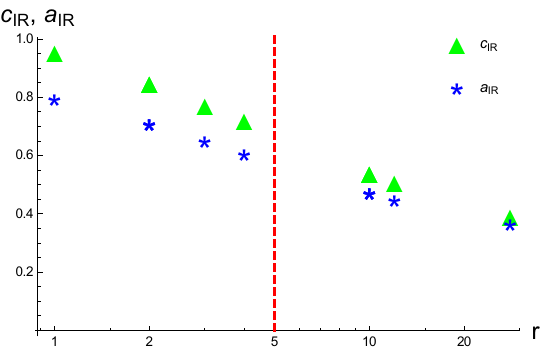}
\caption{$D_4$}
\label{MNcD4}
\end{subfigure}
\hspace{12pt} \begin{subfigure}[b]{0.45\textwidth}
\includegraphics[width=\textwidth]{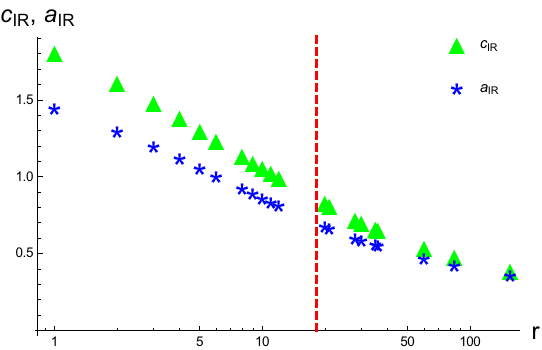}
\caption{$E_6$}
\label{MNcE6}
\end{subfigure}
\begin{subfigure}[b]{0.45\textwidth}
\includegraphics[width=\textwidth]{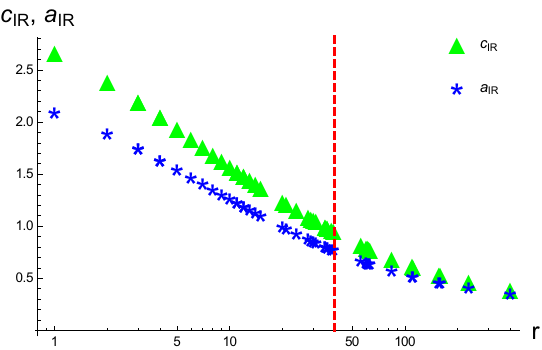}
\caption{$E_7$}
\label{MNcE7}
\end{subfigure}
\hspace{12pt} \begin{subfigure}[b]{0.45\textwidth}
\includegraphics[width=\textwidth]{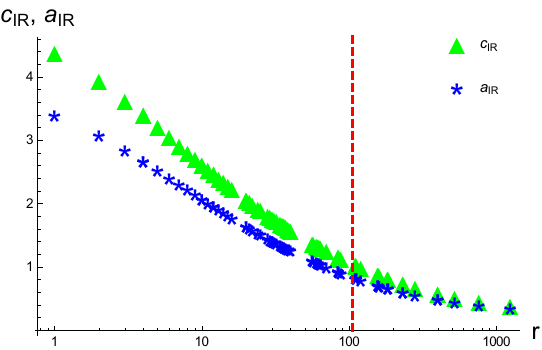}
\caption{$E_8$}
\label{MNcE8}
\end{subfigure}
\caption{Plots of $a_{\mathrm{IR}}$ (blue stars) and $c_{\mathrm{IR}}$ (green
triangles) vs embedding index $r$ for the different probe D3-brane
theories. The red vertical dashed line denotes the largest value of $r$ before
the Coulomb branch operator $Z$ decouples.
Anything to the right of this line has a single emergent $U(1)$
to rescue the Coulomb branch operator. The plots are log-scaled on the x-axis
for presentation purposes due to the fact that the region of deformed theories
is denser around lower values of $r$ and becomes more sparse as $r$
increases.}%
\label{MNc}%
\end{figure}

The tables are organized as follows. For the top tables, first we list
the Bala-Carter label of the deformation, or simply the partition of the
fundamental representation's splitting in the case of $\SO(8)$. The second
column gives the value of the embedding index $r$. The following three columns
give the anomalies $a_{\mathrm{IR}}$ and $c_{\mathrm{IR}}$, as well as the
value of the parameter $t$ after re-doing any $a$-maximization if necessary.
Whenever fields decouple (because they first hit the unitarity bound and are
rescued by emergent $U(1)$'s) then we can look at the interacting part versus the
complete contribution to $a_{\mathrm{IR}}$ and $c_{\mathrm{IR}}$. Indeed,
whenever an operator decouples it contributes a factor of $1/48$ or $1/24$ to
$a_{\mathrm{IR}}$ and $c_{\mathrm{IR}}$ respectively, and we separately 
report these values in our tables. The first number in
columns 3 and 4 is only the interacting piece, while the second number also
includes the contribution from any free multiplets that decoupled. Thus those
numbers only differ by an integer $n$ times $1/48$ (or $1/24$), where $n$ is
equal to the number of multiplets generators that have decoupled and become
free. If there is no emergent $U(1)$ introduced and no field decouples then
there is only an interacting piece and only the first number makes sense and
is listed. Finally, the last two columns give the scaling dimension of the
Coulomb branch parameter $Z$ and the lowest scaling dimension of the mesons
$\mathcal{O}$'s.

For the bottom tables we first list the Bala-Carter label of the deformation,
followed by the residual flavor symmetry. The following four columns
correspond to the flavor central charges $k_{\mathrm{IR}}$ taken with
respect to the residual flavor symmetry. For each we list their value with only the
interacting part of the theory or including the free fields which decoupled in
seperate columns. Finally, we note that there are separate values for each of
the subgroups in the product decomposition of the residual flavor, hence the
multiple values listed in each column. For the theories with exceptional flavor
symmetry we only list values that have rational anomalies.

Furthermore, as it is impractical to list all the other values in a single
table we provide plots of $a_{\mathrm{IR}}$ and $c_{\mathrm{IR}}$ as functions
of the embedding index $r$:

\begin{figure}[ptb]
\centering
\begin{subfigure}[b]{0.45\textwidth}
\includegraphics[width=\textwidth]{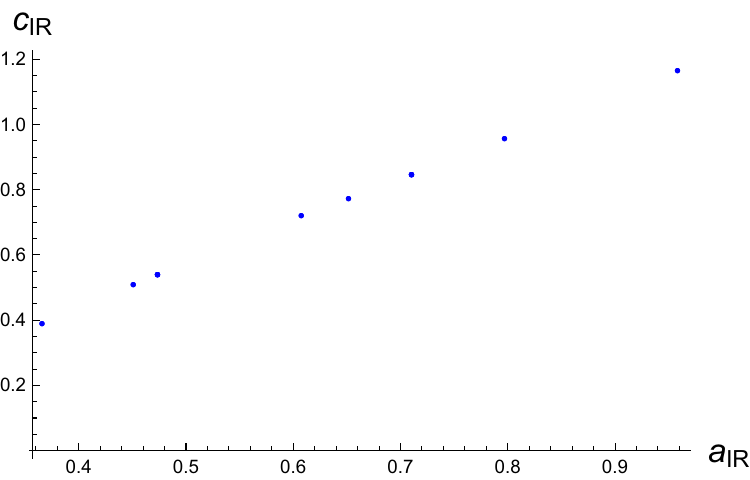}
\caption{$D_4$}
\label{MNratioD4}
\end{subfigure}
\hspace{12pt} \begin{subfigure}[b]{0.45\textwidth}
\includegraphics[width=\textwidth]{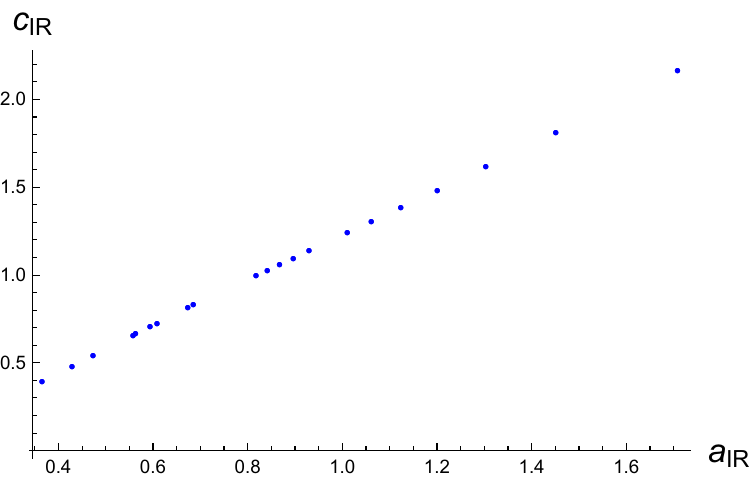}
\caption{$E_6$}
\label{MNratioE6}
\end{subfigure}
\begin{subfigure}[b]{0.45\textwidth}
\includegraphics[width=\textwidth]{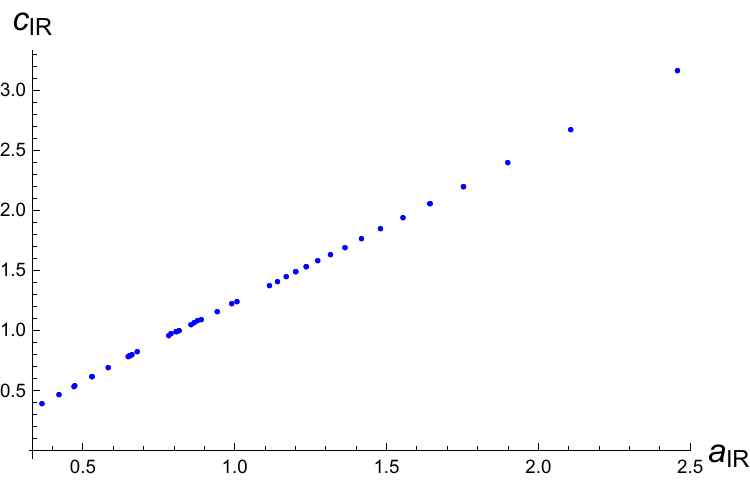}
\caption{$E_7$}
\label{MNratioE7}
\end{subfigure}
\hspace{12pt} \begin{subfigure}[b]{0.45\textwidth}
\includegraphics[width=\textwidth]{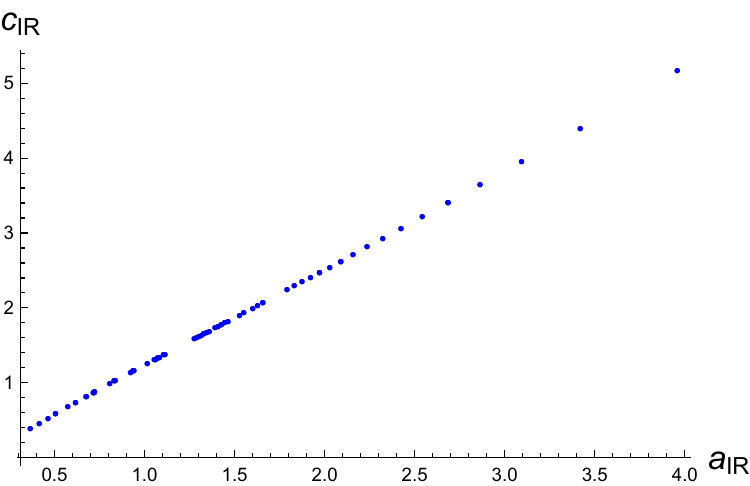}
\caption{$E_8$}
\label{MNratioE8}
\end{subfigure}
\caption{Plots of $c_{\mathrm{IR}}$ vs. $a_{\mathrm{IR}}$ for plain nilpotent
mass deformations of the different probe D3-brane theories.}%
\label{MNratio}%
\end{figure}

As we can see, as $r$ increases, the anomalies
decrease. Whenever an additional deformation is introduced the embedding index
increases. Physically, this translates in a flow to a lower IR theory down the
Hasse diagram of possible RG flows. As a result we expect the degrees of
freedom to decrease, that is $a_{\mathrm{IR}}$ should decrease along this
Hasse diagram. The fact that $a_{\mathrm{IR}}$ is a monotonically decreasing
function of $r$ is an easy consistency check. We also note that the
interacting piece of the anomaly (first value of columns 3) also decreases the
same way.

It is also interesting to note that for a given UV $\mathcal{N} = 2$ fixed point,
the ratio of anomalies $a_{\mathrm{IR}}%
$/$c_{\mathrm{IR}}$ remains roughly constant over the entire nilpotent network. Reference
\cite{Maruyoshi:2018nod} noticed a similar effect.
We also determine the overall statistical spread in the value of the ratio
$a_{\mathrm{IR}}/c_{\mathrm{IR}}$ for plain mass deformations of the probe D3-brane theories.
By inspection of the plots in figure \ref{MNc},
we see that there is a roughly constant value for each theory.
We also calculate the mean and standard deviation by sweeping
over all such theories, the results of which
are shown in table \ref{tab:nilpD3}. Quite remarkably, the standard deviation
is on the order of $1\%$ to $5\%$, indicating a remarkably stable value across
the entire network of flows. Another curious feature is that the mean value of
$a_{\mathrm{IR}}/c_{\mathrm{IR}}$ decreases as we increase to larger flavor symmetries.
Precisely the opposite behavior is observed in the nilpotent networks of
4D\ conformal matter.

\begin{table}[H]
\centering
\begin{tabular}
[c]{|c|c|c|c|c|}\hline
  & $D_{4}$ & $E_{6}$ & $E_{7}$ & $E_{8}$\\\hline
Mean & $0.86$ & $0.83$ & $0.82$ & $0.81$\\\hline
Std. Dev. & $0.03$ & $0.03$ & $0.03$ & $0.03$\\\hline
Max & $0.94$ & $0.94$ & $0.94$ & $0.94$\\\hline
Min & $0.82$ & $0.79$ & $0.78$ & $0.77$\\\hline
\end{tabular}
\caption{Table of means and standard deviations for the ratio $a_{\mathrm{IR}}/c_{\mathrm{IR}}$ across
the entire nilpotent network defined by plain mass deformations of probe
D3-brane theories. We also display the maximum and minimum values.}
\label{tab:nilpD3}
\end{table}

\subsection{Flipper Field Deformations}

Consider next flipper field deformations of the probe D3-brane theories.
As one would expect, we recover the results from \cite{Agarwal:2016pjo}.
In Appendix \ref{completeTables} we present all our results for $D_{4}$
flavor symmetry and only list the values with rational anomalies for the
exceptional flavors $E_{6,7,8}$. Furthermore, we highlight cases where we obtain
known enhancements to $\mathcal{N}=2$ theories such as $H_{0}$, $H_{1}$, and
$H_{2}$ (as already pointed out in \cite{Agarwal:2016pjo}), and we find
an enhancement of the $E_{7}$ Minahan-Nemeschansky
theory to the Argyres-Douglas theory $H_{1}$, in agreement with \cite{Giacomelli:2017ckh, Giacomelli:2018ziv, CartaGiacomelliSavelli}. 
It is associated with the Bala-Carter label $E_6$ which has embedding index $r = 156$.
In such cases we can compute the embedding index $r_{F}$ of
the residual flavor symmetry and see that not only $a_{\mathrm{IR}}$ and
$c_{\mathrm{IR}}$ match the known values but $k_{\mathrm{IR}} r_{F}$ also
yields the proper value for the flavor central charge of these theories. It is
noteworthy that in those particular cases, the chiral multiplets,
$M_{j_{(a)},-j_{(a)}}$, that survive transform trivially under the residual
flavor symmetry and therefore do not introduce any additional contributions to
the flavor central charge. This is however not true in general.

We also again plot $a_{\mathrm{IR}}$ and $c_{\mathrm{IR}}$ as functions of
the embedding index $r$ for each of the above cases.
\begin{figure}[ptb]
\centering
\begin{subfigure}[b]{0.45\textwidth}
\includegraphics[width=\textwidth]{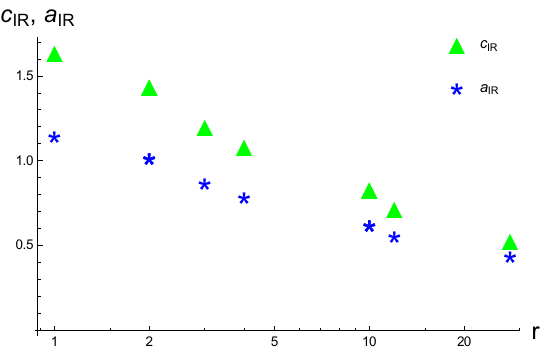}
\caption{$D_4$}
\label{MSMNcD4}
\end{subfigure}
\hspace{12pt} \begin{subfigure}[b]{0.45\textwidth}
\includegraphics[width=\textwidth]{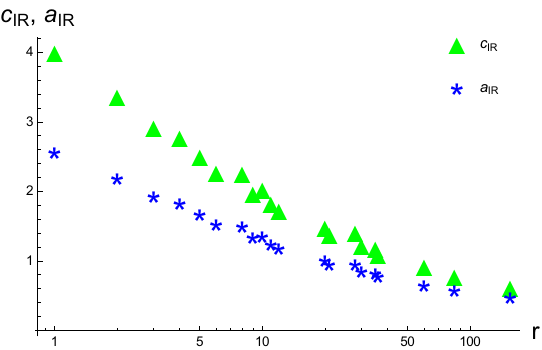}
\caption{$E_6$}
\label{MSMNcE6}
\end{subfigure}
\begin{subfigure}[b]{0.45\textwidth}
\includegraphics[width=\textwidth]{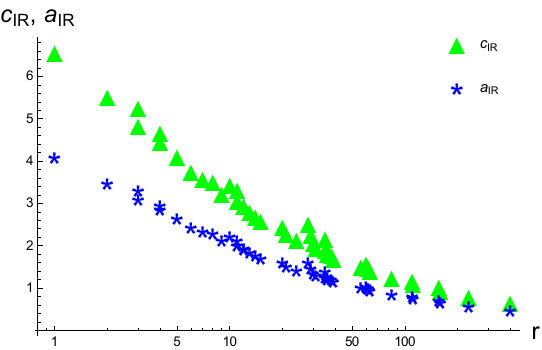}
\caption{$E_7$}
\label{MSMNcE7}
\end{subfigure}
\hspace{12pt} \begin{subfigure}[b]{0.45\textwidth}
\includegraphics[width=\textwidth]{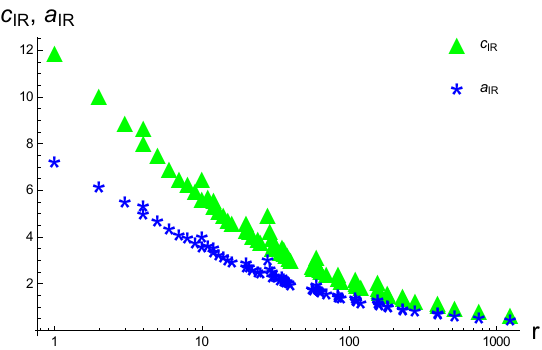}
\caption{$E_8$}
\label{MSMNcE8}
\end{subfigure}
\caption{Plots of $a_{\mathrm{IR}}$ (blue stars) and $c_{\mathrm{IR}}$ (green
triangles) vs embedding index $r$ for the different flipper field
deformations of probe D3-brane theories.}
\label{MSMNc}%
\end{figure}

This time we see that the central charges do not exactly decrease
as the embedding index $r$ increases. However, they do decrease along the
flows defined by the Hasse-diagrams, as expected. Another interesting feature of
these Hasse diagram flows is that the number of flipper field deformations which actually participate in
a flow can vary wildly from orbit to orbit (since the number of $\mathfrak{su}(2)_D$ irreducible representations
also jumps a fair amount). Of course, such fields must be included in computing
various anomalies, even if they serve to decouple mesonic operators which drop below the unitarity bound.
Doing so, we find that $a_{\mathrm{IR}}$ indeed decreases monotonically along a flow.

This raises the question of alternative numerical invariants instead of the embedding index which might be used to order RG~flows in this class of theories. We have chosen the embedding index because this is the quantity which naturally appears in the construction of the infrared R-symmetry (see equations \ref{IRanomalies}). Additionally, it is numerically simple to obtain and often a useful proxy for the ordering of the RG~flows. We are not aware of any other quantity which could provide a better trade off between accuracy and the complexity to compute it. Looking at the Hasse-diagram of the corresponding nilpotent orbits, one would expect that a more accurate description requires more parameters than just one. This would turn the presented plots into higher dimensional ones. For instance, the $x$-axis would need to be replaced by a series of branches corresponding to the full Hasse diagrams. The resulting plots would be much more complex than they need to be. Especially given how closely the embedding index gets to properly ordering the RG~flows. Hence, we continue to rely on this physical parameter rather than try and introduce a less natural quantity.

\begin{figure}[ptb]
\centering
\begin{subfigure}[b]{0.45\textwidth}
\includegraphics[width=\textwidth]{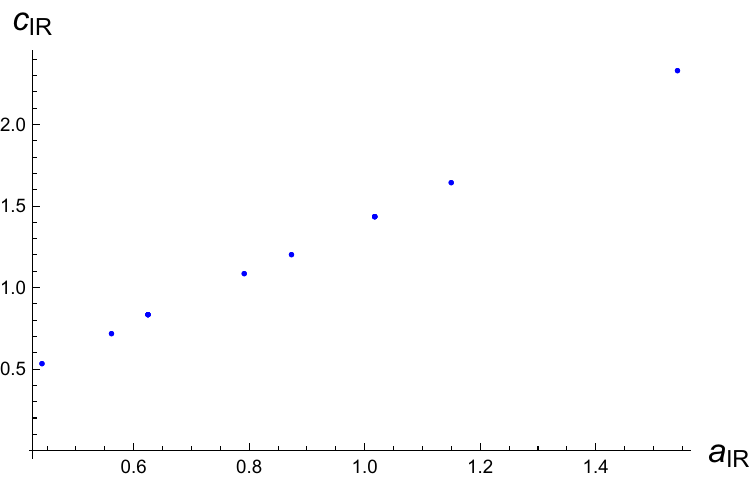}
\caption{$D_4$}
\label{MNMSratioD4}
\end{subfigure}
\hspace{12pt} \begin{subfigure}[b]{0.45\textwidth}
\includegraphics[width=\textwidth]{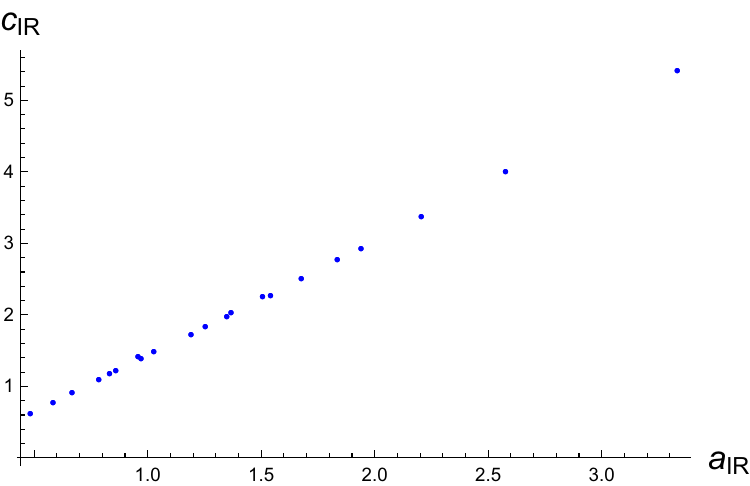}
\caption{$E_6$}
\label{MNMSratioE6}
\end{subfigure}
\begin{subfigure}[b]{0.45\textwidth}
\includegraphics[width=\textwidth]{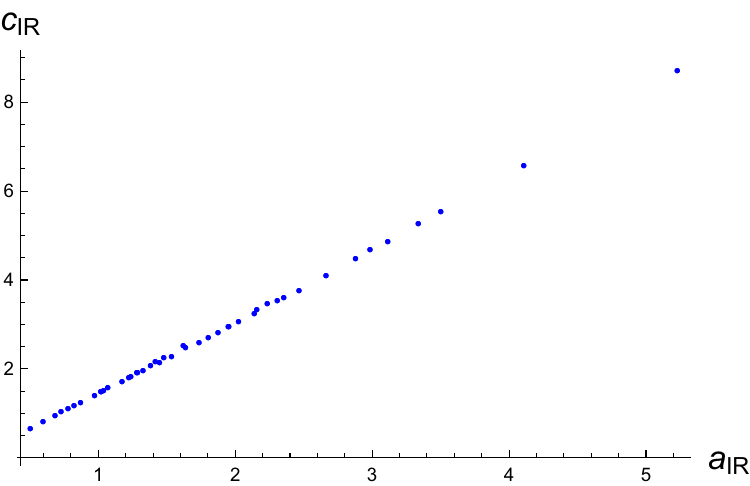}
\caption{$E_7$}
\label{MNMSratioE7}
\end{subfigure}
\hspace{12pt} \begin{subfigure}[b]{0.45\textwidth}
\includegraphics[width=\textwidth]{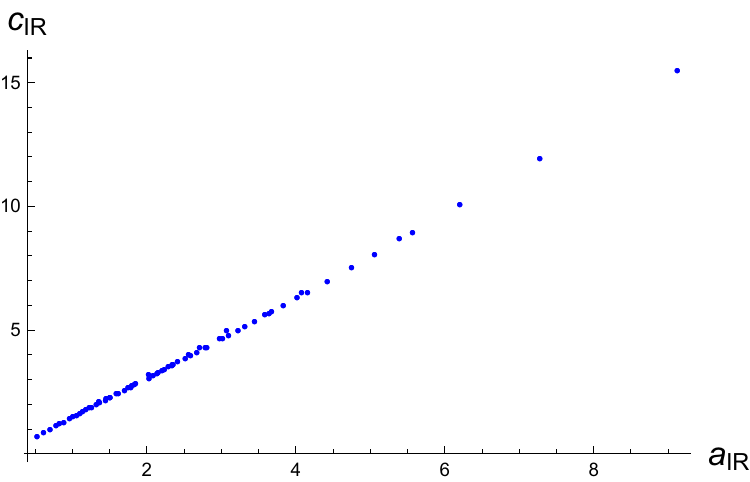}
\caption{$E_8$}
\label{MNMSratioE8}
\end{subfigure}
\caption{Plots of $c_{\mathrm{IR}}$ vs. $a_{\mathrm{IR}}$ for the different
flipper field deformations of probe D3-brane theories.}
\label{MNMSratio}%
\end{figure}

Finally, another interesting feature of our analysis is that the ratio $a_{\mathrm{IR}}$%
/$c_{\mathrm{IR}}$ is roughly constant for a fixed deformation, given a flavor
symmetry $\mathfrak{g}_{\mathrm{UV}}$ in the UV (see figure \ref{MNMSratio}).
Much as for the plain nilpotent mass deformations,
the overall statistical spread in the value of the ratio
$a_{\mathrm{IR}}/c_{\mathrm{IR}}$ is also remarkably small, and is
on the order of $1\%$ to $5\%$, indicating a remarkably stable value across
the entire network of flows. Another curious feature is that the mean value of
$a_{\mathrm{IR}}/c_{\mathrm{IR}}$ decreases as we increase to larger flavor symmetries.
Precisely the opposite behavior is observed in the nilpotent networks of
4D\ conformal matter. See table \ref{tab:flipD3} for the specific values.

\begin{table}[H]
\centering
\begin{tabular}
[c]{|c|c|c|c|c|}\hline
& $D_{4}$ & $E_{6}$ & $E_{7}$ & $E_{8}$\\\hline
Mean & $0.73$ & $0.69$ & $0.67$ & $0.65$\\\hline
Std. Dev. & $0.04$ & $0.04$ & $0.03$ & $0.03$\\\hline
Max & $0.83$ & $0.78$ & $0.77$ & $0.75$\\\hline
Min & $0.66$ & $0.62$ & $0.6$ & $0.59$\\\hline
\end{tabular}
\caption{Table of means and standard deviations for the ratio $a_{\mathrm{IR}}/c_{\mathrm{IR}}$ across
the entire nilpotent network defined by flipper field deformations of probe
D3-brane theories. We also display the maximum and minimum values.}
\label{tab:flipD3}
\end{table}

\section{4D Conformal Matter Theories \label{sec:CM}}

In this section we turn to the case of 4D conformal matter theories. In
F-theory terms, these are obtained from a pair of intersecting 7-branes each
with gauge group $G$ which intersect along a common $T^{2}$, namely we have
the compactification of 6D\ conformal matter to an $\mathcal{N}=2$ theory.
Some properties of these theories such as the anomaly polynomial were
determined in \cite{Ohmori:2015pua, Ohmori:2015pia}, and their role as building blocks in
generalized quiver gauge theories was studied in \cite{Apruzzi:2017iqe, Apruzzi:2018oge}.

Now, in this case, the interpretation of the mass parameters is somewhat
different from the D3-brane case. The reason is that the 4D conformal matter
defines a current which couples to the gauge fields of the 7-brane.
More precisely, from the $(0,1)$ connection and the adjoint valued $(2,0)$-form, it
is now the pullback of the $(0,1)$ connection $\mathbb{A}_{(0,1)}$ onto the
$T^{2}$ which actually couples to the 4D conformal matter. A mass deformation
then corresponds to switching on a zero mode for this connection along the
curve. Now in the case where the associated Wilson loop is not unipotent (so
that the zero mode is not nilpotent), this would be an element of the Deligne cohomology
$\mathcal{D}_{2,2}(CY_{4})$ for the associated elliptically fibered Calabi-Yau fourfold
of the F-theory model (see \cite{Aspinwall:1998he} as well as
\cite{Anderson:2013rka}). This can also be viewed as a T-brane deformation of
sorts, because in the limit where the mass parameter is nilpotent, this
deformation is \textquotedblleft invisible\textquotedblright\ in
the associated moduli space problem.\footnote{More precisely, the moduli space
can develop singularities, and as explained in \cite{Anderson:2013rka}, the gauge
theory on the 7-brane serves to complete the moduli space in these singular
limits.} Clearly, it is also natural to promote these background parameters to
a dynamical field, as will happen if we wrap these 7-branes on compact
K\"{a}hler surfaces, and some examples of weakly gauging flavor symmetries in
this way were studied in \cite{Apruzzi:2018oge}. To get a stringy
embedding of the flipper fields, however, we must take a suitable limit
where the gauge fields become IR\ free, but the chiral superfields remain dynamical.

Our plan in the remainder of this section will be to discuss some further
aspects of these conformal matter theories. We begin by reviewing some aspects
of the original $\mathcal{N}=2$ theories, and then turn to an analysis of the
resulting nilpotent network of $\mathcal{N} = 1$ fixed points. When we turn 
to the plots and statistics for these networks, we treat the nilpotent orbit with 
$G_L \leftrightarrow G_R$ interchanged as distinct.

\subsection{Summary of UV $\mathcal{N}=2$ Fixed Points}

We now review some aspects of $\mathcal{N} = 2$ $(G,G)$
4D conformal matter obtained from compactification of
$(G,G)$ 6D conformal matter on a $T^2$. We present in
table \ref{CMUVscaling} the values for the central
charges and flavor symmetries, together with the dimensions and multiplicities
of the Coulomb branch operators. We give further details on how those results
are obtained in Appendix~\ref{CMappendix}.

\begin{table}[h]
\centering
{\renewcommand{\arraystretch}{1.2}
\begin{tabular}
[c]{|c||c|c|c|c|}\hline
$(G_L,G_R)$ & $(D_k , D_k)$ & $(E_{6},E_6)$ & $(E_{7},E_7)$ & $(E_{8},E_8)$\\\hline\hline
$a_{\mathrm{UV}}$ & $\frac{1}{24}\left(  k(14k-19)-53\right)  $ & $\frac
{613}{24}$ & $\frac{817}{12}$ & $\frac{1745}{8}$\\\hline
$c_{\mathrm{UV}}$ & $\frac{1}{6}\left(  k(4k-5)-13\right)  $ & $\frac{173}{6}$
& $\frac{442}{6}$ & $\frac{457}{2}$\\\hline
$k^{flav}_{L},k^{flav}_{R}$ & $4k-4$ & 24 & 36 & 60\\\hline
$\Delta(Z_{i})$ & $\{6_{1},...,(2k-2)_{1}\}$ &
\parbox{2cm}{$\{6_1,8_1,9_1,$ $12_2\}$} &
\parbox{2.5cm}{$\{6_1,8_1,10_1,$ $12_2,14_2,18_3\}$} &
\parbox{3cm}{$\{6_1,8_1,12_2,14_2,$ $18_3,20_3,24_4,30_5\}$}\\\hline
\end{tabular}
}\caption{Anomalies and scaling dimensions for 4D $\mathcal{N} = 2$ $(G,G)$
conformal matter. In the last row, the subscripts are the multiplicities, i.e. the number of Coulomb
branch operators with that specific scaling dimension.}%
\label{CMUVscaling}%
\end{table}

The dimension of the Coulomb branch for the different conformal matter
theories on $T^{2}$ are
\begin{align}
\mathrm{dim}_{\mathbb{C}}\left(  \mathrm{Coul}\left[  (D_k,D_k) \right]  \right)   &  =k-3,\\
\mathrm{dim}_{\mathbb{C}}\left(  \mathrm{Coul}\left[  (E_{6},E_6) \right]  \right)
&  =5,\\
\mathrm{dim}_{\mathbb{C}}\left(  \mathrm{Coul}\left[  (E_{7},E_7)\right]  \right)
&  =10,\\
\mathrm{dim}_{\mathbb{C}}\left(  \mathrm{Coul}\left[  (E_{8},E_8)\right]  \right)
&  =21.
\end{align}
which matches the expectation from 6D \cite{DelZotto:2014hpa}:
\begin{equation}
\mathrm{dim}_{\mathbb{C}}\left(  \mathrm{Coul}\left[  G\right]  \right)
)=h_{G}^{\vee}-r_{G}-1\,,
\end{equation}
where $r_{G}$ is the rank of $G$ and $h_{G}^{\vee}$ is
the dual Coxeter number of $G$. In order to extract the dimensions of the
Coulomb branch operators for the different conformal matter theories, we read
off the scaling dimension of the deformations from the mirror geometries of the
elliptic threefold of the F-theory geometry. The mirror
geometries for $(E_n,E_n)$ theories were provided in
\cite{DelZotto:2015rca} and the $(D_k, D_k)$ case can be obtained from
the curve in equation (5.4) of reference \cite{Ohmori:2015pua}.

\subsection{Plain Mass Deformations}

The computations for conformal matter follow the
general procedure outlined in previous sections.
We now have two flavor groups, so two nilpotent orbits labelled
by corresponding Bala-Carter labels. Each one comes with an embedding index
$r_{L}$ and $r_{R}$.

We have actually already encountered the ($D_{4}$, $D_{4}$) 4D conformal matter theory: it is
simply the rank one $E_8$ Minahan-Nemeschansky theory (it can still be
accessed with the code described in Appendix~\ref{completeTables}). It mainly serves as a
cross-check on the general procedure, and we find perfect agreement for those
deformations which live in an $\mathfrak{so}(8) \times \mathfrak{so}(8)$ subalgebra.
Thus, we simply list in Appendix \ref{completeTables} the rational theories
in the case where the parent 4D conformal matter theory has exceptional flavor symmetry.
Due to their large size the tables are also split
in their length. The top half contains the Bala-Carter labels, embedding
indices, anomalies and $t_{*}$. The bottom half repeats the Bala-Carter labels
and $t_{*}$ before providing scaling dimensions. Finally, the tables for the
flavor central charges are too large to include here. So, we refer the reader
to the companion \texttt{Mathematica} code for those results.

We also provide contour plots of $a_{\mathrm{IR}}$ vs. the embedding indices
of the right and left flavors. We hasten to add that while the partial
ordering of nilpotent orbits enforces a corresponding ordering for the associated
embedding indices, the converse is not true (the Hasse diagram has more fine structure).
This is an unfortunate artifact of displaying all of our data with respect to a two-dimensional
contour plot.
\begin{figure}[ptb]
\centering
\begin{subfigure}[b]{0.495\textwidth}
\includegraphics[width=\textwidth]{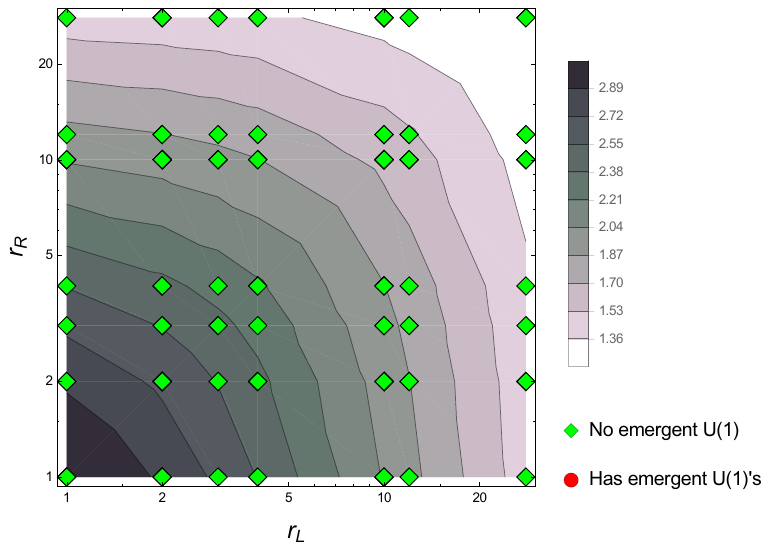}
\caption{($D_4$, $D_4$)}
\label{CMcontourD4}
\end{subfigure}
\begin{subfigure}[b]{0.495\textwidth}
\includegraphics[width=\textwidth]{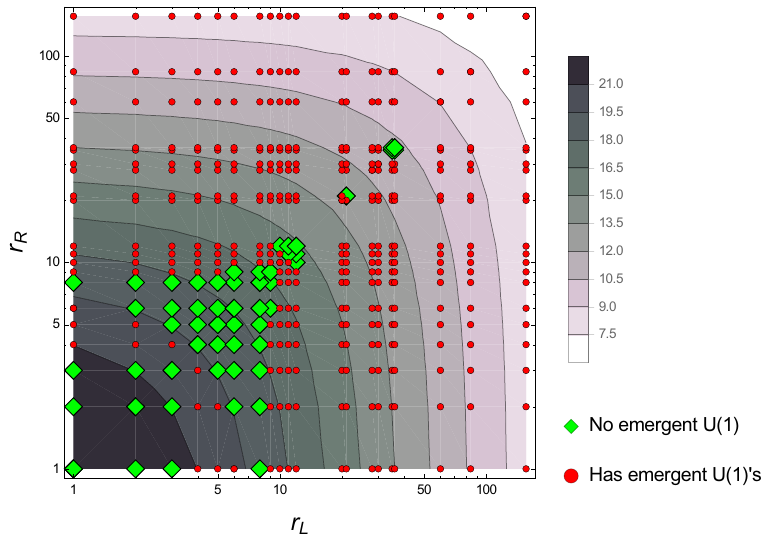}
\caption{($E_6$, $E_6$)}
\label{CMcontourE6}
\end{subfigure}
\begin{subfigure}[b]{0.495\textwidth}
\includegraphics[width=\textwidth]{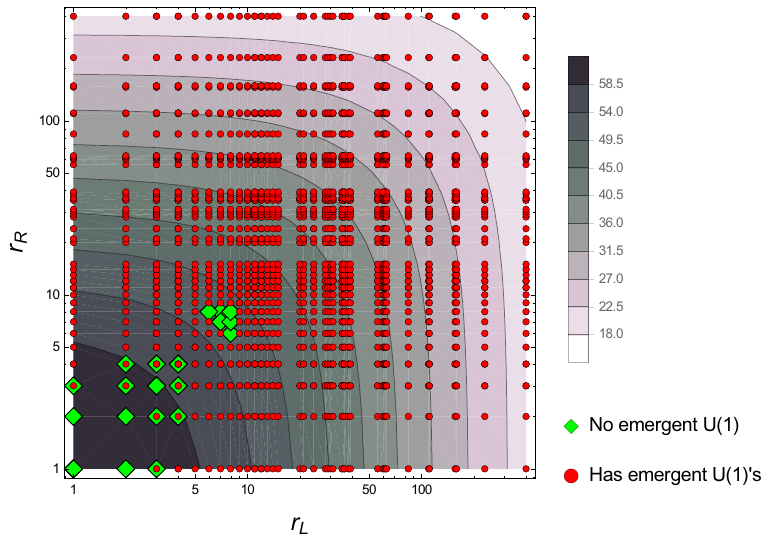}
\caption{($E_7$, $E_7$)}
\label{CMcontourE7}
\end{subfigure}
\begin{subfigure}[b]{0.495\textwidth}
\includegraphics[width=\textwidth]{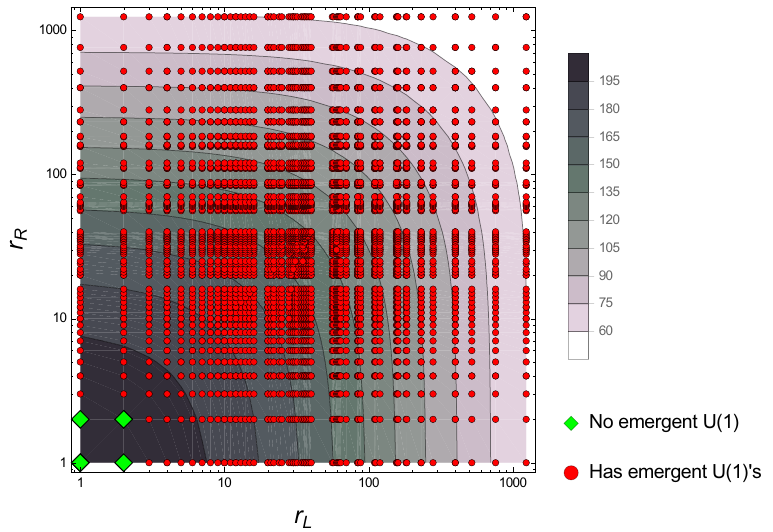}
\caption{($E_8$, $E_8$)}
\label{CMcontourE8}
\end{subfigure}
\caption{Plots of $a_{\mathrm{IR}}$ vs left and right embedding indices for the
different plain mass nilpotent
deformations of 4D conformal matter theories. The contour plots are obtained by
extrapolating between the actual data points which are labelled in green diamonds and
red circles. The green diamonds correspond to deformations where all operators remain
above the unitarity bound and no emergent $U(1)$ appears. The red circles
correspond to deformations where some operators hit the unitarity bound and
emergent $U(1)$'s are present. We emphasize that sometimes different nilpotent orbits can have the
same embedding index. A log-scale is used to spread the dense region
at low values of the embedding indices.}%
\label{CMcontour}%
\end{figure}
Of course, the plots (just like the tables) are symmetric under
the interchange of $r_{L}$ with $r_{R}$. We also see that for any fixed value
of $r_{L}$ the value of $a_{\mathrm{IR}}$ decreases as the deformation on the
right increases (along the Hasse diagram) when the interacting piece plus free
decoupled fields are considered, as well as when central charges of only the
interacting piece are analyzed (the plots for only the interacting piece would
look very similar).

Furthermore, if we simultaneously increase both $r_{L}$ and $r_{R}$ while
keeping $r_{L}=r_{R}$ (along the Hasse diagram), then $a_{\mathrm{IR}}$ monotonically decreases. This is
again consistent with the expectation that the number of degrees of freedom should
decrease as the deformations becomes larger along the RG flows.

Another interesting feature of our numerical sweep is that we sometimes encounter theories where an operator
decouples, but further down the Hasse diagram, we see no apparent unitarity bound violations. This does
not contradict the general structure implied by the nilpotent cone, since deeper down in the Hasse diagram it
often happens that the top spin operator of $\mathfrak{su}(2)_D$ may not be a top-spin operator deeper down in the nilpotent
cone. As we have already explained, the lower spin operators are trivial in the chiral ring of the IR fixed point, so
it is neither here nor there to see a jump in the number of emergent $U(1)$'s as we proceed deeper into the IR.

\begin{figure}[ptb]
\centering
\begin{subfigure}[b]{0.45\textwidth}
\includegraphics[width=\textwidth]{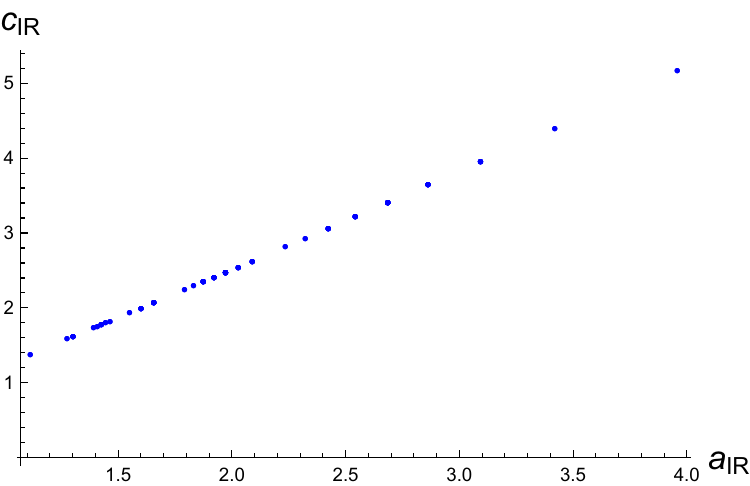}
\caption{$(D_4 , D_4)$}
\label{CMratioD4}
\end{subfigure}
\hspace{12pt} \begin{subfigure}[b]{0.45\textwidth}
\includegraphics[width=\textwidth]{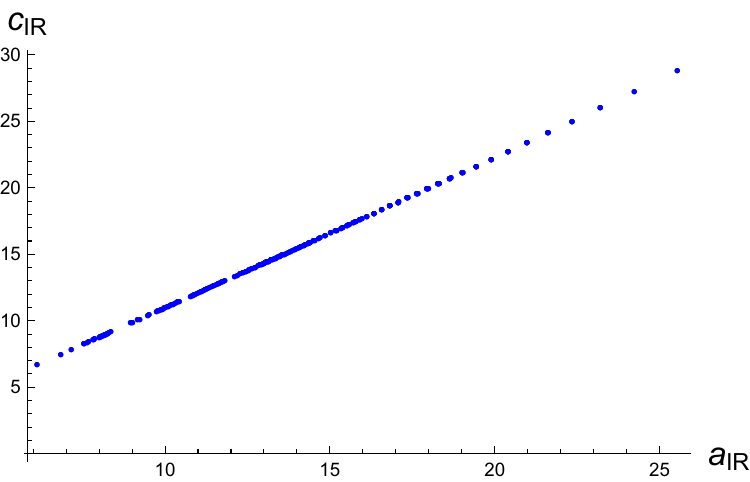}
\caption{$(E_6 , E_6)$}
\label{CMratioE6}
\end{subfigure}
\begin{subfigure}[b]{0.45\textwidth}
\includegraphics[width=\textwidth]{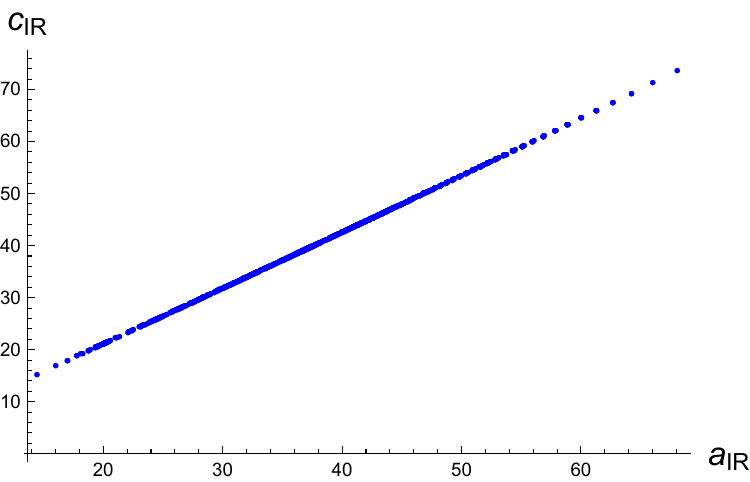}
\caption{$(E_7 , E_7)$}
\label{CMratioE7}
\end{subfigure}
\hspace{12pt} \begin{subfigure}[b]{0.45\textwidth}
\includegraphics[width=\textwidth]{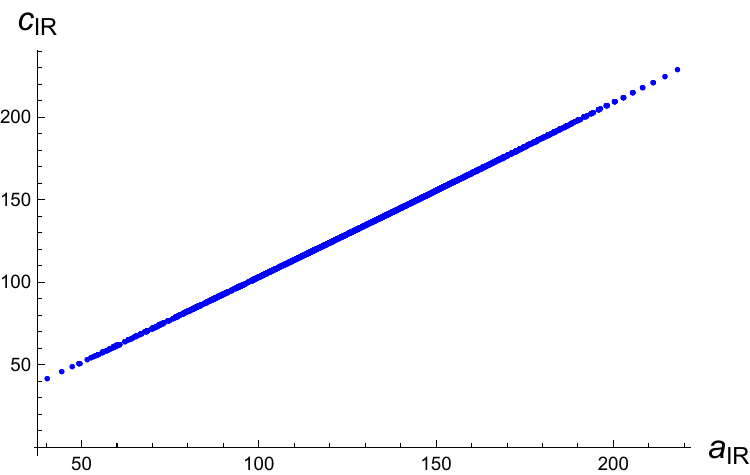}
\caption{$(E_8 , E_8)$}
\label{CMratioE8}
\end{subfigure}
\caption{Plots of $c_{\mathrm{IR}}$ vs $a_{\mathrm{IR}}$ for the different plain mass nilpotent
deformations of 4D conformal matter theories.}%
\label{CMratio}%
\end{figure}

We also determine the overall statistical spread in the value of the ratio
$a_{\mathrm{IR}}/c_{\mathrm{IR}}$ for plain mass deformations of the probe D3-brane theories.
By inspection of the plots in figure \ref{CMratio},
we see that there is a roughly constant value for each theory. We
also calculate the mean and standard deviation by sweeping
over all such theories. Just as in the case of the probe D3-brane theories,
we find that the standard deviation
is on the order of $1\%$ to $5\%$, indicating a remarkably stable value across
the entire network of flows. The specific values are displayed in
table \ref{tab:plainCM}. Another curious feature is that the mean value of
$a_{\mathrm{IR}}/c_{\mathrm{IR}}$ increases as we go to larger UV flavor symmetries.
Precisely the opposite behavior is observed in the nilpotent networks of
probe D3-brane theories.

\begin{table}[H]
\centering
\begin{tabular}
[c]{|c|c|c|c|c|}\hline
& $(D_{4},D_{4})$ & $(E_{6},E_{6})$ & $(E_{7},E_{7})$ & $(E_{8},E_{8}%
)$\\\hline
Mean & $0.80$ & $0.91$ & $0.94$ & $0.97$\\\hline
Std. Dev. & $0.01$ & $0.01$ & $0.004$ & $0.003$\\\hline
Max & $0.81$ & $0.91$ & $0.95$ & $0.98$\\\hline
Min & $0.77$ & $0.89$ & $0.92$ & $0.96$\\\hline
\end{tabular}
\caption{Table of means and standard deviations for the ratio $a_{\mathrm{IR}}/c_{\mathrm{IR}}$ across
the entire nilpotent network defined by plain mass deformations of 4D conformal matter.
We also display the maximum and minimum values.}
\label{tab:plainCM}
\end{table}

\subsection{Flipper Field Deformations}

Finally, we come to flipper field deformations of conformal matter. The analysis is
simplified by the fact that we do not need to worry about mesons decoupling since they are automatically
rescued (if they drop below the unitarity bound) by the flipper fields $M$ to which they couple.

As before, the results with rational values are tabulated in Appendix \ref{completeTables},
and more general deformations can be accessed via the \texttt{Mathematica} code.

Finally, we provide contour plots of $a_{\mathrm{IR}}$ vs. the left and
right embedding indices $r_{L}$ and $r_{R}$. Again, we emphasize that what
really needs to be monotonic is the flow down the Hasse diagram, which in most cases (though not all)
aligns with the increase of the embedding indices $r_L$ and $r_R$. Quite remarkably,
even this coarse data based on the embedding indices (though there are a few exceptions)
usually is enough to establish monotonicity.

\begin{figure}[ptb]
\centering
\begin{subfigure}[b]{0.475\textwidth}
\includegraphics[width=\textwidth]{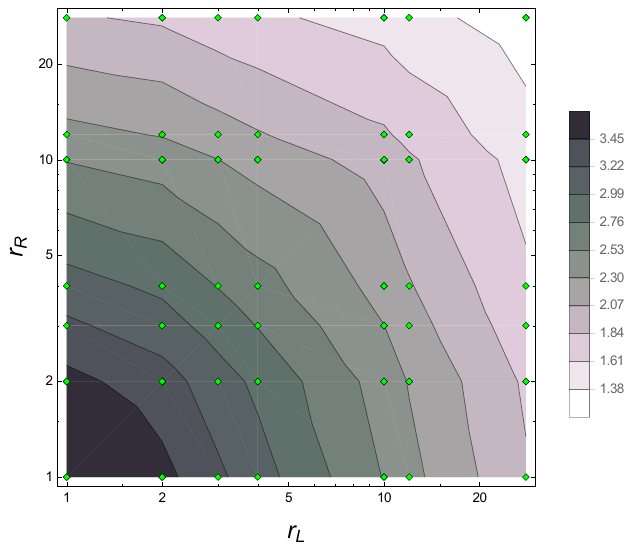}
\caption{($D_4$, $D_4$)}
\label{MSCMcontourD4}
\end{subfigure}
\hspace{12pt} \begin{subfigure}[b]{0.475\textwidth}
\includegraphics[width=\textwidth]{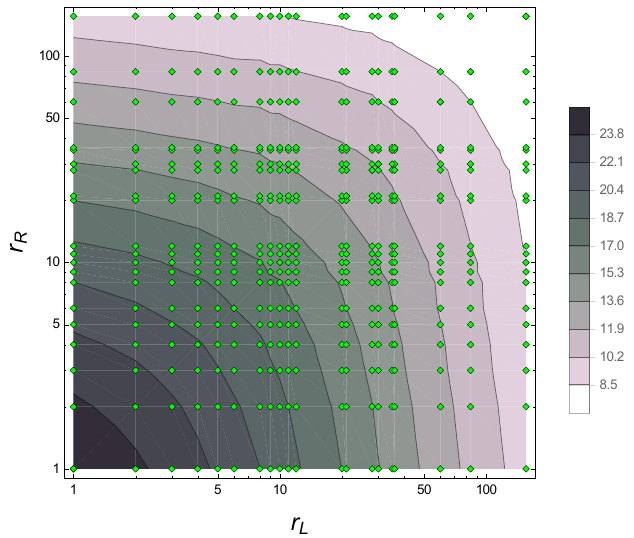}
\caption{($E_6$, $E_6$)}
\label{MSCMcontourE6}
\end{subfigure}
\par
\bigskip
\par
\bigskip\begin{subfigure}[b]{0.475\textwidth}
\includegraphics[width=\textwidth]{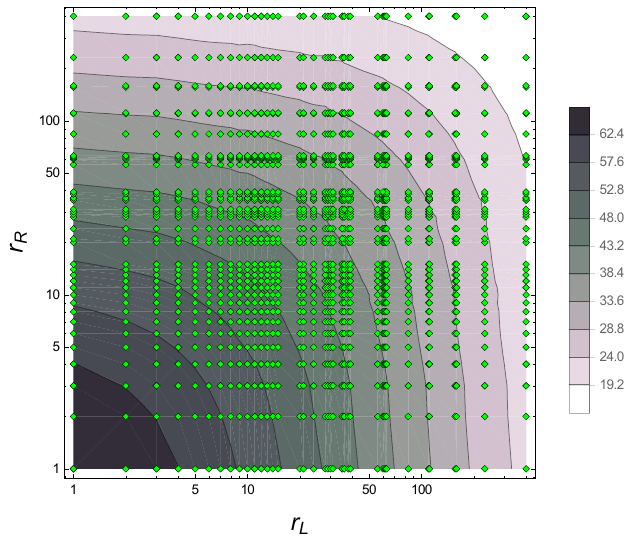}
\caption{($E_7$, $E_7$)}
\label{MSCMcontourE7}
\end{subfigure}
\hspace{12pt} \begin{subfigure}[b]{0.475\textwidth}
\includegraphics[width=\textwidth]{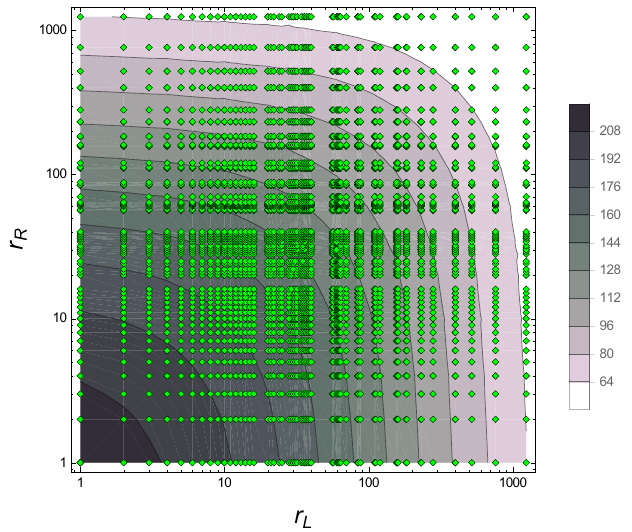}
\caption{($E_8$, $E_8$)}
\label{MSCMcontourE8}
\end{subfigure}
\caption{Plots of $a_{\mathrm{IR}}$ vs left and right embedding indices for the
different conformal matter theories, with flipper field deformations. The
contour plots are obtain by extrapolating between the actual data points which
are labelled in green.}%
\label{MSCMcontour}%
\end{figure}

\begin{figure}[ptb]
\centering
\begin{subfigure}[b]{0.45\textwidth}
\includegraphics[width=\textwidth]{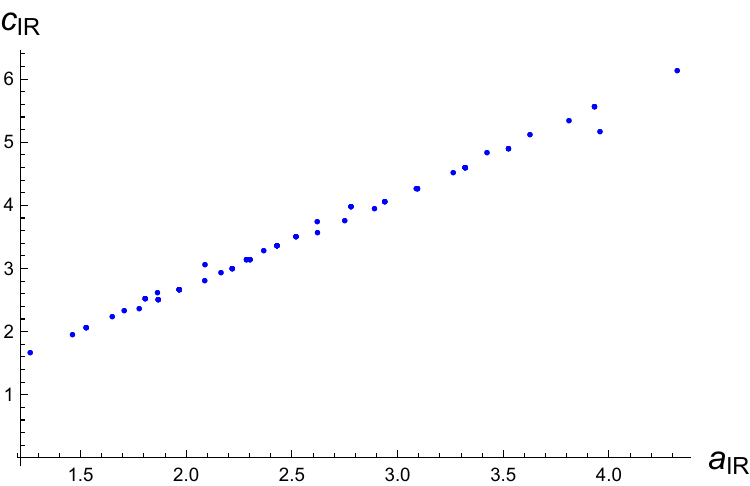}
\caption{$(D_4,D_4)$}
\label{CMMSratioD4}
\end{subfigure}
\hspace{12pt} \begin{subfigure}[b]{0.45\textwidth}
\includegraphics[width=\textwidth]{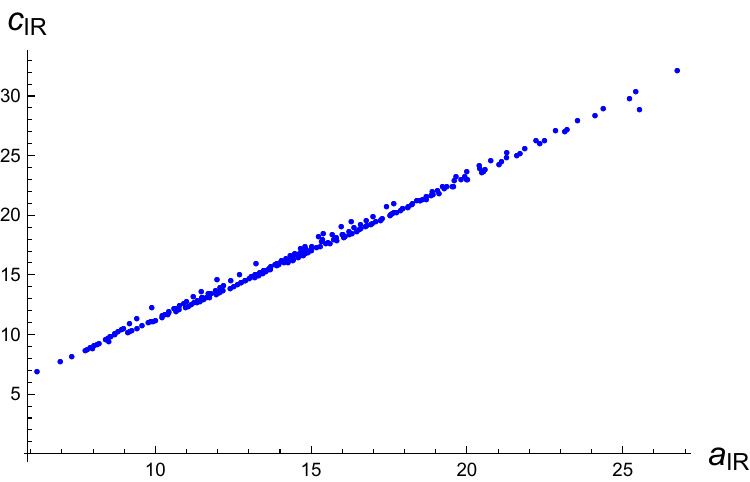}
\caption{$(E_6,E_6)$}
\label{CMMSratioE6}
\end{subfigure}
\begin{subfigure}[b]{0.45\textwidth}
\includegraphics[width=\textwidth]{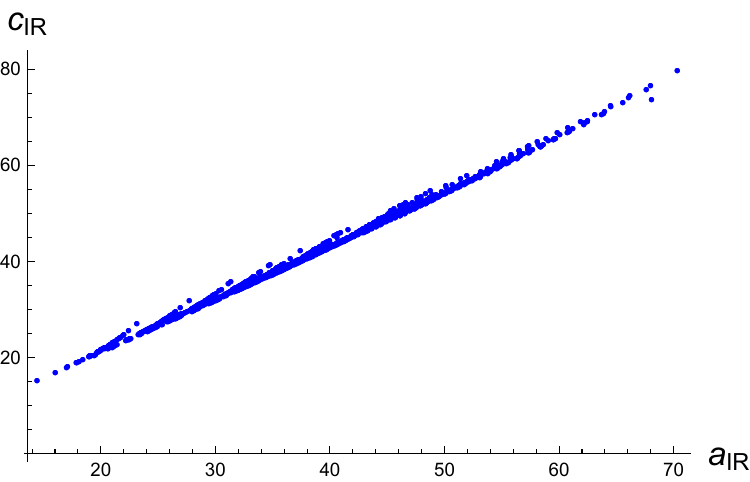}
\caption{$(E_7,E_7)$}
\label{CMMSratioE7}
\end{subfigure}
\hspace{12pt} \begin{subfigure}[b]{0.45\textwidth}
\includegraphics[width=\textwidth]{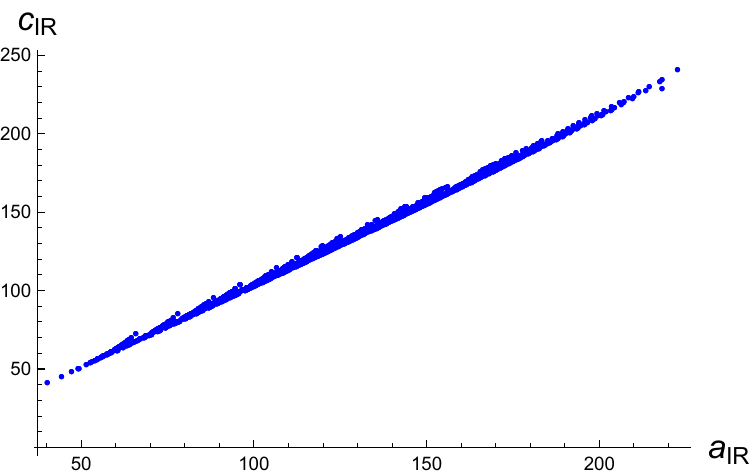}
\caption{$(E_8,E_8)$}
\label{CMMSratioE8}
\end{subfigure}
\caption{Plots of $c_{\mathrm{IR}}$ vs. $a_{\mathrm{IR}}$ for flipper field
deformations of 4D conformal matter.}%
\label{CMMSratio}%
\end{figure}

We also determine the overall statistical spread in the value of the ratio
$a_{\mathrm{IR}}/c_{\mathrm{IR}}$ for flipper field deformations of 4D conformal matter.
By inspection of the plots in figure \ref{CMMSratio},
we see that there is a roughly constant value for each theory. We also
calculate the mean and standard deviation by sweeping
over all such theories, displaying the results
in table \ref{tab:flipCM}. As in all the other cases we have considered, the standard deviation
is on the order of $1\%$ to $5\%$, indicating a remarkably stable value across
the entire network of flows. Another curious feature is that the mean value of
$a_{\mathrm{IR}}/c_{\mathrm{IR}}$ increases as we increase to larger flavor symmetries.
Precisely the opposite behavior is observed in the nilpotent networks of
probe D3-brane theories.

\begin{table}[H]
\centering
\begin{tabular}
[c]{|c|c|c|c|c|}\hline
& $(D_{4},D_{4})$ & $(E_{6},E_{6})$ & $(E_{7},E_{7})$ & $(E_{8},E_{8}%
)$\\\hline
Mean & $0.73$ & $0.87$ & $0.92$ & $0.96$\\\hline
Std. Dev. & $0.01$ & $0.02$ & $0.01$ & $0.01$\\\hline
Max & $0.80$ & $0.91$ & $0.95$ & $0.98$\\\hline
Min & $0.68$ & $0.81$ & $0.86$ & $0.91$\\\hline
\end{tabular}
\caption{Table of means and standard deviations for the ratio $a_{\mathrm{IR}}/c_{\mathrm{IR}}$ across
the entire nilpotent network defined by flipper field deformations of 4D conformal matter.
We also display the maximum and minimum values.}
\label{tab:flipCM}
\end{table}

\newpage

\section{Conclusions \label{sec:CONC}}

One of the important open issues in the study of conformal field theories is
to better understand the totality of fixed points, and their
network of flows under deformations. In this paper we have
shown that a great deal of information on the structure of RG\ flows for
4D\ SCFTs can be extracted in the special case of nilpotent mass deformations.
Starting from a UV\ $\mathcal{N}=2$ SCFT, we have presented a general analysis of the
resulting $\mathcal{N}=1$ fixed points, both in the case of plain mass
deformations, as well as in the generalization to flipper field deformations,
where these parameters are treated as background vevs for a dynamical adjoint
valued $\mathcal{N}=1$ chiral superfield of the parent theory. In addition to
presenting a general analysis of the resulting fixed points, we have performed
an explicit sweep over all possible nilpotent deformations for the
$\mathcal{N}=2$ theories defined by D3-branes probing a $D$- or $E$-type 7-brane,
as well as the nilpotent deformations of 4D $(G,G)$ conformal matter. In both
cases, we have found strong evidence that the mathematical partial ordering
defined by the nilpotent cone of the associated Lie algebras is obeyed in the
physical theories as well. Moreover, the directed graph of this partially
ordered set also lines up with the possible relevant deformations of the
physical theory, providing a very detailed picture of the possible RG\ flows
from one fixed point to another.
The structure of the Hasse diagrams obtained provides a partially ordered set, which cleanly matches to physical 4D RG~flows. We can then take advantage of this fact (even in a more general setting) whenever there is a flavor symmetry present and we activate a breaking pattern generated by a nilpotent orbit.
In addition to presenting the full sweep
over theories in a companion \texttt{Mathematica} program, we have also observed a
number of intriguing \textquotedblleft phenomenological\textquotedblright%
\ features, including the appearance of several theories with
rational anomalies. We have also seen that for a given UV $\mathcal{N} = 2$ fixed point,
the ratio $a_{\mathrm{IR}}/c_{\mathrm{IR}}$ is roughly constant over the entire nilpotent network.
In the remainder of this section we discuss some avenues of further investigation.

One item left open by our analysis is a full treatment of the full network of
RG\ flows in cases where mesonic operators decouple from the new IR\ fixed
point. As we have already explained, such mesonic operators are often
necessary to perform further perturbations deeper down in the Hasse diagram,
so the absence of these operators could a priori pose some issues in the
context of matching the full network defined by the Hasse diagram to corresponding RG\ flows.
Even so, we have not found an explicit example which demonstrates that any
links are in fact \textquotedblleft broken.\textquotedblright\ It would be
most illuminating to further understand this class of theories.

Even within the class of theories considered here, there are some additional
relevant deformations we could contemplate switching on. This includes the
possibility of mass deformations which are semi-simple, namely their matrix
representatives are diagonalizable. Since such diagonal elements can also be
presented as the sum of two nilpotent elements, it is quite likely that the
analysis presented here may implicitly cover such cases as well, and may
actually help to \textquotedblleft explain\textquotedblright\ the appearance
of our rational theories. It would be interesting to analyze this issue further.

The bulk of this paper has focussed on determining various properties of the
new infrared fixed points generated by nilpotent mass deformations, including
the operator scaling dimensions of various operators. Another tractable
quantity to potentially extract is the superconformal index.
This could shed additional light on the IR\ properties
of these theories. Additionally, it would be quite interesting to see whether
there is a corresponding partial ordering for these indices, as induced by the
partial ordering on nilpotent orbits.

Much of our analysis has focussed on the case of a single D3-brane probing an
F-theory 7-brane, as well as the case of \textquotedblleft rank one 6D
conformal matter,\textquotedblright\ namely (in M-theory terms) a single
M5-brane probing an ADE singularity. It would be quite natural to extend the
analysis presented here to the case of additional branes. While the anomalies
for the case of multiple D3-branes have already been determined \cite{Aharony:2007dj},
the corresponding statements for multiple M5-branes probing an
ADE\ singularity, and the resulting 4D\ anomaly polynomial are apparently
unknown. With this result in hand, it would then be possible to study
nilpotent mass deformations for this class of theories as well.

Another natural class of theories involves the compactification of
6D\ conformal matter on more general Riemann surfaces in the presence of
background fluxes and punctures. In this case, even before switching on
nilpotent mass deformations, we expect from the general procedure outlined in
\cite{Benini:2009mz} to get a 4D $\mathcal{N}=1$ SCFT, as in
references \cite{Gaiotto:2015usa,
Ohmori:2015pua, DelZotto:2015rca, Franco:2015jna, Ohmori:2015pia, Coman:2015bqq,
Morrison:2016nrt, Heckman:2016xdl, Razamat:2016dpl, Bah:2017gph, Apruzzi:2017iqe,
Kim:2017toz, Hassler:2017arf, Bourton:2017pee, Kim:2018bpg, Apruzzi:2018oge}.
Many of these theories admit a weakly coupled Lagrangian description \cite{Kim:2018bpg, Razamat:2018gro}, so
studying the possible nilpotent deformation and comparing the central charges
with the class of theories studied here might lead to
Lagrangian descriptions for some of the resulting IR\ fixed points.

Finally, we have also seen a number of numerical coincidences, including the
appearance of rational theories, as well as a relatively constant value for
$a_{\mathrm{IR}}/c_{\mathrm{IR}}$ over an entire nilpotent network.
It would be very interesting to understand whether these
coincidences have a simple top down interpretation.

\section*{Acknowledgements}

We thank J. Kaidi, M. Fazzi, S. Giacomelli, T. Rudelius and F. Xu for useful discussions.
FA, JJH and TBR also thank the 2018 Summer Workshop at the Simons Center for Geometry and Physics
for hospitality during the completion of this work. The work of FA, FH, JJH, and TBR is supported
by NSF CAREER grant PHY-1756996. The work of FA and FH is also supported by NSF grant
PHY-1620311.

\newpage

\appendix

\section{The Embedding Index \label{app:EMBED}}

\label{index} The embedding index $r$ here refers to that of a splitting of
the group $G=D_{4}$, or $E_{6,7,8}$ into irreducible representations (irreps) of $\SU(2)$.
There are two equivalent ways of computing this embedding index $r$. The first method is by
computing the sum of the indices of the $\SU(2)$ irreps divided by the index
of the representation of the group $G$ being split. That is, given a
representation $\rho(G)$ of $G$ and the branching $\rho(G) \rightarrow m_{1}
\mathbf{n}_{1} + m_{2} \mathbf{n}_{2} + \dots$ where $m_{(a)}$ are
multiplicities and $\mathbf{n}_{(a)}$ are $\SU(2)$ irreps, the embedding index
is given by:
\begin{align}
\label{rindex1}r  & = \frac{\sum_{(a)} m_{(a)} \cdot\mathrm{ind}%
(\mathbf{n}_{(a)})}{\mathrm{ind}(\rho(G))}.
\end{align}
For instance the splitting of $D_{4}$ according to the partition $[5,3]$ gives:
$\mathbf{28} \rightarrow3(\mathbf{3})+(\mathbf{5})+2(\mathbf{7})$ so that
\begin{align}
r=\frac{3\times4+ 20 +2 \times56}{12} = 12
\end{align}

As we can see, this definition of the embedding index is representation
independent. However it requires that we know the branching rule of splitting
of $G$ to $\SU(2)$ caused by the deformation of interest.

For this reason, we turn to the second method which makes use of the decorated Dynkin diagrams provided in
 \cite{Chacaltana:2012zy} for the exceptional groups.
Their labels specify a vector $v$ in the Cartan subalgebra which then yields
the projection matrix $\mathbb{P} = v \cdot C_{\mathfrak{g}}^{-1}$. $C_{\mathfrak{g}}$
is the Cartan matrix of the Lie algebra ${\mathfrak{g}}$, and $\mathbb{P}$ is the
projection matrix of the weights of ${\mathfrak{g}}$ into the $SU(2)_{D}$ nilpotent subalgebra.
As a result the decorated Dynkin diagrams can be directly used to obtain the branching rules and the embedding indices,
\begin{equation} \label{rindex}
r  = \frac{1}{2} \, \mathrm{Tr}(v\cdot C_{\mathfrak{g}}^{-1} \cdot v^{T})
\end{equation}
where the $\frac{1}{2}$ coefficient is simply a normalization factor.

Now, for $D_{4}$ we do not have the decorated Dynkin diagrams readily
available to us, so we need to compute them. We start with the 12 possible
partitions of $\SO(8)$ provided by \cite{rakotoarisoa2017bala}. Following
this procedure along with \cite{Haouzi:2016yyg} one can obtain the vectors
$v$ for $\SO(2k)$ in the same form as the ones provided by
\cite{Chacaltana:2012zy} for the exceptional groups. In summary the procedure
is as follows:

We begin by listing the possible partitions of $\SO(2k)$: $p_{i} = \{n_{l}\}$
where $i$ runs over the number of possible nilpotent deformations of $\SO(2k)$
and $n_{l}$ are integers summing to $2k$. The nilpotent deformation defines an
$\SU(2)$ subalgebra $[H,X]=2X$, $[H,X^{\dagger}]=-2X^{\dagger}$,
$[X,X^{\dagger}]=H$ where $X$ is the nilpotent orbit/deformation. $X$ is
directly constructed from the partitions: $X$ is a $2k \times2k$ matrix filled
on the first superdiagonal by the Jordan blocks corresponding to the $\SU(2)$
irreps defined by the partitions. Namely $\sqrt{j(j+1)-m(m+1)}$ where $-j \leq
m \leq j-1$. For instance, the $\SO(10)$ partition $\{7,3\}$ yields two Jordan
blocks. $X$ is zero everywhere except on the first super diagonal which is
given by the list $(\sqrt{6},\sqrt{10},\sqrt{12},\sqrt{12},\sqrt{10},\sqrt
{6},0,\sqrt{2},\sqrt{2})$ where for the first block (which defines the first 6
entries) we have $j=3$ and for the second block (which defines the last 2
entries) we have $j=1$.

Then the corresponding Cartan matrix $H$ is given by $[X,X^{\dagger}]=H$,
which is a diagonal matrix whose entries are then sorted in increasing order.
Furthermore, $\SO(2k)$ has $k$ Cartan matrices $H_{q}$ with $q=1,\cdots,k$.
The projection matrix (or just vector here) is $\alpha= \{\alpha_{i}\}$ given
by solving the linear equations:
\begin{align}
\label{H}\sum_{i=1}^{k} \alpha_{i} H_{i} = H
\end{align}
and the decorated Dynkin diagrams are given by the vector $v = \alpha\cdot C_{\SO(2k)}$.
Each partition yields a different $H$ and therefore a different set of
equations \eqref{H} and Dynkin labels $v$.

We should note that this analysis makes extensive use of the \texttt{LieArt} 
package of reference \cite{Feger:2012bs}.

\subsection*{$SO(8)$ Example}

To illustrate we work out an example with $\SO(8)$ in detail:

One partition of $\SO(8)$ is given by $[5,3]$. So the raising operator matrix
is:
\begin{align}
X =
\begin{pmatrix}
0 & 2 & 0 & 0 & 0 & 0 & 0 & 0\\
0 & 0 & \sqrt{6} & 0 & 0 & 0 & 0 & 0\\
0 & 0 & 0 & \sqrt{6} & 0 & 0 & 0 & 0\\
0 & 0 & 0 & 0 & 2 & 0 & 0 & 0\\
0 & 0 & 0 & 0 & 0 & 0 & 0 & 0\\
0 & 0 & 0 & 0 & 0 & 0 & \sqrt{2} & 0\\
0 & 0 & 0 & 0 & 0 & 0 & 0 & \sqrt{2}\\
0 & 0 & 0 & 0 & 0 & 0 & 0 & 0
\end{pmatrix}
\end{align}
and the corresponding Cartan matrix $H=[X,X^{\dagger}] = \text{diag}%
(4,2,2,0,0,-2,-2,-4)$ after sorting out the entries.

The 4 Cartans of $\SO(8)$ are given by:
\begin{align}
\label{cartansSO8}H_{1} & =\text{diag}(1,-1,0,0,0,0,1,-1)\\
H_{2} & =\text{diag}(0,1,-1,0,0,1,-1,0)\\
H_{3} & =\text{diag}(0,0,1,-1,1,-1,0,0)\\
H_{4} & =\text{diag}(0,0,1,1,-1,-1,0,0)
\end{align}
where we are using the mathematician's conventions to be consistent with the
use of the \texttt{LieArt} package.

The projection matrix $\alpha=(\alpha_{1}, \alpha_{2}, \alpha_{3}, \alpha
_{4})$ is then obtained by solving the equation:
\begin{align}
\label{projectionsolve}\alpha_{1} H_{1} + \alpha_{2} H_{2} + \alpha_{3} H_{3}
+ \alpha_{4} H_{4} = H
\end{align}
which yields:
\begin{align}
\label{projection}\alpha= (4,6,4,4).
\end{align}
Thus given the Cartan matrix:
\begin{align}
\label{CartanMatrix}C_{\SO(8)} =
\begin{pmatrix}
2 & -1 & 0 & 0\\
-1 & 2 & -1 & -1\\
0 & -1 & 2 & 0\\
0 & -1 & 0 & 2
\end{pmatrix}
\end{align}
the decorated Dynkin diagram specifies a vector $v = \alpha\cdot C_{\SO(8)}$ given by:
\begin{align}
\label{label}v = (2,0,2,2)
\end{align}

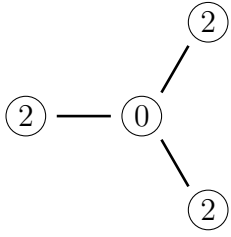
\begin{figure}
\centering
\begin{tikzpicture}[]
\node[circle,draw=black, fill=white, inner sep=0pt,minimum size=15pt]   (p1)                 {2};
\node[circle,draw=black, fill=white, inner sep=0pt,minimum size=15pt]   (p2) [right=1.0cm of p1]  {0};
\node[circle,draw=black, fill=white, inner sep=0pt,minimum size=15pt]   (p3) [above right=0.866cm and 0.5cm of p2] {2};
\node[circle,draw=black, fill=white, inner sep=0pt,minimum size=15pt]   (p4) [below right=0.866cm and 0.5cm of p2] {2};
\draw[-, line width=1pt] (p1.east) -- (p2.west);
\draw[-, line width=1pt] (p2.north east) -- (p3.south west);
\draw[-, line width=1pt] (p2.south east) -- (p4.north west);
\end{tikzpicture}
\caption{Decorated Dynkin diagram for the $[5,3]$ partition of $\SO(8)$ }%
\label{DynkinLabel}%
\end{figure}

This procedure is repeated for every partition of $SO(2k)$ so as to obtain all
of the necessary decorated Dynkin diagrams and projection matrices.


\section{From 6D to 4D Conformal Matter \label{CMappendix}}

In this Appendix we collect some features of 6D conformal matter and its compactification on a $T^2$.
At long distances, this yields a 4D $\mathcal{N} = 2$ SCFT. Here, we review both the scaling dimensions
of Coulomb branch operators and the anomalies of these theories.

\subsection*{Coulomb Branch Operators}

In this subsection we calculate the scaling dimension of the operators parameterizing the Coulomb branch.
This data follows directly from the analysis of references \cite{Ohmori:2015pua, DelZotto:2015rca, Ohmori:2015pia}.
Our main task here is to extract from this analysis the corresponding scaling dimensions. References \cite{Ohmori:2015pua,
Ohmori:2015pia} implicitly give this information by showing that 4D $\mathcal{N} = 2$ $(G,G)$ conformal matter is actually a
compactification of a class $\mathcal{S}$ theory, specifying the corresponding Gaiotto curve as well. In reference \cite{DelZotto:2015rca}
the corresponding Seiberg-Witten curve is obtained via the mirror to the elliptically fibered Calabi-Yau threefold of
the F-theory background used to produce the 6D SCFT. Observe that F-theory compactified on a $T^2$ yields IIA on the same elliptic threefold,
and mirror symmetry takes us to type IIB. The advantage of the IIB presentation is that now the Coulomb branch is parameterized in
terms of the complex structure of this mirror geometry.

We opt to use the explicit Calabi-Yau geometries presented in reference \cite{DelZotto:2015rca}.
To aid comparison with the results of this reference, we refer to the theory of 6D conformal
matter with $(G,G)$ flavor symmetry given by $N$ M5-branes probing an ADE singularity as $\mathcal{T}(G,N)$.
In this paper we focus exclusively on the case $N = 1$.

We now use the results of reference \cite{DelZotto:2015rca} on the associated
mirror geometries to compute  the scaling dimensions of the Coulomb branch for the
theories $\mathcal T(E_{6,7,8},1)$, on $T^2$. This method has been used before
for $\mathcal{N} = 2$ SCFTs, and is essentially adapted from the technique presented
in reference \cite{Argyres:1995xn}.

The IIB mirror geometry for $\mathcal T(E_{6},1)$ on $T^2$ is given by the following local Calabi-Yau threefold:
\be
 \begin{cases}
 &f=w^2+x_1^3+x_2^2\rho+\rho^2+(m_1+m_1^{'} y_1 )x_1x_2^2+(m_2+m_2^{'} y_1) x_1x_2+  (m_3 +u_1 y_1+m_3^{'}y_1^2) x_2^2 \nonumber\\
 & ~~~+ (m_4+u_2 y_1+m_4^{'} y_1^2) x_1+(m_5+u_3 y_1+m_5^{'}y_1^2) x_2+(m_6+u_4y_1+u_5y_1^2+m_6^{'}y_1^3)=0  \nonumber\\
& \rho=(1+y_1+y_2),\\
& x^2_2=\rho.
\end{cases}
\label{eq:E6mirrorcurve}
\ee
where $y_1$ is a $\mathbb C^*$ coordinate, $x_1,x_2,w,\rho$ are complex coordinates, $m_i$ are general mass parameters and $u_i$ are the coulomb branch operator vevs,
\begin{equation}
u_i \equiv \langle Z_i \rangle,
\end{equation}
$f$ is a homogeneous polynomial in the complex coordinates and it scales as follows:
\begin{equation}
f(\lambda^ax_1,\lambda^b x_2, \lambda^c \rho, \lambda^d w,y_1)= \lambda^e f(x_1,x_2,\rho,w,y_1).
\end{equation}
The holomorphic three-form is defined as follows
\begin{equation}
\Omega = \frac{dx_1 \wedge dx_2}{w}\wedge \frac{dy_1}{y_1}
\end{equation}
By fixing the scale of $\Omega(\lambda^ax_1,\lambda^b x_2, \lambda^c \rho, \lambda^d w,y_1) =\lambda \Omega(x_1,x_2,\rho,w,y_1)$ to the unity, i.e. $[\Omega]=1$, the first four monomials of $f$ uniquely fix the other scalings
\begin{equation}
[x_1]=a=4, \quad [x_2]=b=3, \quad [\rho]=c=[w]=d=6, \quad [f]=e=12.
\end{equation}
Recalling that $y_1$ does not scale since it is just a phase, we obtain the scaling dimension of the Coulomb branch parameters,
\begin{equation}
[u_1]=6, \quad [u_2]=8, \quad [u_3]=9, \quad [u_4]=[u_5]=12.
\end{equation}
This agrees with the scaling dimensions of the Coulomb branch operators for the class $\mathcal{S}$
trinions with two minimal and one maximal puncture in \cite{Chacaltana:2014jba}.

The IIB mirror Calabi-Yau for $\mathcal T(E_{7},1)$ on $T^2$ is described by
\be
 \begin{cases}
 &f = x_1^2+x_2^3\rho+\rho^3+(m_1+m_1^{'}y_1) x_2\rho^2+(m_2+u_1y_1+m_2^{'} y_1^2)\rho^2 + \nonumber\\
& ~~(m_3+u_2y_1+m_3^{'} y_1^2) x_2\rho +(m_4+u_3 y_1+m_4^{'} y_1^2)x_2^2 +(m_7+u_4 y_1+u_5y_1^2+m_4^{'} y_1^3)\rho +\nonumber\\
&(m_6+u_6 y_1+u_7y_1^2+m_5^{'} y_1^3)x_2 +(m_7+u_8y_1+u_9y_1^2+u_{10}y_1^3+m_7^{'}y_1^4)=0.   \nonumber\\
&\rho=(1+y_1+y_2).
 \end{cases}
 \ee
where again $y_1$ is a $\mathbb C^*$ coordinate, and $x_1,x_2,\rho$ are complex coordinates.
The homogeneous polynomial $f$ scales as follows:
\begin{equation} \label{eq:homo}
f(\lambda^ax_1,\lambda^b x_2, \lambda^c \rho,y_1)= \lambda^e f(x_1,x_2,y_1).
\end{equation}
The holomorphic three-form reads
\begin{equation} \label{eq:holo3form}
\Omega = \frac{dx_2 \wedge d\rho}{x_1}\wedge \frac{dy_1}{y_1}
\end{equation}
and we impose that it scales like $[\Omega]=1$. The first three monomials again fix the scaling of the complex coordinates and of $f$:
\begin{equation}
[x_1]=a=9, \quad [x_2]=b=4, \quad [\rho]=c=6, \quad [f]=e=18.
\end{equation}
By looking at the scaling of the other monomials involving the Coulomb branch vevs, the scaling dimensions of the Coulomb branch parameters are assigned
\begin{align}
& [u_1]=6, \quad [u_2]=8, \quad [u_3]=10, \quad [u_4]=[u_5]=12, \nonumber \\
& \quad [u_6]=[u_7]=14, \quad [u_8]=[u_9]=[u_{10}]=18.
\end{align}
This agrees with the scaling dimensions of the Coulomb branch operators for the class $\mathcal{S}$
trinions with two minimal and one maximal puncture in \cite{Chacaltana:2017boe}.

The IIB mirror Calabi-Yau for $\mathcal T(E_{8},1)$ on $T^2$ is described by
 \be
 \begin{cases}
 &f = x_1^2+x_2^3+\rho^5+(m_1+m_1^{'}y_1) x_2\rho^3+(uy_1^2)\rho^4+(m_2+u_1y_1+m_2^{'} y_1^2)x_2\rho^2+ \nonumber\\
& (m_3+u_2y_1+u_3y_1^2+m_3^{'} y_1^3) \rho^3 +(m_4+u_4 y_1+u_5y_1^2+m_4^{'} y_1^3) x_2\rho+\nonumber\\
&(m_5+u_6 y_1+u_7y_1^2+u_8y_1^3+m_5^{'} y_1^4) \rho^2+(m_6+u_9y_1+u_{10}y_1^2+u_{11}y_1^3+m_6^{'}y_1^4)x_2+ \nonumber\\
&(m_7+u_{12}y_1+u_{13}y_1^2+u_{14}y_1^3+u_{15}y_1^4+m_7^{'}y_1^5)\rho+  \nonumber\\
&((m_8+u_{16}y_1+u_{17}y_1^2+u_{18}y_1^3+u_{19}y_1^4+u_{20}y_1^5+m_8^{'}y_1^6)=0; \nonumber\\
&\rho=(1+y_1+y_2).
 \end{cases}
 \ee
where again $y_1$ is a $\mathbb C^*$ coordinate, and the $x_1,x_2,\rho$ are complex
coordinates. The homogeneous polynomial $f$ scales as in equation \eqref{eq:homo}.
The holomorphic three-form is analogous to the $E_7$ case, \eqref{eq:holo3form}.
By imposing $[\Omega]=1$, the first three monomials of $f$ fix the scaling of the coordinates,
\begin{equation}
[x_1]=a=15, \quad [x_2]=b=10, \quad [\rho]=c=6, \quad [f]=e=30.
\end{equation}
The other monomials involving the Coulomb branch vevs automatically assign the following scaling dimensions
\begin{align}
& [u]=6, \quad [u_1]=8, \quad [u_2]=[u_3]=12,\quad [u_4]=[u_5]=14, \quad [u_6]=[u_7]=[u_8]=18, \nonumber \\
&[u_9]=[u_{10}]=[u_{11}]=20, \quad [u_{12}]=[u_{13}]=[u_{14}]=[u_{15}]=24, \nonumber  \\
&[u_{16}]=[u_{17}]=[u_{18}]=[u_{19}]=[u_{20}]=30.
\end{align}
This agrees with the scaling dimensions of the Coulomb branch operators for the class $\mathcal{S}$ trinions with two minimal and one maximal puncture in \cite{Chcaltana:2018zag}.

Finally, for the $D_{k}$ conformal matter theories $\mathcal T(SO(2k),1)$ with $k>2$ on $T^2$ the scaling dimensions of the
Coulomb branch operators can be read off in a similar way from the curve (5.4) in \cite{Ohmori:2015pua}.

\subsection*{Anomaly Polynomials}

Given the importance of the UV anomalies we now review
how they were obtained in table \ref{CMUVscaling}. When studying an M5-brane
probing $D$- and $E$-type singularities we obtain 6D SCFTs also called ($G$,$G$)
6D conformal matter with anomaly polynomial:
\begin{align}
\label{eqn:I8CM}I_{8}=\alpha c_{2}(R_{6D})^{2}+\beta c_{2} (R_{6D}%
)p_{1}(T)+\gamma p_{1}(T)^{2}+\delta p_{2}(T)+\kappa_{L} p_{1}(T) \frac{\,
\mathrm{Tr}(F_{L}^{2})}{4}+\kappa_{R} p_{1}(T) \frac{\, \mathrm{Tr}(F_{R}%
^{2})}{4} + \dots
\end{align}
where the explicit expression for the 6D anomaly polynomial coefficients were
computed in \cite{Ohmori:2014kda}, and are listed in table~\ref{anoPolyCoeff}.
\begin{table}[h]
\centering
{\renewcommand{\arraystretch}{1.2}
\begin{tabular}
[c]{|c||c|c|c|c|}\hline
$(G,G)$ & $(D_k,D_k)$ & $(E_{6},E_{6})$ & $(E_{7},E_{7})$ & $(E_{8},E_{8})$\\\hline
$24 \alpha$ & $10k^{2}-57k+81$ & 319 & 1670 & 12489\\\hline
$48 \beta$ & $-(2k^{2}-3k-9)$ & -89 & -250 & -831\\\hline
$\frac{5760}{7} \gamma$ & $k(2k-1)+1$ & 79 & 134 & 249\\\hline
$\frac{5760}{4} \delta$ & $-\left( k(2k-1)+1\right) $ & -79 & -134 &
-249\\\hline
$24\kappa_{L}=24\kappa_{R} $ & $2k-2$ & 12 & 18 & 30\\\hline
\end{tabular}
}\caption{Coefficients of 6D anomaly polynomial \eqref{eqn:I8CM}}%
\label{anoPolyCoeff}%
\end{table}

In order to obtain a 4D $\mathcal{N}=2$ SCFT, we compactify these theories on
$T^{2}$ and consider the general anomaly polynomial for a 4D theory
\begin{align}
\label{eqn:I6}I_{6} = \frac{k_{RRR}}{6}c_{1}(R)^{3}-\frac{k_{R}}{24}%
p_{1}(T)c_{1}(R)+k_{RG_{L}G_{L}}\frac{\, \mathrm{Tr}(F^{2}_{G_{L}})}{4}%
c_{1}(R)+k_{RG_{R}G_{R}}\frac{\, \mathrm{Tr}(F^{2}_{G_{R}})}{4}c_{1}(R)+
\dots\,,
\end{align}
where $R=R_{\mathrm{UV}}$ is the R-symmetry of the UV $\mathcal{N} = 2$ SCFT,
viewed as an $\mathcal{N} =1$ SCFT, $T$ is the
formal tangent bundle, $F$ is the field strength of $G_{L}$ or $G_{R}$ flavor
symmetries, and the dots indicate possible abelian flavor symmetries and mixed
contributions. Moreover we have the following relations
\begin{align}
\, \mathrm{Tr}(R^{3})=k_{RRR}, &  & \, \mathrm{Tr}(R)=k_{R}, &  & \,
\mathrm{Tr}(RF^{A}_{G_{L,R}}F^{B}_{G_{L,R}})=-\frac{k_{RG_{L,R}G_{L,R}}}%
{2}\delta^{AB} \,. &
\end{align}
From them, the definition of $R$, and
\begin{align}
\, \mathrm{Tr}\left( R_{\mathcal{N}=2}F^{A}_{G_{L,R}}F^{B}_{G_{L,R}}\right)
=-\frac{k_{L,R}}{2} \delta^{AB}%
\end{align}
we read off the anomalies
\begin{align}
a_{\mathrm{UV}}  & = \frac{9}{32}k_{RRR}-\frac{3}{32}k_{R}\\
c_{\mathrm{UV}}  & = \frac{9}{32}k_{RRR}-\frac{5}{32}k_{R}\\
k_{L}  & = 3k_{RG_{L}G_{L}}\\
k_{R}  & = 3k_{RG_{R}G_{R}}\,.
\end{align}
In terms of the 6D anomaly polynomial coefficients
\cite{Ohmori:2015pua,Ohmori:2015pia}, we finally identify
\begin{align}
a_{\mathrm{UV}}  & = 24\gamma-12\beta-18\delta\\
c_{\mathrm{UV}}  & = 64\gamma-12\beta-8\delta\\
k_{L}  & = 48\kappa_{L}\\
k_{R}  & = 48\kappa_{R}\,.
\end{align}
Once evaluated at the values of table \ref{anoPolyCoeff} the above equations
yield exactly the UV values of table \ref{CMUVscaling}, as expected.


\section{Accessing the Complete Tables \label{completeTables}}

Included with the \texttt{arXiv }submission is a set of \texttt{Mathematica}
scripts which can be used to access the full set of theories generated by
nilpotent deformations of the $\mathcal{N}=2$ theories considered in this
paper. Indeed, due to the rather large size of the dataset it is impractical to list all
of our results in the format of a paper.

Instead we have written a \texttt{Mathematica} script which outputs the
complete list of all possible nilpotent deformations for the theories
described above. The necessary files are attached to this paper. To access
them, first proceed to the \texttt{arXiv} abstract page for this paper. On the
righthand side, there is a box with the title \textquotedblleft
Download.\textquotedblright\ Click on \textquotedblleft Other
formats\textquotedblright\ and then download the source files for the
\texttt{arXiv} submission.

To access the full database, one simply needs to download the following six
files and store them in the same folder: \textquotedblleft
ProbeD3brane.m\textquotedblright, \textquotedblleft
ConformalMatter.m\textquotedblright, \textquotedblleft
ProbeD3braneFlavorK.m\textquotedblright, \textquotedblleft
ConformalMatterFlavorK.m\textquotedblright, \textquotedblleft
NilpotentDeformations.m\textquotedblright, \textquotedblleft
Results.nb\textquotedblright. Essentially, the first file contains all of the
information for nilpotent deformations of the probe D3-brane theories (with and without
flipper field deformations), except for the flavor central charge. The second
file stores all of the information for the nilpotent deformations of 4D conformal matter (with
and without flipper field deformations), except for the flavor central
charge. The next two files contain all of the information about the flavor
central charges for the Minahan-Nemeshansky and conformal matter theories
respectively. The file \textquotedblleft
NilpotentDeformations.m\textquotedblright\ does all of the formatting, and
finally the code \textquotedblleft Results.nb\textquotedblright\ loads the
previous three packages and outputs the results. Thus the only file the user
needs to run and worry about is the last one: \textquotedblleft
Results.nb\textquotedblright. When running this file the user is provided
with a list of options:

\begin{enumerate}
\item First one can choose between the four kinds of deformations:
probe D3-brane theories with plain mass deformations, probe D3-brane theories with
flipper field deformations, 4D conformal matter with plain mass deformations,
and 4D conformal matter with flipper field deformations.

\item Secondly one can choose between the $a_{\mathrm{IR}}$, $c_{\mathrm{IR}}$
anomalies and operator scaling dimensions or the tables with the flavor central charges.

\item Then the user should select the flavor groups: $D_{4}$, $E_{6}$, $E_{7}%
$, or $E_{8}$ for deformations of the probe D3-brane theories, and $(D_{4},D_{4})$,
$(E_{6},E_{6})$, $(E_{7},E_{7})$, or $(E_{8},E_{8})$ for deformations of 4D conformal matter.

\item If a probe D3-brane theory is selected then the user can
choose from two options:

\begin{enumerate}

\item select a single deformation by choosing the Bala-Carter label (or
partition of $D_{4}$) of the flavor group from the provided popup menu below.

\item select the whole table.
\end{enumerate}

\item If instead a 4D conformal matter theory is selected the user has three
options:

\begin{enumerate}

\item select a single deformation chosen by selecting the left and right
Bala-Carter labels (or partitions of $D_{4}$) for the breaking of the left and
right flavors.

\item select all of the deformations with a given left (or right) deformation,
by selecting a single Bala-Carter label (or partition of $D_{4}$).

\item select the whole table.
\end{enumerate}

\item The resulting table is then outputted. We also provide for the probe D3-brane theories
the branching rules from the adjoint of $G$ to the $\SU(2)$ irreps for the selected deformations.
\end{enumerate}

Finally, due to the form of the general equations used to compute the central
charges it is clear that all of our results are algebraic numbers. However not all are
rational. To differentiate the two in the tables we list the rational values
exactly (by keeping their rational form) while we only give numerical values
for the ones with irrational central charges.

For the convenience of the reader, in the following
subsections we list the explicit tables for all of the
nilpotent deformations of the probe D3-brane theory with $SO(8)$ flavor symmetry,
but only the rational theories for the other nilpotent networks.

As a point of notation, here we make reference to $K_{\mathrm{IR}}$ as well as $k_{\mathrm{IR}}$. 

\subsection*{Nilpotent Network for $SU(2)$ with Four Flavors}

\begin{table}[H]
  \centering
  \begin{adjustbox}{center}
\scalebox{1.0}{{\renewcommand{\arraystretch}{1.2} $\begin{array}{|c|c|c|c|c|c|c|}
 \hline
  \text{[B-C]} & \text{r} & a_{\text{IR}} & c_{\text{IR}} & t_* & \Delta _{\text{IR}}\text{(Z)} & \text{Min(}\Delta _{\text{IR}}\text{($\cO$'s))} \\
  \hline
  \left[1^8\right] & 0       & \frac{23}{24} & \frac{7}{6} &  \frac{2}{3} & \numprint{2.} & \numprint{2.} \\
  \left[2^2,1^4\right]       & 1  & \numprint{0.797038} & \numprint{0.955454} &  \numprint{0.506932} & \numprint{1.5208} & \numprint{1.4792}\\
  \left.\text{[3,}1^5\right] & 2  & \numprint{0.710272} & \numprint{0.846192} &  \numprint{0.434945} & \numprint{1.30483} & \numprint{1.69517} \\
  \left[2^4\text{]II}\right. & 2  & \numprint{0.710272} & \numprint{0.846192} &  \numprint{0.434945} & \numprint{1.30483} & \numprint{1.69517} \\
  \left[2^4\text{]I}\right.  & 2  & \numprint{0.710272} & \numprint{0.846192} &  \numprint{0.434945} & \numprint{1.30483} & \numprint{1.69517} \\
  \text{[3,}2^2\text{,1]}    & 3  & \numprint{0.651529} & \numprint{0.773372} &  \numprint{0.389898} & \numprint{1.16969} & \numprint{1.53788} \\
  \left[3^2,1^2\right]       & 4  & \numprint{0.607635} & \numprint{0.71946} &  \numprint{0.357838} & \numprint{1.07351} & \numprint{1.38973} \\
  \left[4^2\text{]I}\right.  & 10 & \{\numprint{0.452668},\numprint{0.473501}\} & \{\numprint{0.498618},\numprint{0.540284}\} &  \numprint{0.247886} & \numprint{1.} & \numprint{1.51269} \\
  \left[4^2\text{]II}\right. & 10 & \{\numprint{0.452668},\numprint{0.473501}\} & \{\numprint{0.498618},\numprint{0.540284}\} &  \numprint{0.247886} & \numprint{1.} & \numprint{1.51269} \\
  \left.\text{[5,}1^3\right] & 10 & \{\numprint{0.452668},\numprint{0.473501}\} & \{\numprint{0.498618},\numprint{0.540284}\} &  \numprint{0.247886} & \numprint{1.} & \numprint{1.51269} \\
  \text{[5,3]}               & 12 & \{\numprint{0.430022},\numprint{0.450856}\} & \{\numprint{0.467234},\numprint{0.508901}\} &  \numprint{0.227913} & \numprint{1.} & \numprint{1.63252} \\
  \text{[7,1]}               & 28 & \{\numprint{0.345121},\numprint{0.365954}\} & \{\numprint{0.348767},\numprint{0.390434}\} &  \numprint{0.151192} & \numprint{1.} & \numprint{1.63928} \\
 \hline
    \end{array}$}}
\end{adjustbox}

  \begin{adjustbox}{center}
\scalebox{0.9}{{\renewcommand{\arraystretch}{1.2} $\begin{array}{|c|c|c|c|}
 \hline
  \text{[B-C]} & \text{SU(2})_{D}\text{$\times $Residual}  & k_{\text{IR}}\text{ interact} & k_{\text{IR}}\text{+free} \\
  \hline
  \text{[$1^8$]} & \text{SO(8)}  & 4 & 4 \\
 \text{[$2^2,1^4$]} & \text{SU(2)}\times \text{SO(4)}\times \text{SU(2)} & \{\numprint{3.04159},\numprint{3.04159}\} &  \{\numprint{3.04159},\numprint{3.04159}\} \\
 \text{[$3,1^5$]} & \text{SU(2)}\times \text{SO(5)}  & \{\numprint{2.60967}\} & \{\numprint{2.60967}\} \\
 \text{[$2^4$]II} & \text{SU(2)}\times \text{Sp(4)}  & \{\numprint{2.60967}\} & \{\numprint{2.60967}\} \\
 \text{[$2^4$]I} & \text{SU(2)}\times \text{Sp(4)}  & \{\numprint{2.60967}\} & \{\numprint{2.60967}\} \\
 \text{[$3,2^2,1$]} & \text{SU(2)}\times \text{SU(2)}  & \{\numprint{2.33939}\} & \{\numprint{2.33939}\} \\
 \text{[$3^2,1^2$]} & \text{SU(2)}\times \text{U(1)}\times \text{U(1)}  & \{\numprint{3.22054},\numprint{1.07351}\} &  \{\numprint{3.22054},\numprint{1.07351}\} \\
 \text{[$5,1^3$]} & \text{SU(2)}\times \text{SU(2)} & \{\numprint{2.97463}\} & \{\numprint{2.97463}\} \\
 \text{[$4^2$]II} & \text{SU(2)}\times \text{SU(2)} & \{\numprint{2.97463}\} & \{\numprint{2.97463}\} \\
 \text{[$4^2$]I} & \text{SU(2)}\times \text{SU(2)}  & \{\numprint{2.97463}\} & \{\numprint{2.97463}\} \\
 \text{[$5,3$]} & \text{SU(2)}  & \{\} & \{\} \\
 \text{[$7,1$]} & \text{SU(2)}  & \{\} & \{\} \\
 \hline
    \end{array}$}}
\end{adjustbox}
  \caption{Plain nilpotent deformations of the probe D3-brane theory with $D_4$ flavor symmetry.
  The top table has the central charges $a_{\mathrm{IR}}$ and $c_{\mathrm{IR}}$ as well as scaling dimensions while the table below contains the information about the flavor central charges.}
  \label{MND4}
\end{table}

\begin{table}[H]
\centering
\begin{adjustbox}{center}
\scalebox{0.9}{{\renewcommand{\arraystretch}{1.2} $\begin{array}{|c|c|c|c|c|c|c|c|}
\hline
\text{[B-C]} & \text{r} & a_{\text{IR}} & c_{\text{IR}} &  t_* & \Delta _{\text{IR}}\text{(Z)} & \text{Min(}\Delta _{\text{IR}}\text{($\cO$'s))}\\
\hline
\left[1^8\right] & 0 & \left\{\frac{23}{24},\frac{37}{24}\right\} & \left\{\frac{7}{6},\frac{7}{3}\right\}  & \frac{2}{3} & \numprint{2.} & \numprint{2.} \\
\left[2^2,1^4\right] & 1       & \{ \numprint{0.962469},\numprint{1.3583}\} & \{ \numprint{1.26671},\numprint{2.05838}\} & \numprint{0.459126} & \numprint{1.37738} & \numprint{1.62262} \\
\left.\text{[3,}1^5\right] & 2 & \{ \numprint{0.80915}, \numprint{1.26748}\} & \{ \numprint{1.0197},\numprint{1.93637}\}  & \numprint{0.37588} &  \numprint{1.12764} & \numprint{1.87236} \\
\left[2^4\text{]II}\right. & 2 & \{ \numprint{0.80915}, \numprint{1.26748}\} & \{ \numprint{1.0197},\numprint{1.93637}\}  & \numprint{0.37588} &  \numprint{1.12764} & \numprint{1.87236} \\
\left[2^4\text{]I}\right. & 2  & \{ \numprint{0.80915}, \numprint{1.26748}\} & \{ \numprint{1.0197},\numprint{1.93637}\}  & \numprint{0.37588} &  \numprint{1.12764} & \numprint{1.87236} \\
\text{[3,}2^2\text{,1]} & 3    & \{ \numprint{0.727701},\numprint{1.20687}\} & \{ \numprint{0.911122},\numprint{1.86946}\}& \numprint{0.344209} & \numprint{1.03263} & \numprint{1.70921} \\
\rowcolor{LightCyan} \left[3^2,1^2\right] & 4 & \left\{\frac{7}{12},\frac{7}{6}\right\} & \left\{\frac{2}{3},\frac{11}{6}\right\} &  \frac{1}{3} & \numprint{1.} & \numprint{1.5} \\
\rowcolor{LightCyan} \left.\text{[5,}1^3\right] & 10 & \left\{\frac{11}{24},\frac{25}{24}\right\} & \left\{\frac{1}{2},\frac{5}{3}\right\} & \frac{2}{9} & \numprint{1.} & \numprint{1.66667} \\
\rowcolor{LightCyan} \left[4^2\text{]II}\right. & 10 & \left\{\frac{11}{24},\frac{25}{24}\right\} & \left\{\frac{1}{2},\frac{5}{3}\right\} & \frac{2}{9} & \numprint{1.} & \numprint{1.66667} \\
\rowcolor{LightCyan} \left[4^2\text{]I}\right. & 10 & \left\{\frac{11}{24},\frac{25}{24}\right\} & \left\{\frac{1}{2},\frac{5}{3}\right\} & \frac{2}{9} & \numprint{1.} & \numprint{1.66667} \\
\text{[5,3]} & 12 & \left\{\frac{6349}{13872},\frac{1769}{1734}\right\} & \left\{\frac{3523}{6936},\frac{5663}{3468}\right\} &  \frac{10}{51} & \numprint{1.} & \numprint{1.82353} \\
\rowcolor{LightCyan} \text{[7,1]} & 28 & \left\{\frac{43}{120},\frac{113}{120}\right\} & \left\{\frac{11}{30},\frac{23}{15}\right\} &  \frac{2}{15} & \numprint{1.} & \numprint{1.8} \\
\hline
\end{array}$}}
\end{adjustbox}
\begin{adjustbox}{center}
\scalebox{0.9}{{\renewcommand{\arraystretch}{1.2} $\begin{array}{|c|c|c|c|}
\hline
\text{[B-C]} & \text{SU(2})_{D}\text{$\times $Residual} & k_{\text{IR}}\text{ interact} & k_{\text{IR}}\text{+free} \\
\hline
\text{[$1^8$]} & \text{SO(8)}  & 4 & 16 \\
\text{[$2^2,1^4$]} & \text{SU(2)}\times \text{SO(4)}\times \text{SU(2)} &  \{\numprint{6.49049},\numprint{6.49049}\} &\{\numprint{10.4905},\numprint{10.4905}\} \\
\text{[$3,1^5$]} & \text{SU(2)}\times \text{SO(5)} & \{\numprint{3.74472}\} & \{\numprint{9.74472}\} \\
\text{[$2^4$]II} & \text{SU(2)}\times \text{Sp(4)} & \{\numprint{3.74472}\} & \{\numprint{9.74472}\} \\
\text{[$2^4$]I} & \text{SU(2)}\times \text{Sp(4)} & \{\numprint{3.74472}\} & \{\numprint{9.74472}\} \\
\text{[$3,2^2,1$]} & \text{SU(2)}\times \text{SU(2)}  & \{\numprint{2.48369}\} & \{\numprint{8.48369}\} \\
\text{[$3^2,1^2$]} & \text{SU(2)}\times \text{U(1)}\times \text{U(1)} & \{3,1\} & \{9,3\} \\
\text{[$5,1^3$]} & \text{SU(2)}\times \text{SU(2)}  & \left\{\frac{8}{3}\right\} &\left\{\frac{32}{3}\right\} \\
\text{[$4^2$]II} & \text{SU(2)}\times \text{SU(2)}  & \left\{\frac{8}{3}\right\} &\left\{\frac{32}{3}\right\} \\
\text{[$4^2$]I} & \text{SU(2)}\times \text{SU(2)}  & \left\{\frac{8}{3}\right\} &\left\{\frac{32}{3}\right\} \\
\text{[$5,3$]} & \text{SU(2)}  & \{\} & \{\} \\
\text{[$7,1$]} & \text{SU(2)}  & \{\} & \{\} \\
\hline
\end{array}$}}
\end{adjustbox}
\caption{Flipper field deformations of the probe D3-brane theory with $D_{4}$ flavor. The
top table has the central charges $a_{\mathrm{IR}}$ and $c_{\mathrm{IR}}$ as well
as scaling dimensions while the table below contains the information about the
flavor central charges. The cyan highlighted entries align with the $H_0$, $H_1$ and $H_2$ Argyres-Douglas theories,
as first noted in \cite{Maruyoshi:2016tqk, Maruyoshi:2016aim}. The other rational entry with partition [5,3] also aligns with \cite{Agarwal:2016pjo}}%
\label{MNMSD4}%
\end{table}

\subsection*{Tables of Rational Theories: Minahan-Nemeschansky Theories}

\begin{table}[H]
  \centering
  \begin{adjustbox}{center}
\scalebox{1}{\renewcommand{\arraystretch}{1.2} $\begin{array}{|c|c|c|c|c|c|c|}
 \hline
  \text{[B-C]} & \text{r} & a_{\text{IR}} & c_{\text{IR}} &  t_* & \Delta _{\text{IR}}\text{(Z)} & \text{Min(}\Delta _{\text{IR}}\text{($\cO$'s))} \\
  \hline
   0 & 0 & \frac{41}{24} & \frac{13}{6} &  \frac{2}{3} & \numprint{3.} & \numprint{2.} \\
   A_2+2A_1 & 6 & \frac{97}{96} & \frac{119}{96} &  \frac{1}{3} & \numprint{1.5} & \numprint{1.5} \\
 \hline
  \end{array}$}
\end{adjustbox}

  \begin{adjustbox}{center}
\scalebox{1}{{\renewcommand{\arraystretch}{1.2} $\begin{array}{|c|c|c|c|}
 \hline
  \text{[B-C]} & \text{SU(2})_{D}\text{$\times $Residual} &  k_{\text{IR}}\text{ interact} & k_{\text{IR}}\text{+free} \\
  \hline
  0 & E_6  & 6 & 6 \\
  \text{$A_2+2A_1$} & \text{SU(2)}\times \text{SU(2)}\times \text{U(1)} & \{18,18\} & \{18,18\} \\
  \hline
    \end{array}$}}
\end{adjustbox}
\caption{Plain nilpotent mass deformations of the Minahan-Nemeschansky theory with $E_6$ flavor. The top table has the central charges $a_{\mathrm{IR}}$ and $c_{\mathrm{IR}}$ as well as scaling dimensions while the table below contains the information about the flavor central charges.}
\label{MNE6}
\end{table}

\begin{table}[H]
  \centering
  \begin{adjustbox}{center}
\scalebox{1}{\renewcommand{\arraystretch}{1.2} $\begin{array}{|c|c|c|c|c|c|c|}
 \hline
  \text{[B-C]} & \text{r} & a_{\text{IR}} & c_{\text{IR}} &  t_* & \Delta _{\text{IR}}\text{(Z)} & \text{Min(}\Delta _{\text{IR}}\text{($\cO$'s))} \\
  \hline
    0 & 0 & \frac{59}{24} & \frac{19}{6} &  \frac{2}{3} & \numprint{4.} & \numprint{2.} \\
    A_1 & 1 & \frac{158}{75} & \frac{401}{150} &  \frac{8}{15} & \numprint{3.2} & \numprint{1.4} \\
    A_2+3A_1 & 7 & \frac{7150}{5043} & \frac{17785}{10086} &  \frac{40}{123} & \numprint{1.95122} & \numprint{1.53659} \\
    A_4+A_2 & 24 & \frac{478}{507} & \frac{1177}{1014} & \frac{8}{39} & \numprint{1.23077} & \numprint{1.46154} \\
    \text{($A_5$)'} & 35 & \frac{7075}{8664} & \frac{4345}{4332} &  \frac{10}{57} & \numprint{1.05263} & \numprint{1.42105} \\
    \text{($A_5$)$\texttt{"}$} & 35 & \frac{7075}{8664} & \frac{4345}{4332} &  \frac{10}{57} & \numprint{1.05263} & \numprint{1.42105} \\
    A_6 & 56 & \left\{\frac{3803}{5776},\frac{5885}{8664}\right\} & \left\{\frac{2253}{2888},\frac{890}{1083}\right\} &  \frac{8}{57} & \numprint{1.} & \numprint{1.52632} \\
    D_6(a_1) & 62 & \left\{\frac{253}{400},\frac{49}{75}\right\} & \left\{\frac{149}{200},\frac{59}{75}\right\} &  \frac{2}{15}  & \numprint{1.} & \numprint{1.4} \\
    E_7(a_3) & 111 & \left\{\frac{659}{1296},\frac{343}{648}\right\} & \left\{\frac{373}{648},\frac{50}{81}\right\} &  \frac{8}{81} & \numprint{1.} & \numprint{1.51852} \\
 \hline
\end{array}$}
\end{adjustbox}

  \begin{adjustbox}{center}
\scalebox{1}{{\renewcommand{\arraystretch}{1.2} $\begin{array}{|c|c|c|c|}
 \hline
  \text{[B-C]} & \text{SU(2})_{D}\text{$\times $Residual}  & k_{\text{IR}}\text{ interact} & k_{\text{IR}}\text{+free} \\
\hline
   0 & \text{E}_7  & 8 & 8 \\
 \text{$A_1$} & \text{SU(2)}\times \text{SO(12)} & \left\{\frac{32}{5}\right\} & \left\{\frac{32}{5}\right\} \\
 \text{$A_2+3A_1$} & \text{SU(2)}\times \text{G}_2 & \left\{\frac{320}{41}\right\} &   \left\{\frac{320}{41}\right\} \\
 \text{$A_4+A_2$} & \text{SU(2)}\times \text{SU(2)}  & \left\{\frac{480}{13}\right\} &   \left\{\frac{480}{13}\right\} \\
 \text{$A_5'$} & \text{SU(2)}\times \text{SU(2)}\times \text{SU(2)} &  \left\{\frac{40}{19},\frac{120}{19}\right\} & \left\{\frac{40}{19},\frac{120}{19}\right\} \\
 \text{$A_5''$} & \text{SU(2)}\times \text{G}_2 &  \left\{\frac{40}{19}\right\} & \left\{\frac{40}{19}\right\} \\
 \text{$A_6$} & \text{SU(2)}\times \text{SU(2)} &  \left\{\frac{224}{19}\right\} & \left\{\frac{224}{19}\right\} \\
 \text{$D_6(a_1)$} & \text{SU(2)}\times \text{SU(2)} & \left\{\frac{8}{5}\right\} & \left\{\frac{8}{5}\right\} \\
  \hline
    \end{array}$}}
\end{adjustbox}
\caption{Plain nilpotent mass deformations of the Minahan-Nemeschansky theory with $E_7$ flavor, only rational values. The top table has the central charges $a_{\mathrm{IR}}$ and $c_{\mathrm{IR}}$ as well as scaling dimensions while the table below contains the information about the flavor central charges.}
\label{MNE7}
\end{table}

\begin{table}[H]
  \centering
\begin{adjustbox}{center}
\scalebox{1}{\renewcommand{\arraystretch}{1.2} $\begin{array}{|c|c|c|c|c|c|c|}
 \hline
  \text{[B-C]} & \text{r} & a_{\text{IR}} & c_{\text{IR}} &  t_* & \Delta _{\text{IR}}\text{(Z)} & \text{Min(}\Delta _{\text{IR}}\text{($\cO$'s))} \\
  \hline
  0 & 0 & \frac{95}{24} & \frac{31}{6} &  \frac{2}{3} & 6. & 2. \\
  A_2+3A_1 & 7 & \frac{223}{96} & \frac{281}{96} &  \frac{1}{3} & 3. & 1.5 \\
  E_8(a_1) & 760 & \left\{\frac{5471}{13872},\frac{120}{289}\right\} & \left\{\frac{2897}{6936},\frac{531}{1156}\right\} & \frac{2}{51} & 1. & 1.58824 \\
 \hline
\end{array}$}
\end{adjustbox}

  \begin{adjustbox}{center}
\scalebox{1}{{\renewcommand{\arraystretch}{1.2} $\begin{array}{|c|c|c|c|}
 \hline
  \text{[B-C]} & \text{SU(2})_{D}\text{$\times $Residual} &  k_{\text{IR}}\text{ interact} & k_{\text{IR}}\text{+free} \\
\hline
  0 & \text{E}_8  & 12 & 12 \\
 \text{$A_2+3A_1$} & \text{SU(2)}\times \text{G}_2\times \text{SU(2)} & \{12,6\} & \{12,6\} \\
  \hline
    \end{array}$}}
\end{adjustbox}
\caption{Plain nilpotent mass deformations of the Minahan-Nemeschansky theory with $E_8$ flavor, only rational values. The top table has the central charges $a_{\mathrm{IR}}$ and $c_{\mathrm{IR}}$ as well as scaling dimensions while the table below contains the information about the flavor central charges.}
\label{MNE8}
\end{table}

\begin{table}[H]
\centering
\begin{adjustbox}{center}
\scalebox{0.9}{{\renewcommand{\arraystretch}{1.2} $\begin{array}{|c|c|c|c|c|c|c|c|}
\hline
\text{[B-C]} & \text{r} & a_{\text{IR}} & c_{\text{IR}} &  t_* & \Delta _{\text{IR}}\text{(Z)} & \text{Min(}\Delta _{\text{IR}}\text{($\cO$'s))}\\
\hline
0 & 0 & \left\{\frac{41}{24},\frac{10}{3}\right\} & \left\{\frac{13}{6},\frac{65}{12}\right\} &  \frac{2}{3} & \numprint{3.} & \numprint{2.} \\
\rowcolor{LightCyan}  D_4 & 28 & \left\{\frac{7}{12},\frac{53}{24}\right\} & \left\{\frac{2}{3},\frac{47}{12}\right\} &  \frac{1}{6} & \numprint{1.} & \numprint{1.5} \\
\rowcolor{LightCyan}  D_5 & 60 & \left\{\frac{11}{24},\frac{25}{12}\right\} & \left\{\frac{1}{2},\frac{15}{4}\right\} &  \frac{1}{9} & \numprint{1.} & \numprint{1.66667} \\
\rowcolor{LightCyan}  E_6 & 156 & \left\{\frac{43}{120},\frac{119}{60}\right\} & \left\{\frac{11}{30},\frac{217}{60}\right\}  & \frac{1}{15} & \numprint{1.} & \numprint{1.8} \\
\hline
\end{array}$}}
\end{adjustbox}
\par
\begin{adjustbox}{center}
\scalebox{0.9}{{\renewcommand{\arraystretch}{1.2} $\begin{array}{|c|c|c|c|}
\hline
\text{[B-C]} & \text{SU(2})_{D}\text{$\times $Residual} & k_{\text{IR}}\text{ interact} & k_{\text{IR}}\text{+free} \\
\hline
0 & \text{E}_6  & 6 & 30 \\
\text{$D_4$} & \text{SU(2)}\times \text{SU(3)}  & \{3\} & \{15\} \\
\text{$D_5$} & \text{SU(2)}\times \text{U(1)}  & \{6\} & \{24\} \\
\hline
\end{array}$}}
\end{adjustbox}
\caption{Flipper field deformations of the Minahan-Nemeschansky theory with $E_{6}$ flavor, only
rational values. The top table has the central charges $a_{\mathrm{IR}}$ and
$c_{\mathrm{IR}}$ as well as scaling dimensions while the table below contains
the information about the flavor central charges. The cyan highlighted
entries align with the $H_0$, $H_1$ and $H_2$ Argyres-Douglas theories,
as first noted in \cite{Maruyoshi:2016aim,Agarwal:2016pjo}.}%
\label{MN-MS-E6}%
\end{table}

\begin{table}[H]
\centering
\begin{adjustbox}{center}
\scalebox{0.9}{{\renewcommand{\arraystretch}{1.2} $\begin{array}{|c|c|c|c|c|c|c|c|}
\hline
\text{[B-C]} & \text{r} & a_{\text{IR}} & c_{\text{IR}} &  t_* & \Delta _{\text{IR}}\text{(Z)} & \text{Min(}\Delta _{\text{IR}}\text{($\cO$'s))}\\
\hline
0 & 0 & \left\{\frac{59}{24},\frac{251}{48}\right\} & \left\{\frac{19}{6},\frac{209}{24}\right\} &  \frac{2}{3} & \numprint{4.} & \numprint{2.}\\
A_2+3A_1 & 7 & \left\{\frac{12163}{8214},\frac{134899}{32856}\right\} & \left\{\frac{121465}{65712},\frac{466453}{65712}\right\} &  \frac{31}{111} & \numprint{1.67568} &
\numprint{1.74324} \\
\rowcolor{LightCyan}  E_6 & 156 & \left\{\frac{11}{24},\frac{155}{48}\right\} & \left\{\frac{1}{2},\frac{145}{24}\right\} &  \frac{2}{27} & \numprint{1.} & \numprint{1.66667} \\
\rowcolor{LightCyan} E_7 & 399 & \left\{\frac{43}{120},\frac{751}{240}\right\} & \left\{\frac{11}{30},\frac{709}{120}\right\}  & \frac{2}{45} & \numprint{1.} & \numprint{1.8} \\
\hline
\end{array}$}}
\end{adjustbox}
\par
\begin{adjustbox}{center}
\scalebox{0.9}{{\renewcommand{\arraystretch}{1.2} $\begin{array}{|c|c|c|c|}
\hline
\text{[B-C]} & \text{SU(2})_{D}\text{$\times $Residual}  & k_{\text{IR}}\text{ interact} & k_{\text{IR}}\text{+free} \\
\hline
0 & \text{E}_7  & 8 & 44 \\
\text{$A_2+3A_1$} & \text{SU(2)}\times \text{G}_2 & \left\{\frac{284}{37}\right\} &\left\{\frac{1246}{37}\right\} \\
\text{$E_6$} & \text{SU(2)}\times \text{SU(2)} & \left\{\frac{8}{3}\right\} & \left\{\frac{44}{3}\right\} \\
\hline
\end{array}$}}
\end{adjustbox}
\caption{Flipper field deformations of the Minahan-Nemeschansky theory with $E_{7}$ flavor, only
rational values. The top table has the central charges $a_{\mathrm{IR}}$ and
$c_{\mathrm{IR}}$ as well as scaling dimensions while the table below contains
the information about the flavor central charges. The cyan highlighted entries
align with the $H_0$ and $H_1$ Argyres-Douglas theories,
as first noted in \cite{Maruyoshi:2016aim}. Compared with reference \cite{Maruyoshi:2016aim},
we also find an additional flipper field deformation which yields the $H_1$ theory for the
$E_6$ Bala-Carter label, with embedding index $r = 156$. The other rational central charges are also in agreement with \cite{AgarwalMaruyoshiSong}.}%
\label{MN-MS-E7}%
\end{table}

\begin{table}[H]
\centering
\begin{adjustbox}{center}
\scalebox{0.9}{{\renewcommand{\arraystretch}{1.2} $\begin{array}{|c|c|c|c|c|c|c|c|}
\hline
\text{[B-C]} & \text{r} & a_{\text{IR}} & c_{\text{IR}} &  t_* & \Delta _{\text{IR}}\text{(Z)} & \text{Min(}\Delta _{\text{IR}}\text{($\cO$'s))}\\
\hline
0 & 0 & \left\{\frac{95}{24},\frac{73}{8}\right\} & \left\{\frac{31}{6},\frac{31}{2}\right\} &  \frac{2}{3} & \numprint{6.} & \numprint{2.} \\
A_3 & 10 & \left\{\frac{497803}{221952},\frac{529689}{73984}\right\} & \left\{\frac{635435}{221952},\frac{939321}{73984}\right\} &  \frac{53}{204} & \numprint{2.33824} &   \numprint{1.44118} \\
A_3+A_1 & 11 & \left\{\frac{139189}{60552},\frac{214667}{30276}\right\} & \left\{\frac{91127}{30276},\frac{95318}{7569}\right\} &  \frac{64}{261} &  \numprint{2.2069} &  \numprint{1.52874}  \\
E_7\left(a_5\right) & 39 & \left\{\frac{445}{324},\frac{2065}{324}\right\} & \left\{\frac{281}{162},\frac{1901}{162}\right\} & \frac{4}{27} & \numprint{1.33333} & \numprint{1.66667} \\
E_7\left(a_4\right) & 63 & \left\{\frac{1691}{1452},\frac{8951}{1452}\right\} & \left\{\frac{541}{363},\frac{4171}{363}\right\}  & \frac{4}{33} & \numprint{1.09091} & \numprint{1.54545} \\
\rowcolor{LightCyan}  E_8 & 1240 & \left\{\frac{43}{120},\frac{221}{40}\right\} & \left\{\frac{11}{30},\frac{107}{10}\right\}  & \frac{2}{75} & \numprint{1.} & \numprint{1.8} \\
\hline
\end{array}$}}
\end{adjustbox}
\begin{adjustbox}{center}
\scalebox{0.9}{{\renewcommand{\arraystretch}{1.2} $\begin{array}{|c|c|c|c|}
\hline
\text{[B-C]} & \text{SU(2})_{D}\text{$\times $Residual} & k_{\text{IR}}\text{ interact} & k_{\text{IR}}\text{+free} \\
\hline
0 & E_8  & 12 & 72 \\
\text{$A_3$} & \text{SU(2)}\times \text{SO(11)} & \{6\} & \{32\} \\
\text{$A_3+A_1$} & \text{SU(2)}\times \text{SO(7)}\times \text{SU(2)} &\left\{\frac{220}{29},\frac{421}{87}\right\} & \left\{\frac{800}{29},\frac{2161}{87}\right\} \\
\text{$E_7(a_5)$} & \text{SU(2)}\times \text{SU(2)} & \left\{\frac{29}{9}\right\} &\left\{\frac{137}{9}\right\} \\
\text{$E_7(a_4)$} & \text{SU(2)}\times \text{SU(2)} & \left\{\frac{31}{11}\right\} &\left\{\frac{152}{11}\right\} \\
\hline
\end{array}$}}
\end{adjustbox}
\caption{Flipper field deformations of the Minahan-Nemeschansky theory with $E_{8}$ flavor, only
rational values. The top table has the central charges $a_{\mathrm{IR}}$ and
$c_{\mathrm{IR}}$ as well as scaling dimensions while the table below contains
the information about the flavor central charges. The cyan highlighted entry
aligns with the $H_0$ Argyres-Douglas theory,
as first noted in \cite{Maruyoshi:2016aim}. The other rational central charges are also in agreement with \cite{AgarwalMaruyoshiSong}.}%
\label{MN-MS-E8}%
\end{table}

\subsection*{Tables of Rational Theories: Conformal Matter}

\begin{table}[H]
\centering
\begin{adjustbox}{left}
\scalebox{1.0}{{\renewcommand{\arraystretch}{1.2}
$\begin{array}{|c|c|c|c|c|c|c|c|}
\hline
\text{[B-C]}_L & \text{[B-C]}_R & r_L & r_R & r_L+r_R & a_{\text{IR}} & c_{\text{IR}} & t_* \\
\hline
0 & 0 & 0 & 0 & 0 & \frac{613}{24} & \frac{173}{6} &  \frac{2}{3} \\
2A_2+A_1 & 2A_2 & 9 & 8 & 17 & \frac{68050}{4107} & \frac{150715}{8214}  & \frac{40}{111} \\
A_5 & 2A_2+A_1 & 35 & 9 & 44 & \left\{\frac{316}{25},\frac{3817}{300}\right\} & \left\{\frac{346}{25},\frac{2101}{150}\right\}  & \frac{4}{15}\\
\hline
\end{array}$}} $\cdots$
\end{adjustbox}
\centering
\begin{adjustbox}{left}
\scalebox{1.0}{{\renewcommand{\arraystretch}{1.2}
$\begin{array}{|c|c|c|c|c|c|}
\hline
\text{[B-C]}_L & \text{[B-C]}_R & t_* & \text{Min(}\Delta _{\text{IR}}\text{(Z's))} & \text{Min(}\Delta_{\text{IR}}\left(\cO_L\text{'s))}\right. & \text{Min(}\Delta _{\text{IR}}\left(\cO_R\text{'s))}\right. \\
0 & 0 & \frac{2}{3} & \numprint{6.} & \numprint{2.} & \numprint{2.} \\
2A_2+A_1 & 2A_2 &\frac{40}{111} & \numprint{3.24324} & \numprint{1.10811} & \numprint{1.37838} \\
A_5 & 2A_2+A_1 & \frac{4}{15} & \numprint{2.4} & \numprint{1.} & \numprint{1.6} \\
\hline
\end{array}$ }}
\end{adjustbox}
\caption{Plain nilpotent mass deformations of ($E_{6}$, $E_{6}$) conformal matter, only rational values.}%
\label{CME6}%
\end{table}

\begin{table}[H]
\centering
\begin{adjustbox}{left}
\scalebox{1.0}{{\renewcommand{\arraystretch}{1.2} $\begin{array}{|c|c|c|c|c|c|c|c|}
\hline
\text{[B-C]}_L & \text{[B-C]}_R & r_L & r_R & r_L+r_R & a_{\text{IR}} & c_{\text{IR}} & t_*  \\
\hline
0 & 0 & 0 & 0 & 0 & \frac{817}{12} & \frac{221}{3} & \frac{2}{3} \\
D_4+A_1 & D_4+A_1 & 29 & 29 & 58 & \left\{\frac{314941}{8400},\frac{105097}{2800}\right\} & \left\{\frac{47843}{1200},\frac{15981}{400}\right\} & \frac{31}{105}  \\
D_5 & \text{(3}A_1\text{)''} & 60 & 3 & 63 & \left\{\frac{233959}{6272},\frac{235135}{6272}\right\} & \left\{\frac{247315}{6272},\frac{249667}{6272}\right\} & \frac{13}{42}  \\
D_5 & \text{(3}A_1\text{)'} & 60 & 3 & 63 & \left\{\frac{233959}{6272},\frac{235135}{6272}\right\} & \left\{\frac{247315}{6272},\frac{249667}{6272}\right\} &  \frac{13}{42}  \\
D_5+A_1 & 0 & 61 & 0 & 61 & \left\{\frac{63612}{1681},\frac{1022835}{26896}\right\} & \left\{\frac{538047}{13448},\frac{271545}{6724}\right\} & \frac{13}{41} \\
D_5+A_1 & D_4\left(a_1\right) & 61 & 12 & 73 & \left\{\frac{27729}{784},\frac{6969}{196}\right\} & \left\{\frac{3663}{98},\frac{14799}{392}\right\} & \frac{2}{7}  \\
D_5+A_1 & A_3+2A_1 & 61 & 12 & 73 & \left\{\frac{27729}{784},\frac{6969}{196}\right\} & \left\{\frac{3663}{98},\frac{14799}{392}\right\} & \frac{2}{7} \\
E_6\left(a_1\right) & A_3 & 84 & 10 & 94 & \left\{\frac{1583}{48},\frac{199}{6}\right\} & \left\{\frac{4177}{120},\frac{2111}{60}\right\}  & \frac{4}{15}\\
E_6 & A_3 & 156 & 10 & 166 & \left\{\frac{995}{36},\frac{1999}{72}\right\} & \left\{\frac{1049}{36},\frac{529}{18}\right\} &  \frac{2}{9} \\
E_7\left(a_1\right) & A_2 & 231 & 4 & 235 & \left\{\frac{2992009}{121104},\frac{187789}{7569}\right\} & \left\{\frac{1576001}{60552},\frac{198577}{7569}\right\} &
\frac{52}{261}\\
E_7\left(a_1\right)  & 4A_1 & 231 & 4 & 235 & \left\{\frac{2992009}{121104},\frac{187789}{7569}\right\} & \left\{\frac{1576001}{60552},\frac{198577}{7569}\right\} & \frac{52}{261} \\
\hline
\end{array}$}} $\cdots$
\end{adjustbox}
\begin{adjustbox}{left}
\scalebox{1.0}{{\renewcommand{\arraystretch}{1.2} $\begin{array}{|c|c|c|c|c|c|}
\hline
\text{[B-C]}_L & \text{[B-C]}_R &  t_* & \text{Min(}\Delta _{\text{IR}}\text{(Z's))} & \text{Min(}\Delta_{\text{IR}}\left(\cO_L\text{'s))}\right. & \text{Min(}\Delta _{\text{IR}}\left(\cO_R\text{'s))}\right. \\
\hline
0 & 0  & \frac{2}{3} & \numprint{6.} & \numprint{2.} & \numprint{2.} \\
D_4+A_1 & D_4+A_1 & \frac{31}{105}     & \numprint{2.65714} & \numprint{1.} & \numprint{1.} \\
D_5 & \text{(3}A_1\text{)''}  & \frac{13}{42} & \numprint{2.78571} & \numprint{1.} & \numprint{2.07143} \\
D_5 & \text{(3}A_1\text{)'}   & \frac{13}{42}  & \numprint{2.78571} & \numprint{1.} & \numprint{1.83929} \\
D_5+A_1 & 0 & \frac{13}{41}   & \numprint{2.85366} & \numprint{1.} & \numprint{2.5243}9 \\
D_5+A_1 & D_4\left(a_1\right) & \frac{2}{7} & \numprint{2.57143} & \numprint{1.} & \numprint{1.28571} \\
D_5+A_1 & A_3+2A_1    & \frac{2}{7} & \numprint{2.57143} & \numprint{1.} & \numprint{1.28571} \\
E_6\left(a_1\right) & A_3  & \frac{4}{15} & \numprint{2.4}     & \numprint{1.} & \numprint{1.4} \\
E_6 & A_3            & \frac{2}{9} & \numprint{2.}      & \numprint{1.} & \numprint{1.66667} \\
E_7\left(a_1\right)  & A_2  &  \frac{52}{261} & \numprint{1.7931} & \numprint{1.} & \numprint{2.10345} \\
E_7\left(a_1\right)  & 4A_1 & \frac{52}{261} & \numprint{1.7931} & \numprint{1.} & \numprint{2.25287} \\
\hline
\end{array}$}}
\end{adjustbox}
\caption{Plain nilpotent mass deformations of ($E_{7}$, $E_{7}$) conformal matter, only rational values.}%
\label{CME7}%
\end{table}\begin{table}[H]
\centering
\begin{adjustbox}{left}
\scalebox{0.85}{{\renewcommand{\arraystretch}{1.2} $\begin{array}{|c|c|c|c|c|c|c|c|}
\hline
\text{[B-C]}_L & \text{[B-C]}_R & r_L & r_R & r_L+r_R & a_{\text{IR}} & c_{\text{IR}} &  t_* \\
\hline
0 & 0 & 0 & 0 & 0 & \frac{1745}{8} & \frac{457}{2} &  \frac{2}{3} \\
A_3+A_2 & 3A_1 & 14 & 3 & 17 & \left\{\frac{2594465245}{14362032},\frac{325018777}{1795254}\right\} & \left\{\frac{1347452419}{7181016},\frac{676568695}{3590508}\right\}  & \frac{824}{1641}\\
D_5 & 0 & 60 & 0 & 60 & \left\{\frac{88198105}{591576},\frac{44209973}{295788}\right\} & \left\{\frac{11389690}{73947},\frac{45780601}{295788}\right\} & \frac{194}{471}\\
E_7\left(a_3\right) & 2A_2+A_1 & 111 & 9 & 120 & \left\{\frac{12055}{96},\frac{12091}{96}\right\} & \left\{\frac{12425}{96},\frac{12497}{96}\right\} & \frac{1}{3} \\
E_8\left(b_5\right) & D_6\left(a_1\right) & 160 & 62 & 222 & \left\{\frac{823817}{8112},\frac{103463}{1014}\right\} & \left\{\frac{422939}{4056},\frac{213413}{2028}\right\} &  \frac{10}{39} \\
D_7 & E_6\left(a_1\text{)+}A_1\right. & 182 & 85 & 267 & \left\{\frac{187823116685}{1971613488},\frac{47191962037}{492903372}\right\} &
\left\{\frac{96328408265}{985806744},\frac{12159142466}{123225843}\right\} & \frac{4588}{19227} \\
E_8\left(b_4\right) & A_2+A_1 & 232 & 5 & 237 & \left\{\frac{1832579}{17328},\frac{76553}{722}\right\} & \left\{\frac{943241}{8664},\frac{157989}{1444}\right\} & \frac{16}{57} \\
\hline
\end{array}$}} $\cdots$
\end{adjustbox}
\begin{adjustbox}{left}
\scalebox{0.85}{{\renewcommand{\arraystretch}{1.2} $\begin{array}{|c|c|c|c|c|c|c|c|c|c|c|c|}
\hline
\text{[B-C]}_L & \text{[B-C]}_R & t_* & \text{Min(}\Delta _{\text{IR}}\text{(Z's))} & \text{Min(}\Delta_{\text{IR}}\left(\cO_L\text{'s))}\right. & \text{Min(}\Delta _{\text{IR}}\left(\cO_R\text{'s))}\right. \\
\hline
0 & 0 & \frac{2}{3} & \numprint{6.} & \numprint{2.} & \numprint{2.} \\
A_3+A_2 & 3A_1 & \frac{824}{1641} & \numprint{4.5192} & \numprint{1.} & \numprint{1.117} \\
D_5 & 0 & \frac{194}{471} & \numprint{3.70701} & \numprint{1.} & \numprint{2.38217} \\
E_7\left(a_3\right) & 2A_2+A_1 & \frac{1}{3} & \numprint{3.} & \numprint{1.} & \numprint{1.25} \\
E_8\left(b_5\right) & D_6\left(a_1\right) & \frac{10}{39} & \numprint{2.30769} & \numprint{1.} & \numprint{1.} \\
D_7 & E_6\left(a_1\text{)+}A_1\right. & \frac{4588}{19227} & \numprint{2.1476} & \numprint{1.} & \numprint{1.} \\
E_8\left(b_4\right) & A_2+A_1 & \frac{16}{57} & \numprint{2.52632} & \numprint{1.} & \numprint{1.73684} \\
\hline
\end{array}$}}
\end{adjustbox}
\caption{Plain nilpotent mass deformations of ($E_{8}$, $E_{8}$) conformal matter, only rational
values.}%
\label{CME8}%
\end{table}

\begin{table}[H]
\centering
\begin{adjustbox}{left}
\scalebox{0.85}{{\renewcommand{\arraystretch}{1.2} $\begin{array}{|c|c|c|c|c|c|c|c|}
\hline
\text{[B-C]}_L & \text{[B-C]}_R & r_L & r_R & r_L+r_R & a_{\text{IR}} & c_{\text{IR}} &  t_* \\
\hline
\left[1^8\right] & \left[1^8\right] & 0 & 0 & 0 & \left\{\frac{95}{24},\frac{41}{8}\right\} & \left\{\frac{31}{6},\frac{15}{2}\right\} & \frac{2}{3} \\
\text{[7,1]} & \left[4^2\text{]I}\right. & 28 & 10 & 38 & \left\{\frac{245399}{107736},\frac{87785}{35912}\right\} & \left\{\frac{95905}{26934},\frac{34961}{8978}\right\} &  \frac{34}{201} \\
\text{[7,1]} & \left[4^2\text{]II}\right. & 28 & 10 & 38 & \left\{\frac{245399}{107736},\frac{87785}{35912}\right\} & \left\{\frac{95905}{26934},\frac{34961}{8978}\right\} &  \frac{34}{201}\\
\text{[7,1]} & \left.\text{[5,}1^3\right] & 28 & 10 & 38 & \left\{\frac{245399}{107736},\frac{87785}{35912}\right\} & \left\{\frac{95905}{26934},\frac{34961}{8978}\right\} &  \frac{34}{201} \\
\hline
\end{array}$}} $\cdots$
\end{adjustbox}
\begin{adjustbox}{left}
\scalebox{0.85}{{\renewcommand{\arraystretch}{1.2} $\begin{array}{|c|c|c|c|c|c|c|c|c|}
\hline
\text{[B-C]}_L & \text{[B-C]}_R & t_* & \text{Min(}\Delta _{\text{IR}}\text{(Z's))} & \text{Min(}\Delta_{\text{IR}}\left(\cO_L\text{'s))}\right. & \text{Min(}\Delta _{\text{IR}}\left(\cO_R\text{'s))}\right. \\
\hline
\left[1^8\right] & \left[1^8\right]  & \frac{2}{3} & \numprint{6.} & \numprint{2.} & \numprint{2.}  \\
\text{[7,1]} & \left[4^2\text{]I}\right.  & \frac{34}{201} & \numprint{1.52239} & \numprint{1.47761} & \numprint{1.98507} \\
\text{[7,1]} & \left[4^2\text{]II}\right. & \frac{34}{201} & \numprint{1.52239} & \numprint{1.47761} & \numprint{1.98507} \\
\text{[7,1]} & \left.\text{[5,}1^3\right] & \frac{34}{201} & \numprint{1.52239} & \numprint{1.47761} & \numprint{1.98507} \\
\hline
\end{array}$}}
\end{adjustbox}
\caption{Flipper field deformations of ($D_{4}$, $D_{4}$) conformal matter, only rational values.}%
\label{MS-CM-D4}%
\end{table}

\begin{table}[H]
\centering
\begin{adjustbox}{left}
\scalebox{0.85}{{\renewcommand{\arraystretch}{1.2} $\begin{array}{|c|c|c|c|c|c|c|c|}
\hline
\text{[B-C]}_L & \text{[B-C]}_R & r_L & r_R & r_L+r_R & a_{\text{IR}} & c_{\text{IR}} & t_* \\
\hline
0 & 0 & 0 & 0 & 0 & \left\{\frac{613}{24},\frac{691}{24}\right\} & \left\{\frac{173}{6},\frac{106}{3}\right\} &  \frac{2}{3} \\
A_3+A_1 & A_1 & 11 & 1 & 12 & \left\{\frac{248983}{13872},\frac{144577}{6936}\right\} & \left\{\frac{137641}{6936},\frac{44453}{1734}\right\}  & \frac{20}{51} \\
D_4 & A_3+A_1 & 28 & 11 & 39 & \left\{\frac{5271}{400},\frac{3223}{200}\right\} & \left\{\frac{2893}{200},\frac{1017}{50}\right\}  & \frac{4}{15} \\
D_5\left(a_1\right) & A_3 & 30 & 10 & 40 & \left\{\frac{15737}{1200},\frac{9631}{600}\right\} & \left\{\frac{8651}{600},\frac{1522}{75}\right\} & \frac{4}{15} \\
D_5\left(a_1\right) & A_3+A_1 & 30 & 11 & 41 & \left\{\frac{1364659}{104976},\frac{836513}{52488}\right\} & \left\{\frac{749681}{52488},\frac{132256}{6561}\right\} & \frac{64}{243} \\
\hline
\end{array}$}} $\cdots$
\end{adjustbox}
\begin{adjustbox}{left}
\scalebox{0.85}{{\renewcommand{\arraystretch}{1.2} $\begin{array}{|c|c|c|c|c|c|c|c|c|}
\hline
\text{[B-C]}_L & \text{[B-C]}_R & t_* & \text{Min(}\Delta _{\text{IR}}\text{(Z's))} & \text{Min(}\Delta_{\text{IR}}\left(\cO_L\text{'s))}\right. & \text{Min(}\Delta _{\text{IR}}\left(\cO_R\text{'s))}\right. \\
\hline
0 & 0 & \frac{2}{3} & \numprint{6.} & \numprint{2.} & \numprint{2.} \\
A_3+A_1 & A_1 & \frac{20}{51} & \numprint{3.52941} & \numprint{1.} & \numprint{1.82353} \\
D_4 & A_3+A_1 & \frac{4}{15} & \numprint{2.4} & \numprint{1.} & \numprint{1.4} \\
D_5\left(a_1\right) & A_3 & \frac{4}{15} & \numprint{2.4} & \numprint{1.} & \numprint{1.4} \\
D_5\left(a_1\right) & A_3+A_1 & \frac{64}{243} & \numprint{2.37037} & \numprint{1.} & \numprint{1.41975} \\
\hline
\end{array}$}}
\end{adjustbox}
\caption{Flipper field deformations of ($E_{6}$, $E_{6}$) conformal matter, only rational values.}%
\label{MS-CM-E6}%
\end{table}

\begin{table}[H]
\centering
\begin{adjustbox}{left}
\scalebox{0.85}{{\renewcommand{\arraystretch}{1.2} $\begin{array}{|c|c|c|c|c|c|c|c|}
\hline
\text{[B-C]}_L & \text{[B-C]}_R & r_L & r_R & r_L+r_R & a_{\text{IR}} & c_{\text{IR}} &  t_*  \\
\hline
0 & 0 & 0 & 0 & 0 & \left\{\frac{817}{12},\frac{589}{8}\right\} & \left\{\frac{221}{3},\frac{339}{4}\right\} &  \frac{2}{3} \\
A_3+A_2+A_1 & 0 & 15 & 0 & 15 & \left\{\frac{1241}{24},\frac{453}{8}\right\} & \left\{\frac{661}{12},\frac{779}{12}\right\} &  \frac{4}{9} \\
A_5+A_1 & A_2+3A_1 & 36 & 7 & 43 & \left\{\frac{3931}{96},\frac{4417}{96}\right\} & \left\{\frac{4163}{96},\frac{5135}{96}\right\} &  \frac{1}{3} \\
E_6\left(a_1\right) & A_2+A_1 & 84 & 5 & 89 & \left\{\frac{235499}{6936},\frac{270757}{6936}\right\} & \left\{\frac{62449}{1734},\frac{40039}{867}\right\}  & \frac{14}{51} \\
E_6 & A_4+A_1 & 156 & 21 & 177 & \left\{\frac{44180297}{1642800},\frac{1096539}{34225}\right\} & \left\{\frac{23344541}{821400},\frac{2649843}{68450}\right\} & \frac{116}{555} \\
\hline
\end{array}$}} $\cdots$
\end{adjustbox}
\begin{adjustbox}{left}
\scalebox{0.85}{{\renewcommand{\arraystretch}{1.2} $\begin{array}{|c|c|c|c|c|c|c|c|c|}
\hline
\text{[B-C]}_L & \text{[B-C]}_R & t_* & \text{Min(}\Delta _{\text{IR}}\text{(Z's))} & \text{Min(}\Delta_{\text{IR}}\left(\cO_L\text{'s))}\right. & \text{Min(}\Delta _{\text{IR}}\left(\cO_R\text{'s))}\right. \\
\hline
0 & 0 & \frac{2}{3} & \numprint{6.} & \numprint{2.} & \numprint{2.} \\
A_3+A_2+A_1 & 0  & \frac{4}{9} & \numprint{4.} & \numprint{1.} & \numprint{2.33333} \\
A_5+A_1 & A_2+3A_1 & \frac{1}{3} & \numprint{3.} & \numprint{1.} & \numprint{1.5} \\
E_6\left(a_1\right) & A_2+A_1 & \frac{14}{51} & \numprint{2.47059} & \numprint{1.} & \numprint{1.76471} \\
E_6 & A_4+A_1 & \frac{116}{555} & \numprint{1.88108} & \numprint{1.} & \numprint{1.43243} \\
\hline
\end{array}$}}
\end{adjustbox}
\caption{Flipper field deformations of ($E_{7}$, $E_{7}$) conformal matter, only rational values.}%
\label{MS-CM-E7}%
\end{table}

\begin{table}[H]
\centering
\begin{adjustbox}{left}
\scalebox{0.85}{{\renewcommand{\arraystretch}{1.2} $\begin{array}{|c|c|c|c|c|c|c|c|}
\hline
\text{[B-C]}_L & \text{[B-C]}_R & r_L & r_R & r_L+r_R & a_{\text{IR}} & c_{\text{IR}} &  t_* \\
\hline
0 & 0 & 0 & 0 & 0 & \left\{\frac{1745}{8},\frac{5483}{24}\right\} & \left\{\frac{457}{2},\frac{1495}{6}\right\} &  \frac{2}{3} \\
A_4+A_2+A_1 & A_4+2A_1 & 25 & 22 & 47 & \left\{\frac{122989}{816},\frac{21657}{136}\right\} & \left\{\frac{63463}{408},\frac{2934}{17}\right\} &  \frac{20}{51} \\
D_5\left(a_1\right) & A_2+3A_1 & 30 & 7 & 37 & \left\{\frac{1200211}{7500},\frac{632293}{3750}\right\} & \left\{\frac{620893}{3750},\frac{342634}{1875}\right\}  & \frac{32}{75} \\
D_5\left(a_1\text{)+}A_2\right. & A_3+A_2+A_1 & 34 & 15 & 49 & \left\{\frac{122683}{816},\frac{64801}{408}\right\} & \left\{\frac{63361}{408},\frac{8785}{51}\right\} &   \frac{20}{51} \\
E_6\left(a_3\text{)+}A_1\right. & A_2+3A_1 & 37 & 7 & 44 & \left\{\frac{237476949}{1527752},\frac{187894873}{1145814}\right\} & \left\{\frac{30711077}{190969},\frac{407681569}{2291628}\right\}  & \frac{180}{437} \\
D_5+A_1 & D_5 & 61 & 60 & 121 & \left\{\frac{1760291}{14700},\frac{948133}{7350}\right\} & \left\{\frac{903893}{7350},\frac{519934}{3675}\right\}  & \frac{32}{105} \\
D_6\left(a_1\right) & D_5\left(a_1\right) & 62 & 30 & 92 & \left\{\frac{1553}{12},\frac{6655}{48}\right\} & \left\{\frac{6385}{48},\frac{7271}{48}\right\} & \frac{1}{3} \\
D_6 & D_4+A_1 & 110 & 29 & 139 & \left\{\frac{25707707}{218886},\frac{27750643}{218886}\right\} & \left\{\frac{52819073}{437772},\frac{60990817}{437772}\right\}  & \frac{172}{573} \\
E_8\left(b_5\right) & D_4\left(a_1\text{)+}A_1\right. & 160 & 13 & 173 & \left\{\frac{49357}{432},\frac{26681}{216}\right\} & \left\{\frac{25463}{216},\frac{7367}{54}\right\} &  \frac{8}{27} \\
\hline
\end{array}$}} $\cdots$
\end{adjustbox}
\begin{adjustbox}{left}
\scalebox{0.85}{{\renewcommand{\arraystretch}{1.2} $\begin{array}{|c|c|c|c|c|c|c|c|c|}
\hline
\text{[B-C]}_L & \text{[B-C]}_R & t_* & \text{Min(}\Delta _{\text{IR}}\text{(Z's))} & \text{Min(}\Delta_{\text{IR}}\left(\cO_L\text{'s))}\right. & \text{Min(}\Delta _{\text{IR}}\left(\cO_R\text{'s))}\right. \\
\hline
0 & 0 & \frac{2}{3} & \numprint{6.} & \numprint{2.} & \numprint{2.} \\
A_4+A_2+A_1 & A_4+2A_1 & \frac{20}{51} & \numprint{3.52941} & \numprint{1.} & \numprint{1.} \\
D_5\left(a_1\right) & A_2+3A_1 & \frac{32}{75} & \numprint{3.84} & \numprint{1.} & \numprint{1.08} \\
D_5\left(a_1\text{)+}A_2\right. & A_3+A_2+A_1 & \frac{20}{51} & \numprint{3.52941} & \numprint{1.} & \numprint{1.} \\
E_6\left(a_3\text{)+}A_1\right. & A_2+3A_1 & \frac{180}{437} & \numprint{3.70709} & \numprint{1.} & \numprint{1.14645} \\
D_5+A_1 & D_5 &  \frac{32}{105} & \numprint{2.74286} & \numprint{1.} & \numprint{1.} \\
D_6\left(a_1\right) & D_5\left(a_1\right) & \frac{1}{3} &  \numprint{3.} & \numprint{1.} & \numprint{1.} \\
D_6 & D_4+A_1 & \frac{172}{573} & \numprint{2.70157} & \numprint{1.} & \numprint{1.} \\
E_8\left(b_5\right) & D_4\left(a_1\text{)+}A_1\right. & \frac{8}{27} & \numprint{2.66667} & \numprint{1.} & \numprint{1.22222} \\
\hline
\end{array}$}}
\end{adjustbox}
\caption{Flipper field deformations of ($E_{8}$, $E_{8}$) conformal matter, only rational values.}%
\label{MS-CM-E8}%
\end{table}

\newpage

\bibliographystyle{utphys}
\bibliography{4DRGNilpotent}

\end{document}